\documentclass[a4paper,10pt]{article}
\usepackage{graphicx}
\usepackage{amssymb}
\def\Vec#1{\mbox{\boldmath $#1$}}
\begin{document}
\begin{titlepage}
\begin{flushright}
October, 2005 \\
\medskip
hep-th/0510252
\end{flushright}
\begin{center}
\vspace{30mm}
{\Large
\textbf{$Sp(4,\Vec{H}) / \Vec{Z}_2$ \,Pair Universe in $E_6$ Matrix Models}
}\\
\vspace{20mm}
{\large
Yuhi Ohwashi
\footnote{E-mail:\,\, y.ohwashi@fuji.waseda.jp}
}\\
\vspace{10mm}
{\itshape
Sony Institute of Higher Education, Shohoku College \\
428 Nurumizu, Atsugi-city, Kanagawa 243-8501, Japan,
\\ \medskip
Tokyo Metropolitan College of Aeronautical Engineering \\
8-52-1 Minami-senju, Arakawa-ku, Tokyo 116-8523, Japan,
\\ \medskip
Department of Radiological Sciences,
Ibaraki Prefectural University of Health Sciences \\
4669-2 Ami, Ami-machi, Inashiki-gun, Ibaraki 300-0394, Japan
}\\
\vspace{5mm}
\end{center}
\begin{abstract}
We construct interacting $Sp(4,\Vec{H})/\Vec{Z}_2$ {\itshape pair} matrix models inside the {\itshape compact} $E_6$ matrix models.
Generally, models based on the {\itshape compact} $E_6$ seem to always include doubly the degrees of freedom that we need physically.
In this paper, we propose one solution to this problem.
A basic idea is that: `we regard that each point of space-time corresponds to the {\itshape center of projection} of two {\itshape fundamental figures} (i.e.\,two internal structures), and assume that the {\itshape projection} of these fundamental figures from each point possesses one transformation group as a whole'.
Namely, we put emphasis on the `analogy' with the projective geometry.
Given that the whole symmetry is {\itshape compact} $E_6 \times Gauge$, the space $( \Vec{\mathfrak{J}_H} \oplus i \Vec{H^3} ) \times \Vec{\mathcal{G}}$ is promising as two subspaces seen from such a viewpoint.
When this situation is seen from the standpoint of {\itshape Klein's Erlangen Program}, each fundamental figure should also have an independent transformation group.
The symmetry corresponding to this is $Sp(4,\Vec{H})/\Vec{Z}_2 \times Gauge$.
This is ensured by the introduction of the Yokota mapping $\mathcal{Y}$.
As a consequence, we result in the picture of interacting {\itshape pair} universes which are being $\frac{\pi}{2}[rad]$-phase-shifted.
This picture is applicable to all the models based on the {\itshape compact} $E_6$, which may be not only matrix models but also field theories.
An interacting bi-Chern-Simons model is provided when this result is applied to our previous matrix model~\cite{Ohwashi:2001mz}.
This paper is one answer of the author to the doubling problem which has been left in the previous paper~\cite{Ohwashi:2001mz}.
\end{abstract}
\end{titlepage}
%
%
%
\section{Introduction}

In this paper, we construct interacting $Sp(4,\Vec{H})/\Vec{Z}_2$ {\itshape pair} matrix models inside the {\itshape compact} $E_6$ matrix models.
This is one solution of the author for the doubling problem of the degrees of freedom that the models based on the {\itshape compact} $E_6$ have.
To put it briefly, as the simplest solution, we will accept existence of two structures of the universe.
By this assumption, we find that the theories based on the {\itshape compact} $E_6$ are divided into two portions which have structure of completely the same fields.
If each universe is viewed as an independent thing, they constitute multiplets of $Sp(4,\Vec{H})/\Vec{Z}_2$.
Of course, since we consider matrix models, there is another gauge symmetry for the $N \times N$ matrices.
 
Einstein gravity is a low-energy effective theory, because it is not perturbatively renormalizable.
If the energy scale is raised, many irrelevant operators ought to contribute.
Therefore, there is no definite promise that the unified theory at the fundamental scale is surely described with the Riemannian geometry itself.
So we would like to advance an argument paying attention to the `analogy' with the {\itshape projective geometry}.
In the author's viewpoint, the mathematics which the universe should possess in the fundamental scale is the geometry similar to the projective geometry.
Of course the projective geometries have been classified mathematically and there is no {\itshape pure} projective geometry that has the {\itshape compact} $E_6$ as its projective transformation group.
Our idea is that: `with an analogy to the projective geometry in mind, we will take a look at the doubling problem.'
Fortunately, since the treatment of the {\itshape information geometry} is possible to the symplectic groups, there is a possibility that we can discuss the metric geometry in the space of matrix-eigenvalues.
Therefore, it is not necessary that the diffeomorphism is directly introduced into the action itself explicitly.
The author cannot say anything because opinions may vary from the researcher to the researcher. 
After all, however, all theories other than the standard model are still theories of conjecture now.
Therefore, as one trial to make a breakthrough in this situation, the author thinks that there may be an approach like this paper.

Today, the low-energy physics is described well by the standard model as a gauge theory. However, this model is incomplete because gravity is not included.
Therefore, the most important subject of current particle physics is to construct a consistent quantum theory incorporating all interactions.
What has to be noticed is the method of formulization for the construction of the unified theory.
Traditionally, physics has used the perturbative formulation willingly.
However, the construction of the theory along the line of perturbative formulation is essentially defective in the sense that physical quantities are represented by the asymptotic series. 
The final theory must be constructed by the non-perturbative formulation from the beginning.
Then, what kind of mathematical expression will the non-perturbative theory be described by?
Probably, it will be good to recall the constructive definition of the field theory by Wilson in order to answer this.
The essence is that the space-time is seen from a viewpoint of combinatorial analysis.
Another example can be seen in the twistor theory by Penrose.
That the theory is described along the line of combinatorial method means that the physical objects have countability.
Therefore, the simplest method to describe nature will be the representation which uses matrices.
From such reason, we pay attention to the {\itshape matrix model} as the formulization of the theory.
In recent years, many studies have been made on matrix models as the constructive definition of string theory. However, the author thinks that there is no need to relate the matrix model to string theory alone because of the reason mentioned above.
Of course, considering a success that the topology expansion of the $2$D gravity was able to be expressed with the double scaling limit of the $\frac{1}{N}$ expansion of the matrix model, it is reasonable to suppose that a corresponding matrix model exists because $10$D superstring theory can be actually expressed in the form of $2$D supergravity.
On the other hand, however, there is a fact that the matrix model is also useful in different fields like loop quantum gravity.
In fact, the matrix model can fill the role of the constructive definition of non-commutative geometry.
In a broad sense, it is playing a linguistic role for describing physics.

Smolin proposed a matrix model based on the real Jordan algebra $\Vec{\mathfrak{J}}$ as one candidate of the unified theory~\cite{Smolin:2001wc}.
His model has been founded on the exceptional Lie group $F_4$ and the trilinear form, which is the invariant of $F_4$, was adopted for the construction of the model.
However, the author pointed out that the complex structure cannot be introduced into his model, and attempted to use the simply connected {\itshape compact} exceptional Lie group $E_6$ instead~\cite{Ohwashi:2001mz}\footnote{Recently, Castro has had a very interesting discussion using quartic $E_7$, $E_8$ invariants further~\cite{Castro:2005pc}. A relevance to the Finslerian-like geometry has been also mentioned. The further development of this direction is expected.}.
The {\itshape cubic form}, which is the $E_6$ invariant, plays an important role in constructing the model based on $E_6$.
It was shown that the matrix Chern-Simons type theory can be derived as the result even when exceptional Lie group $E_6$ is used.
Furthermore, it was shown that {\itshape the compactness condition} of $E_6$ implies {\itshape the postulate of positive definite metric} of our model.
This previous model based on $E_6$ will be introduced briefly in the next section.
We would like to point out that the method using single exceptional Jordan algebra is applicable only to the $F_4$ and $E_6$\footnote{Since $E_7$ and $E_8$ are large, the introduction of spaces like the Freudenthal $\Vec{C}$-vector space $\Vec{\mathfrak{F}^c} = \Vec{\mathfrak{J}^c} \oplus \Vec{\mathfrak{J}^c} \oplus \Vec{C} \oplus \Vec{C}$ is indispensable.}.
The following summarizes the definitions of some groups of type $F_4$ and $E_6$.
\begin{eqnarray}
F_4 &=& \{ \alpha \in Iso_R ( \Vec{\mathfrak{J}} , \Vec{\mathfrak{J}} ) \, | \, \  tr ( \alpha A , \alpha B , \alpha C ) = tr ( A , B , C ) \, , \, ( \alpha A , \alpha B ) = ( A , B ) \} \  , \nonumber \\
F_4{}\Vec{^c} &=& \{ \alpha \in Iso_C ( \Vec{\mathfrak{J}^c} , \Vec{\mathfrak{J}^c} ) \, | \, \  tr ( \alpha X , \alpha Y , \alpha Z ) = tr ( X , Y , Z ) \, , \, ( \alpha X , \alpha Y ) = ( X , Y ) \} \  , \nonumber
\end{eqnarray}
\begin{eqnarray}
E_{6(-26)} &=& \{ \alpha \in Iso_R ( \Vec{\mathfrak{J}} , \Vec{\mathfrak{J}} ) \, | \, \  ( \alpha A , \alpha B , \alpha C ) = ( A , B , C ) \} \  , \nonumber \\
compact \ E_6 &=& \{ \alpha \in Iso_C ( \Vec{\mathfrak{J}^c} , \Vec{\mathfrak{J}^c} ) \, | \, \  ( \alpha X , \alpha Y , \alpha Z ) = ( X , Y , Z ) \, , \, \langle \alpha X , \alpha Y \rangle = \langle X , Y \rangle \} \  , \nonumber \\
E_6{}\Vec{^c} &=& \{ \alpha \in Iso_C ( \Vec{\mathfrak{J}^c} , \Vec{\mathfrak{J}^c} ) \, | \, \  ( \alpha X , \alpha Y , \alpha Z ) = ( X , Y , Z ) \} \  . \nonumber
\end{eqnarray}
Here, $tr ( \ast , \ast , \ast )$ represents the {\itshape trilinear form}, and $( \ast , \ast , \ast )$ represents the {\itshape cubic form}. 
We cannot define $E_6$ using trilinear form only.
$E_6$ must be defined by cubic form.
The cubic form is very different from the trilinear form, and they are given explicitly in the Appendix.
The complex exceptional Jordan algebra $\Vec{\mathfrak{J}^c}$ is the complexification of the exceptional Jordan algebra $\Vec{\mathfrak{J}}$.
The complexification of the exceptional Lie group should not be confused with that of the exceptional Jordan algebra.

As an advantage to use $E_6$ instead of $F_4$, there are the following points other than the introduction of the complex structure mentioned above.
For example, in $E_6$ model, there is no necessity of the change of variables in order to make the theory of the Chern-Simons type.
The Chern-Simons type action can be directly derived from the invariant on $E_6$. 
Moreover, there is an advantage of containing $Spin(10)$ in $E_6$.
In addition, $E_6$ is interesting also from the viewpoint of phenomenology.
As mentioned in the previous paper~\cite{Ohwashi:2001mz}, $E_6$ is considered to be a good group in the sense that the final theory must contain the standard model.
Furthermore, what must not be forgotten is that the compactness condition of $E_6$ automatically implies a positive definite metric for our model~\cite{Ohwashi:2001mz}.
What is called {\itshape the postulate of positive definite metric} is, from the viewpoint of our model, merely an immediate consequence of the fact that our universe is {\itshape compact}.

On the other hand, there are problems in the model based on the {\itshape compact} $E_6$.
Main problems are two: one is the doubling problem of the degrees of freedom, and another is the interpretation problem of the fermion.
In this paper, the author would like to propose one interpretation about the doubling problem of the degrees of freedom.
Another one problem about the fermion is that the anticommuting c-numbers cannot be seen apparently when the action was expanded and expressed by the elements of $\Vec{\mathfrak{J}^c}$. 
In the stage of the complex exceptional Jordan algebra $\Vec{\mathfrak{J}^c}$ itself before the expanding by coefficients of the elements, there is a possibility of realization of the fermion commutation relation in $\Vec{\mathfrak{J}^c}$ algebraically as has been pointed out by Rios~\cite{Rios:2005pk}\footnote{See also \cite{Catto:2003hb}. Rios has pointed out an interesting relevance with the twistor string theory further~\cite{Rios:2005mp}.}.
However, when we take the position that the action is constructed by using the invariant of a certain group, the cubic form expanded by fields is described by the scalars only and there is no anticommuting c-number.
Therefore, we cannot define the formal functional integral to the fermion.
Of course this problem will not arise if we suppose the existence of the complex exceptional Jordan {\itshape super}-algebra.
However, the Jordan {\itshape super}-algebra is extremely artificial algebra.
The author is still hesitant about taking the plunge into that direction.
One possibility is to assume that the approach of the quantum Hall effect is useful.
Another possibility is that the definition of the fermion itself needs to be reexamined.

The reader might think that $E_6$ is the group for phenomenology and that it is not a symmetry for matrix models which deal with the space-time together.
The author, however, thinks that this view is incorrect.
The comparison of the supersymmetry between field theories and matrix models is a good example to illustrate this point.
Usually, when we consider the field theory, the space-time is given from the beginning.
In this case, therefore, the supersymmetry means the exchange of bosons for fermions.
In the matrix model, however, the space-time itself is embedded into the matrix.
Therefore, if a certain matrix model possesses the supersymmetry, it is more precise to say that the symmetry which exchanges the space-time for the matter exists\footnote{Recall the supersymmetry of the IIB matrix model as an example.}.
That is, the meaning of supersymmetry is upgraded in the matrix model.
Therefore, an important point to emphasize is that {\itshape if the $E_6$ symmetry exists behind the matter it must also exist behind the space-time}.
\begin{center}
\begin{tabular}{|l||l|}
\hline
SUSY of Field Theory & Exchange of Bosons for Fermions \\
\hline
SUSY of Matrix Model & Exchange of Space-time (and Bosons) for Matter \\
\hline
\end{tabular}
\end{center}

After being argued about the $F_4$ model~\cite{Smolin:2001wc} by Smolin, and our model~\cite{Ohwashi:2001mz} based on $E_6$, Ramond argued about a similar concept from the viewpoint of supergravity~\cite{Ramond:2001ud}.
Ramond paid attention to the coset $\Vec{\mathfrak{C}}P_2 = F_4/Spin(9)$ of $F_4$ in his paper.
Of course it is also possible to consider this coset in Smolin's $F_4$ model.
In order to realize it, what is necessary is just to add the following constraint to the fields $J$ in Smolin's $F_4$ model.
\begin{eqnarray}
\Vec{\mathfrak{C}}P_2 &=& \{ J \in \Vec{\mathfrak{J}} \, | \, \  J^2 = J , \ tr(J) = 1 \}
\end{eqnarray}
It is known from of old that Jordan type algebras can define the projective spaces (See Appendix~\ref{KP_m}. ).
Because there is no {\itshape pure} projective space which contains the Graves-Cayley projective plane $\Vec{\mathfrak{C}}P_2$, the research of this space itself is very interesting in the sense that it is the highest dimensional space employing $\Vec{\mathfrak{C}}$. 
However, the projective transformation group of $\Vec{\mathfrak{C}}P_2$ is not $F_4$ but $E_{6(-26)}$.
Therefore, in order to make the theory `closed', we have to construct the $E_{6(-26)}$ model by using the {\itshape cubic form} instead of the {\itshape trilinear form}.

Now, because we consider {\itshape compact} $E_6$, of course, different analyses of the theory from Smolin or Ramond are needed.
The point to observe is the relevance between the cubic form, which is the $E_6$ invariant, and the geometry. 
As mentioned in the previous paper, the existence of an unknown geometry similar to the projective geometry can be seen off and on behind the Freudenthal multiplication and the cubic form.
We will take up this point a little in the section \ref{geometry}.
The number on which we focus attention in this paper is the complex Graves-Cayley algebra $\Vec{\mathfrak{C}^c}$.
The first scholar to pay attention to real-octonions was Graves.
However, it was not accepted in people because of the non-associativity.
Afterward, the real-octonions were rediscovered by Cayley (See \cite{Zahlen:1951gc} for example.).
The {\itshape complex} Graves-Cayley algebra $\Vec{\mathfrak{C}^c}$, which we employ in our model, is the complexification of this Graves-Cayley algebra $\Vec{\mathfrak{C}}$.
We use the complex-octonion as the fundamental number.
The introduction of the complex-octonions is extremely essential when dealing with the exceptional Lie groups of type $E_6$, $E_7$, $E_8$.
We would like to note that this $\Vec{\mathfrak{C}^c}$ contains the quaternion field in four-fold as a clue which solves the doubling problem of the {\itshape compact} $E_6$.
This idea is the basics of this paper\footnote{In the final section, we will point out that the same prescription is applicable to the split-quaternions formally.}.
The greatest advantage to make a complex-octonion result in four quaternions is that they own the associativity and it becomes easy to deal with them.
This is not to say that the essential mathematics behind our universe is not non-associative algebra.
The author thinks that the non-associative algebra offers the key to an understanding of the universe.
As the result of that, the picture of two internal structures of the universe is automatically produced as seen in this paper.

This article is organized as follows.
In the next section we briefly review our previous model based on the {\itshape compact} $E_6$ and point out the problems which this model possesses.
The problems discussed here are common to all the models based on the {\itshape compact} $E_6$.
Then, in section 3, our strategy for solving the doubling problem of the degrees of freedom is discussed.
In Accordance with the `analogy' to the projective geometry, we will adopt a solution which follows the Klein's Erlangen Program.
In section 4, a concrete example of our idea is presented.
As a consequence, we result in the picture of interacting $Sp(4,\Vec{H})/\Vec{Z}_2$ {\itshape pair} universe. 
This picture is applicable to all the models based on the {\itshape compact} $E_6$, which may be not only matrix models but also field theories.
In section 5, this picture is applied to our previous matrix model.
The resulting theory is the interacting bi-Chern-Simons model.
This result is interesting because it resembles the bi-Chern-Simons gravity~\cite{Borowiec:2003rd}.
In section 6, the author would like to dare to mention about the unknown geometry.
Although an argument here is open issue and completely incomplete, it offers the way of one approach toward the unknown geometry.
The final section is devoted to the conclusions and discussions.
One answer of the author to the doubling problem is stated.
Also, it is pointed out that the same prescription is applicable to the split-quaternions $\Vec{H'}$ formally.
In the Appendix, elementary descriptions of the complex exceptional Jordan algebra $\Vec{\mathfrak{J}^c}$ are presented, which are needed for this paper.
If the reader has a great interest in the exceptional linear Lie groups further, a series of excellent reports presented by I.~Yokota (\cite{Yokota:1990e6},\cite{Yokota:1990e7},\cite{Yokota:1991e8}) are recommended.

Today, many researchers are content with the studies of five string theories, or the studies of effective theories of M-theory.
The researcher who grope for the possibility of a new theory is extremely few.
However, string theories are the theories asymptotically expanded around the specific backgrounds of what is called the unknown M-theory, and the $11$D supergravity is a `low-energy' effective theory of the M-theory.
One can never restore the original theory from the theory which is represented by the asymptotic expansion because of the lack of one-to-one correspondence.
Also, one can never restore the original theory from the effective theory which is not renormalizable because of the information loss.
Therefore, these are the theories which are far from the true theory.
That is, the final theory must be the theory which has more physical degrees of freedom.
Since it is not known how the future is opened, the author hopes freer standpoints of the researchers are allowed a little.

\section{A model based on the {\itshape compact} $E_6$, and its main problems}

In this section, we briefly review our previous model~\cite{Ohwashi:2001mz} based on the simply connected {\itshape compact} exceptional Lie group $E_6$.
An effective action of the matrix string type can be derived from this model, as mentioned in the previous paper. 
Recently, Castro has pointed out that there is a correspondence between this model and Chern-Simons branes in the large $N$ limit and that it is invariant under volume-preserving reparametrizations of the $3$D world-volume which leave invariant the Nambu-Poisson brackets~\cite{Castro:2005pc}.

In the previous paper, we adopted a following definition for the simply connected {\itshape compact} exceptional Lie group $E_6$ on the occasion of the construction of our model:
\begin{eqnarray}
E_6 = \{ \alpha \in Iso_C ( \Vec{\mathfrak{J}^c} , \Vec{\mathfrak{J}^c} ) \  | \  \  ( \alpha X , \alpha Y , \alpha Z ) = ( X , Y , Z ) \  , \  \langle \alpha X , \alpha Y \rangle = \langle X , Y \rangle \}\label{E6_def1} \  ,
\end{eqnarray}
where \,$( X , Y , Z ) \  \  ( X,Y,Z \in \Vec{\mathfrak{J}^c} )$ is the {\itshape cubic form}.
The second condition, $\langle \alpha X , \alpha Y \rangle = \langle X , Y \rangle$, is added to the definition in order to make the group {\itshape compact}. 

We then defined a model by the action
\begin{eqnarray}
S &=& \Bigl( \  \mathcal{P}^2 ( \mathcal{M}^{[A} ) \  , \  \mathcal{P} ( \mathcal{M}^B ) \  , \  \mathcal{M}^{C]} \  \Bigr) \  \  f_{ABC} \  \  \  , \label{eq:E_6}
\end{eqnarray}
where the $\mathcal{M}^A$ are elements of the complex exceptional Jordan algebra $\Vec{\mathfrak{J}^c}$, \,$\mathcal{P} ( \mathcal{M} )$ is the cycle mapping of $\mathcal{M}$, \,and the $f_{ABC}$ are the structure constants of $\Vec{\mathcal{G}}$ which is the Lie algebra of the gauge group for the $N \times  N$ matrices. The notation $[ \cdots ]$ denotes the {\itshape weight}-1 anti-symmetrization on indices. The degrees of freedom of this model live in $\Vec{\mathfrak{J}^c} \times \Vec{\mathcal{G}}$. The specific components of $\mathcal{M}^A \in \Vec{\mathfrak{J}^c}$ are written as follows: 
\begin{eqnarray}
\mathcal{M}^A
&=&
\left(
      \begin{array}{ccc}
      \mathcal{A}_1{}^A & \varphi_3{}^A & \bar{\varphi}_2{}^A \\
      \bar{\varphi}_3{}^A & \mathcal{A}_2{}^A & \varphi_1{}^A \\
      \varphi_2{}^A & \bar{\varphi}_1{}^A & \mathcal{A}_3{}^A
      \end{array}
\right) \label{eq:components-1} \\
& &
\quad\quad\quad \mathcal{A}_I{}^A \in \Vec{C} \  , \  \  \varphi_I{}^A \in \Vec{\mathfrak{C}^c} \  , \quad (I=1,2,3) \nonumber \\
&\equiv&
\mathcal{M}^A(\mathcal{A},\varphi) \  .
\end{eqnarray}
Here, the $\mathcal{A}_I{}^A$ are complex numbers, and the $\varphi_I{}^A$ are elements of the complex Graves-Cayley algebra.

After decomposing the action in terms of the variables defined in~{(\ref{eq:components-1})}, we get 
\begin{eqnarray}
S &=& \frac{1}{4} \  f_{ABC} \  \epsilon^{IJK} \  \mathcal{A}_I{}^A \mathcal{A}_J{}^B \mathcal{A}_K{}^C + \frac{3}{2} \  f_{ABC} \  \epsilon^{IJK} \  \mathcal{A}_I{}^A ( \varphi_J{}^B , \varphi_K{}^C ) \nonumber \\
& & {} - 3 \  f_{ABC} \  \Vec{Re^c} ( \varphi_3{}^A \varphi_2{}^B \varphi_1{}^C ) + f_{ABC} \sum_{I=1}^{3} ( \  \Vec{Re^c} ( \varphi_I{}^A \varphi_I{}^B \varphi_I{}^C ) \  ) \label{S:1}\\
&=& \frac{1}{4} \  f_{ABC} \  \epsilon^{IJK} \  \mathcal{A}_I{}^A \mathcal{A}_J{}^B \mathcal{A}_K{}^C + \frac{3}{4} \  f_{ABC} \  \epsilon^{IJK} \  \mathcal{A}_I{}^A ( \bar{\varphi}_J{}^B \varphi_K{}^C + \bar{\varphi}_K{}^C \varphi_J{}^B ) \nonumber \label{S:2}\\
& & {} - \frac{3}{2} \  f_{ABC} \  ( \varphi_3{}^A (\varphi_2{}^B \varphi_1{}^C) + (\bar{\varphi}_1{}^C \bar{\varphi}_2{}^B) \bar{\varphi}_3{}^A ) \nonumber \\
& & {} + \frac{1}{2} \  f_{ABC} \sum_{I=1}^{3} ( \varphi_I{}^A (\varphi_I{}^B \varphi_I{}^C) + (\bar{\varphi}_I{}^C \bar{\varphi}_I{}^B) \bar{\varphi}_I{}^A ) \  .
\end{eqnarray}
The first term of this action is identical to the matrix Chern-Simons theory. What should be noted here is that: although the action was originally constructed from the {\itshape algebraic} invariant on the $E_6$ mapping (i.e. the cubic form), it automatically has a Chern-Simons type form. Therefore, it is quite likely that one of the properties of the theory based on $E_6$ is that it is {\itshape topological}.

If we introduce the following notation:
\begin{eqnarray}
\Vec{\mathcal{A}}_I = \mathcal{A}_I{}^A \  \mathbf{T}_A \  , \quad \Vec{\varphi}_{0I} = \varphi_{0I}{}^A \  \mathbf{T}_A \  , \quad \Vec{\varphi}_{iI} = \varphi_{iI}{}^A \  \mathbf{T}_A \  , \label{notation}
\end{eqnarray}
these quantities enable us to write the equations of motion of this theory as 
\begin{eqnarray}
\left\{
\begin{array}{r}
\epsilon^{IJL} \  \Bigl( \  \frac{1}{2} \  [ \Vec{\mathcal{A}}_I , \Vec{\mathcal{A}}_J ] + [ \Vec{\varphi}_{0I} , \Vec{\varphi}_{0J} ] + [ \Vec{\varphi}_{iI} , \Vec{\varphi}_{iJ} ] \  \Bigr) = 0 \\
\epsilon^{IJL} \  \Bigl( \  2 \  [ \Vec{\mathcal{A}}_I , \Vec{\varphi}_{0J} ] + [ \Vec{\varphi}_{0I} , \Vec{\varphi}_{0J} ] - [ \Vec{\varphi}_{iI} , \Vec{\varphi}_{iJ} ] \  \Bigr) = 0 \\
\epsilon^{IJL} \  \Bigl( \  [ \Vec{\mathcal{A}}_I , \Vec{\varphi}_{lJ} ] - [ \Vec{\varphi}_{0I} , \Vec{\varphi}_{lJ} ] \  \Bigr) + \sigma_{ijl} \  \Bigl( \  [ \Vec{\varphi}_{i(L+1)} , \Vec{\varphi}_{j(L+2)} ] - [ \Vec{\varphi}_{iL} , \Vec{\varphi}_{jL} ] \  \Bigr) = 0
\end{array}
\right. \  .
\end{eqnarray}
In the last equation, the index $L$ \,is $\Vec{mod \  3}$, \,and the summation convention is not used with respect to this $L$.
Also, by the introduction of the notation~{(\ref{notation})}, we can express the action~{(\ref{S:2})} as follows:
\begin{eqnarray}
S &=& - \frac{3i}{2} \  \epsilon^{IJK} \  tr \Bigl( \  \frac{1}{3} \Vec{\mathcal{A}}_I [ \Vec{\mathcal{A}}_J , \Vec{\mathcal{A}}_K ] + 2 \Vec{\varphi}_{\tilde{i}I} [ \Vec{\mathcal{A}}_J , \Vec{\varphi}_{\tilde{i}K} ] \  \Bigr) \nonumber \\
& & {} + 6i \  tr \Bigl( \  \Vec{\varphi}_{01} [ \Vec{\varphi}_{03} , \Vec{\varphi}_{02} ] - \Vec{\varphi}_{i1} [ \Vec{\varphi}_{03} , \Vec{\varphi}_{i2} ] - \Vec{\varphi}_{i1} [ \Vec{\varphi}_{i3} , \Vec{\varphi}_{02} ] - \Vec{\varphi}_{01} [ \Vec{\varphi}_{i3} , \Vec{\varphi}_{i2} ] \nonumber \\
& & {} \qquad\quad\quad - \sigma_{ijk} \  ( \Vec{\varphi}_{i1} [ \Vec{\varphi}_{j3} , \Vec{\varphi}_{k2} ] ) + \frac{1}{3} \sigma_{ijk} \sum_{I=1}^{3} ( \Vec{\varphi}_{iI} [ \Vec{\varphi}_{jI} , \Vec{\varphi}_{kI} ] ) \  \Bigr) \  , \label{eq:E_6decompose}
\end{eqnarray}
where $( \tilde{i} = 0,\cdots,7 )$ and $( i = 1,\cdots,7 )$.
Namely, this model has a property that a Chern-Simons type action is coupling with other fields automatically.
It constitutes one multiplet based on $E_6$ as a whole.
This model possesses the following symmetries: the global $E_6$ symmetry resulting from the cubic form, the gauge symmetry resulting from $f_{ABC}$, the cycle mapping $\mathcal{P}$ with respect to the fields, and the matrix translation symmetry with respect to the diagonal parts of the fields.

In addition, an important point to emphasize is the existence of the {\itshape compactness condition} of $E_6$, $\langle \alpha X , \alpha Y \rangle = \langle X , Y \rangle$.
As has been discussed in the previous paper~\cite{Ohwashi:2001mz}, due to this compactness condition, one invariant under the $E_6$ mapping is introduced into the theory.
\begin{eqnarray}
invariant_{E_6 \times Gauge} &=& \langle \mathcal{M}^A , \mathcal{M}\Vec{'}^A \rangle \quad \in \  \Vec{C} \\
&=& tr_{N \times N} \Bigl( \  ( \mathcal{A}_I^\ast{}^A \mathcal{A}\Vec{'}_I{}^B + 2 \varphi_{\tilde{i}I}^\ast{}^A \varphi\Vec{'}_{\tilde{i}I}{}^B ) \  \mathbf{T}_A \mathbf{T}_B \  \Bigr) \  , \\
& & \nonumber \\
\langle \mathcal{M}\Vec{'}^A , \mathcal{M}^A \rangle &=& \langle \mathcal{M}^A , \mathcal{M}\Vec{'}^A \rangle {}^\ast \  . \label{eq:invariant}
\end{eqnarray}
This invariant is, in general, a complex number. The expression~{(\ref{eq:invariant})} indicates that $\langle \mathcal{M}\Vec{'}^A , \mathcal{M}^A \rangle$ and $\langle \mathcal{M}^A , \mathcal{M}\Vec{'}^A \rangle$ are complex conjugates of each other.
What is especially important is the case of $\mathcal{M}\Vec{'}^A = \mathcal{M}^A$ as follows:
\begin{eqnarray}
invariant_{E_6 \times Gauge} &=& \langle \mathcal{M}^A , \mathcal{M}^A \rangle \quad \in \  \Vec{R} \\
&\ge& 0 \  .
\end{eqnarray}
The crucial point to observe here is that {\itshape this invariant satisfies a positivity condition}: \,in general $\langle \mathcal{M}^A , \mathcal{M}^A \rangle \ge 0$, and it vanishes if and only if $\mathcal{M}^A = 0$. Although this invariant is the quantity defined on the product space $\Vec{\mathfrak{J}^c} \times \Vec{\mathcal{G}}$, the resulting structure is, in a sense, quite similar to that of the physical Hilbert space.
Since the internal rotational symmetry caused by $E_6$ mappings exists, this space is, as it were, the Hilbert space which is embedding a kind of spinor structure. 
In the matrix model, the space-time itself is embedded into the matrix.
Therefore, if we take the position that the physical Hilbert space is the {\itshape product space} composed of two parts $\Vec{\mathfrak{J}^c}$ and $\Vec{\mathcal{G}}$, what is called {\itshape the postulate of positive definite metric} is merely an immediate consequence of the fact that our universe is {\itshape compact}.

Moreover, this model has some interesting features.
For example, the theory expanded around one of the classical solutions of this model has a effective action similar to the matrix string theory.
However, there are two main problems in the models based on the {\itshape compact} $E_6$, as mentioned before.
One is the doubling problem of the degrees of freedom, and another is the problem concerning how to include anticommuting c-numbers in the theory.
Especially the former problem is crucial.
Because the cubic form itself has very interesting properties, we have an impulse to discuss them immediately.
However, even if we advance the analysis of the $E_6$ model being left the doubling problem, we will have to return to this problem finally after all.
Therefore, in analyzing the theory, we consider that tackling about this doubling problem first is the top priority.
The following is devoted to one answer of the author to this problem.

\section{Strategy}

In this section, we will discuss our strategy to solve the doubling problem.
We proceed with our arguments on the premise that the true theory have the {\itshape compact} $E_6 \times Gauge$ symmetry.
It seems that this is a valid supposition because $E_6$ is not inconsistent with the standard model.
Under this supposition, we narrow down the answer with the highest probability as the actual universe for this problem, taking account of the existing circumstantial evidences and the demands which is considered to be appropriate.
That is, the `elimination' is adopted.
We consider that what remains at the end will probably be near the truth, after erasing undesirable things.
Although the picture of the universe which we finally choose in this paper is the $( \Vec{\mathfrak{J}_H} \oplus i \Vec{H^3} ) \times \Vec{\mathcal{G}}$ pair universe, some considerations which performed up to the final conclusion are briefly introduced here.

\begin{quotation}
\ \ [{\itshape Consideration} $1$]

$E_6$ contains $Spin(10)$.
Because the critical dimension of the superstring theory is $10$, we have an impulse to adopt a background with respect to $Spin(10)$ immediately.
However, our main purpose is to give an insight to the doubling problem of the degrees of freedom.
Granted that we left this doubling problem unsolved and advance the analysis of the model, we will certainly face it later.
If fields as fluctuations expanded around a background remain after all as garbage being two-fold degrees of freedom, it will be a total loss for us.
Of course one may say that such unnecessary fields are path-integrated out first.
However, the resulting theory is an effective theory.
We are seeking for a solution which has a good visibility in the level of the fundamental scale.
Therefore, it seems that we had better investigate other possibility.
\end{quotation}

\begin{quotation}
\ \ [{\itshape Consideration} $2$]

Additionally, we cannot throw away the impression that the adoption of $Spin(10)$ implies the space-structure which is limited to only the {\itshape flat} from the beginning\footnote{This might be over-considering. If the diffeomorphism is realized in the space of matrix-eigenvalues, it should not matter at all even if the symmetry of the action itself is a flat. Besides, it has been pointed out that a possibility of the introduction of the diffeomorphism exists by changing the interpretation of the matrices to the differential operators on a commutative space~\cite{Hanada:2005vr}.}.
Since we are of course searching for the theory including gravity, we would like to leave a symmetry other than simple flat spaces for the action if we can. 
For that purpose, it seems that it is more realistic to pursue structures other than $Spin(10)$.
\end{quotation}

\begin{quotation}
\ \ [{\itshape Consideration} $3$]

Next, we have an impulse to consider the $\Vec{\mathfrak{C}^c}P_2$ in $E_6$, as Ramond considered the coset $\Vec{\mathfrak{C}}P_2 \simeq F_4 / Spin(9)$ in $F_4$.
\,It is known mathematically that $\Vec{\mathfrak{C}^c}P_2$ is the following:
\begin{eqnarray}
\Vec{\mathfrak{C}^c}P_2 \simeq E_6 / ( ( U(1) \times Spin(10) ) / Z_4 ) \  .
\end{eqnarray}
That is, $\Vec{\mathfrak{C}^c}P_2$ is the Hermitian symmetric space of type $E_6$ and it is the complexification of the Graves-Cayley projective plane.
It is very interesting to analyze this space\footnote{The attraction of this space has been pointed out also in~\cite{Castro:2005pc}.}.
However, as mentioned in the previous paper~\cite{Ohwashi:2001mz}, one of the purposes of our study is to pursue the relevance between the models based on $E_6$ and the geometry.
From this viewpoint, although $\Vec{\mathfrak{C}^c}P_2$ can be defined algebraically, it can never possess the {\itshape pure} (i.e. Desarguesian) projective geometry of the usual meaning.
In addition, since our first priority in this paper is the solution of the doubling problem of degrees of freedom, this space will not be considered this time.
\end{quotation}

\begin{quotation}
\ \ [{\itshape Consideration} $4$]

As a solution of the doubling problem, the simplest idea is that: `we naively acknowledge that there are two universes, and assume that we live in only one side', like a parallel-brane world.
Given that this idea is correct, the structure of $\mathcal{M}^A \in \Vec{\mathfrak{J}^c}$ must be divided into two equal portions.
However, the method of this division is not unique.
Various possibilities exist unfortunately.
A method, which are thought of immediately from the inside of those many candidates, is the following:
\begin{eqnarray}
\mathcal{M}^A
&\equiv&
\mathcal{M}^A(\mathcal{A},\varphi) \\
& &
\quad\quad\quad \mathcal{M}^A \in \Vec{\mathfrak{J}^c} \nonumber \\
&=&
\left(
      \begin{array}{ccc}
      \mathcal{A}_1{}^A & \varphi_3{}^A & \bar{\varphi}_2{}^A \\
      \bar{\varphi}_3{}^A & \mathcal{A}_2{}^A & \varphi_1{}^A \\
      \varphi_2{}^A & \bar{\varphi}_1{}^A & \mathcal{A}_3{}^A
      \end{array}
\right) \\
& &
\quad\quad\quad \mathcal{A}_I{}^A \in \Vec{C} \  , \  \  \varphi_I{}^A \in \Vec{\mathfrak{C}^c} \  , \quad (I=1,2,3) \nonumber \\
&=&
\left(
      \begin{array}{ccc}
      Q_1{}^A & \phi_3{}^A & \bar{\phi}_2{}^A \\
      \bar{\phi}_3{}^A & Q_2{}^A & \phi_1{}^A \\
      \phi_2{}^A & \bar{\phi}_1{}^A & Q_3{}^A
      \end{array}
\right)
+
i
\left(
      \begin{array}{ccc}
      P_1{}^A & \pi_3{}^A & \bar{\pi}_2{}^A \\
      \bar{\pi}_3{}^A & P_2{}^A & \pi_1{}^A \\
      \pi_2{}^A & \bar{\pi}_1{}^A & P_3{}^A
      \end{array}
\right) \\
& &
\quad\quad\quad Q_I{}^A, P_I{}^A \in \Vec{R} \  , \  \  \phi_I{}^A, \pi_I{}^A \in \Vec{\mathfrak{C}} \  , \quad (I=1,2,3) \nonumber \\
&\equiv&
\mathcal{M_R}{}^A(Q,\phi) + \ i \ \mathcal{M_I}{}^A(P,\pi) \  . \\
& &
\quad\quad\quad \mathcal{M_R}{}^A, \mathcal{M_I}{}^A \in \Vec{\mathfrak{J}} \nonumber
\end{eqnarray}
Namely, this is an idea of dividing $\Vec{\mathfrak{J}^c}$ merely into real part and imaginary part.
After such kind of division, we have a feeling that we would like to think that real part $\mathcal{M_R}{}^A$ and imaginary part $\mathcal{M_I}{}^A$ express the two different universes which consist of, for example, the sets of D-instantons and anti-D-instantons, respectively.
However, we hesitate to consider the above division for the following reasons.

Firstly, if we divide like above expression, we had better construct a theory based on the $E_{6(-26)}$ which has a half degrees of freedom from the beginning, by using the following definition:
\begin{eqnarray}
E_{6(-26)} &=& \{ \alpha \in Iso_R ( \Vec{\mathfrak{J}} , \Vec{\mathfrak{J}} ) \, | \, \  ( \alpha A , \alpha B , \alpha C ) = ( A , B , C ) \} \  .
\end{eqnarray}
Because $E_{6(-26)}$ is the projective transformation group of $\Vec{\mathfrak{C}}P_2$, it is true that the research of the models based on the $E_{6(-26)}$ is very interesting.
However, the mathematical meaning of {\itshape the postulate of positive definite metric} in physics is lost in that case.
As discussed in the previous paper~\cite{Ohwashi:2001mz}, the postulate of positive definite metric is directly derived from the compactness condition of $E_6$.
It is considered to be an immediate consequence of the fact that our universe is {\itshape compact}.
The author cannot throw away this beautiful property.

Secondly, we have to remind ourselves of the reason why we used the compact $E_6$.
As also discussed in the previous paper~\cite{Ohwashi:2001mz}, 
The greatest reason why we used the {\itshape compact} $E_6$ is that there was the necessity of introducing the complex structure into the theory from the beginning in order to include the standard model.
However, when the universe is divided like an above expression, the individual universe of $\mathcal{M_R}{}^A$ and $\mathcal{M_I}{}^A$ does not contain the imaginary unit `$i$' in the inside, respectively.
This defect is fatal.
It is not clear anymore for what purpose the model was originally constructed using the {\itshape compact} $E_6$.
Therefore, as a solution over the doubling problem of the degrees of freedom, if the possibility that two universes exist is pursued seriously, {\itshape we have to grope for a combination to which the individual universe contains the imaginary unit `$i$' in each inside}.
\end{quotation}

\begin{quotation}
\ \ [{\itshape Consideration} $5$]

Well, what kind of division should be adopted?
Granted that $\mathcal{M}^A \in \Vec{\mathfrak{J}^c}$ is divided into two basic structures, many combinations exist.
Therefore, in order to point out a promising candidate out of them, it is thought appropriate to form a certain policy.
Now, we would like to pay attention to the fact that the existence of the projective geometry can be seen off and on behind the Freudenthal multiplication and the cubic form.
Of course there is no {\itshape pure} projective geometry that has the {\itshape compact} $E_6$ as its projective transformation group.
We will utilize the `analogy' with the projective geometry.
In accordance with this analogy, we use the {\itshape Klein's Erlangen Program} as our main policy.
It is hypothesized that `the individual fundamental structure itself should possess a symmetry {\itshape respectively}.'
Namely, given that the individual fundamental structure obtained by dividing has no transformation group inside itself, it is not beautiful physically and also mathematically.
Although we are assuming that the whole symmetry which acts on united two fundamental structures is the {\itshape compact} $E_6$, it is desirable that there is a definite symmetry which each possesses when each fundamental structure is seen independently.
Of course such symmetry must be the subgroup of $E_6$.
We would like to advance the argument by using this postulate as our policy.
\end{quotation}

Now, does the answer which satisfies the above requests exist?
One answer to this question is this paper.
At least one promising picture of the universe exists.
The author thinks that the following division is most consistent.
Therefore, in this paper, matrix models will be constructed according to the following methods.
We decompose $\mathcal{M}^A$ into the following two fundamental structures.
\begin{eqnarray}
\mathcal{M}^A
&\equiv&
\mathcal{M}^A(\mathcal{A},\varphi) \\
& &
\quad\quad\quad \mathcal{M}^A \in \Vec{\mathfrak{J}^c} \nonumber \\
&=&
\left(
      \begin{array}{ccc}
      \mathcal{A}_1{}^A & \varphi_3{}^A & \bar{\varphi}_2{}^A \\
      \bar{\varphi}_3{}^A & \mathcal{A}_2{}^A & \varphi_1{}^A \\
      \varphi_2{}^A & \bar{\varphi}_1{}^A & \mathcal{A}_3{}^A
      \end{array}
\right) \label{eq:components} \\
& &
\quad\quad\quad \mathcal{A}_I{}^A \in \Vec{C} \  , \  \  \varphi_I{}^A \in \Vec{\mathfrak{C}^c} \  , \quad (I=1,2,3) \nonumber \\
&=&
\Biggl(
\left(
      \begin{array}{ccc}
      A_1{}^A & \Phi_3{}^A & \tilde{\Phi}_2{}^A \\
      \tilde{\Phi}_3{}^A & A_2{}^A & \Phi_1{}^A \\
      \Phi_2{}^A & \tilde{\Phi}_1{}^A & A_3{}^A
      \end{array}
\right)
+
i
\left(
      \begin{array}{ccc}
      0 & \Psi_3{}^A e_4 & - \Psi_2{}^A e_4 \\
      - \Psi_3{}^A e_4 & 0 & \Psi_1{}^A e_4 \\
      \Psi_2{}^A e_4 & - \Psi_1{}^A e_4 & 0
      \end{array}
\right)
\Biggr) \nonumber \\
& &
{}+ \ \ 
i \ 
\Biggl(
\left(
      \begin{array}{ccc}
      a_1{}^A & \phi_3{}^A & \tilde{\phi}_2{}^A \\
      \tilde{\phi}_3{}^A & a_2{}^A & \phi_1{}^A \\
      \phi_2{}^A & \tilde{\phi}_1{}^A & a_3{}^A
      \end{array}
\right)
+
i
\left(
      \begin{array}{ccc}
      0 & \psi_3{}^A e_4 & - \psi_2{}^A e_4 \\
      - \psi_3{}^A e_4 & 0 & \psi_1{}^A e_4 \\
      \psi_2{}^A e_4 & - \psi_1{}^A e_4 & 0
      \end{array}
\right)
\Biggr) \label{eq:f-component} \\
& &
\quad\quad\quad A_I{}^A, a_I{}^A \in \Vec{R} \  , \  \  \Phi_I{}^A, \Psi_I{}^A, \phi_I{}^A, \psi_I{}^A \in \Vec{H} \  , \quad (I=1,2,3) \nonumber \\
&\equiv&
\Vec{\mathcal{F_R}}{}^A(A,\Phi;\Psi) + \ i \ \Vec{\mathcal{F_I}}{}^A(a,\phi;\psi) \  . \label{eq:figure} \\
& &
\quad\quad\quad \Vec{\mathcal{F_R}}{}^A, \Vec{\mathcal{F_I}}{}^A \in ( \Vec{\mathfrak{J}_H} \oplus i \Vec{H^3} ) \nonumber
\end{eqnarray}
Here, we have introduced the following notation to the complex Graves-Cayley numbers $\varphi_I{}^A \in \Vec{\mathfrak{C}^c} \  (I = 1,2,3)$.
\begin{eqnarray}
\varphi_I{}^A &=& \varphi_{0I}{}^A + \sum_{i=1}^{7} \varphi_{iI}{}^A e_{i} \  \quad \in \Vec{\mathfrak{C}^c} \  \quad (I = 1,2,3) \nonumber \\
&=& \{ \varphi_{0I}{}^A + \varphi_{1I}{}^A e_1 + \varphi_{2I}{}^A e_2 + \varphi_{3I}{}^A e_3 \} + \{ \varphi_{4I}{}^A + \varphi_{5I}{}^A e_1 - \varphi_{6I}{}^A e_2 + \varphi_{7I}{}^A e_3 \} e_4 \nonumber \\
& &
\quad\quad\quad\quad \varphi_{\tilde{i}I}{}^A \in \Vec{C} \  \quad (\tilde{i} = 0,\cdots,7) \nonumber \\
&\stackrel{\mathrm{def}}{\equiv}& \{ (\Phi_{0I}{}^A + i \phi_{0I}{}^A) + (\Phi_{1I}{}^A + i \phi_{1I}{}^A) e_1 + (\Phi_{2I}{}^A + i \phi_{2I}{}^A) e_2 + (\Phi_{3I}{}^A + i \phi_{3I}{}^A) e_3 \} \nonumber \\
& & {} + \{ (- \psi_{0I}{}^A + i \Psi_{0I}{}^A) + (- \psi_{1I}{}^A + i \Psi_{1I}{}^A) e_1 + (- \psi_{2I}{}^A + i \Psi_{2I}{}^A) e_2 + (- \psi_{3I}{}^A + i \Psi_{3I}{}^A) e_3 \} e_4 \nonumber \\
& &
\quad\quad\quad\quad \Phi_{\hat{i}I}{}^A, \Psi_{\hat{i}I}{}^A, \phi_{\hat{i}I}{}^A, \psi_{\hat{i}I}{}^A \in \Vec{R} \  \quad (\hat{i} = 0,\cdots,3) \nonumber \\
&=& \{ \Phi_I{}^A + i \phi_I{}^A \} + \{ - \psi_I{}^A + i \Psi_I{}^A \} e_4 \nonumber \\
& &
\quad\quad\quad\quad \Phi_I{}^A, \Psi_I{}^A, \phi_I{}^A, \psi_I{}^A \in \Vec{H} \nonumber \\
&=& (\Phi_I{}^A + i \Psi_I{}^A e_4) + i (\phi_I{}^A + i \psi_I{}^A e_4) \  \quad \in \  (\Vec{H} \oplus i \Vec{H} e_4) \oplus i (\Vec{H} \oplus i \Vec{H} e_4) \  .
\end{eqnarray}
Therefore, the {\itshape octonionic conjugation} $\bar{\varphi}_I{}^A$ is the following:
\begin{eqnarray}
\bar{\varphi}_I{}^A &\equiv& \varphi_{0I}{}^A - \sum_{i=1}^{7} \varphi_{iI}{}^A e_{i} \nonumber \\
&=& \{ (\Phi_{0I}{}^A + i \phi_{0I}{}^A) - (\Phi_{1I}{}^A + i \phi_{1I}{}^A) e_1 - (\Phi_{2I}{}^A + i \phi_{2I}{}^A) e_2 - (\Phi_{3I}{}^A + i \phi_{3I}{}^A) e_3 \} \nonumber \\
& & {} - \{ (- \psi_{0I}{}^A + i \Psi_{0I}{}^A) + (- \psi_{1I}{}^A + i \Psi_{1I}{}^A) e_1 + (- \psi_{2I}{}^A + i \Psi_{2I}{}^A) e_2 + (- \psi_{3I}{}^A + i \Psi_{3I}{}^A) e_3 \} e_4 \nonumber \\
&=& \{ \tilde{\Phi}_I{}^A + i \tilde{\phi}_I{}^A \} - \{ - \psi_I{}^A + i \Psi_I{}^A \} e_4 \nonumber \\
&=& (\tilde{\Phi}_I{}^A - i \Psi_I{}^A e_4) + i (\tilde{\phi}_I{}^A - i \psi_I{}^A e_4) \  .
\end{eqnarray}
Here, $\tilde{\Phi}_I{}^A$ or $\tilde{\phi}_I{}^A$ represents the {\itshape quaternionic conjugation} of $\Phi_I{}^A$ or $\phi_I{}^A$, respectively.
The quaternionic conjugation is defined as follows as usual:
\begin{eqnarray}
\Phi_I{}^A &=& \Phi_{0I}{}^A + \Phi_{1I}{}^A e_1 + \Phi_{2I}{}^A e_2 + \Phi_{3I}{}^A e_3 \nonumber \\
\Longrightarrow \quad \tilde{\Phi}_I{}^A &\equiv& \Phi_{0I}{}^A - \Phi_{1I}{}^A e_1 - \Phi_{2I}{}^A e_2 - \Phi_{3I}{}^A e_3 \  .
\end{eqnarray}
Additionally, we will refer to the fundamental structures $\Vec{\mathcal{F_R}}{}^A$ and $\Vec{\mathcal{F_I}}{}^A$ in (\ref{eq:figure}), as `{\itshape real figure}' and `{\itshape imaginary figure}', respectively. 
The term `figure' is often used to describe the subspace inside a projective space.
When this situation is seen from the standpoint of the matrix model, the each structure of the universe formed by these is $( \Vec{\mathfrak{J}_H} \oplus i \Vec{H^3} ) \times \Vec{\mathcal{G}}$ individually.
Here, $\Vec{\mathfrak{J}_H}$ is defined by,
\begin{eqnarray}
\Vec{\mathfrak{J}_H} &\stackrel{\mathrm{def}}{\equiv}& \Vec{\mathfrak{J}}(3,\Vec{H}) \\
&=& \{ a \in M(3,\Vec{H}) \  | \  a^\sharp = a \} \  . \quad \bigl( a^\sharp \equiv (\tilde{a})^T \bigr)
\end{eqnarray}
Namely, this is the usual quaternionic edition of the {\itshape real} Jordan algebra.
It is the following if the definition of $\Vec{\mathfrak{J}}$ is also written down for comparison.
\begin{eqnarray}
\left(
\begin{array}{ll}
\Vec{\mathfrak{J}} & \stackrel{\mathrm{def}}{\equiv} \ \ \Vec{\mathfrak{J}}(3,\Vec{\mathfrak{C}}) \\
 & \ = \ \ \{ A \in M(3,\Vec{\mathfrak{C}}) \  | \  A^\ddagger = A \} \  . \quad \bigl( A^\ddagger \equiv (\bar{A})^T \bigr) \\
\end{array}
\right)
\end{eqnarray}
Furthermore, when considering the quaternionic edition of the {\itshape complex} Jordan algebra, the following notation will be used,
\begin{eqnarray}
\Vec{\mathfrak{J}_H^c} &\stackrel{\mathrm{def}}{\equiv}& \Vec{\mathfrak{J}}(3,\Vec{H^c}) \\
&=& \{ x \in M(3,\Vec{H^c}) \  | \  x^\sharp = x \} \  , \quad \bigl( x^\sharp \equiv (\tilde{x})^T \bigr) \\
&=& \{ a + ib \  | \  a,b \in \Vec{\mathfrak{J}_H} \ , \  i^2=-1 \} \  .
\end{eqnarray}
In the same way, the definition of $\Vec{\mathfrak{J}^c}$ is also written down for the comparison below.
\begin{eqnarray}
\left(
\begin{array}{ll}
\Vec{\mathfrak{J}^c} & \stackrel{\mathrm{def}}{\equiv} \ \ \Vec{\mathfrak{J}}(3,\Vec{\mathfrak{C}^c}) \\
 & \ = \ \ \{ X \in M(3,\Vec{\mathfrak{C}^c}) \  | \  X^\ddagger = X \} \  , \quad \bigl( X^\ddagger \equiv (\bar{X})^T \bigr) \\
 & \ = \ \ \{ A + iB \  | \  A,B \in \Vec{\mathfrak{J}} \ , \  i^2=-1 \} \  . \\
\end{array}
\right)
\end{eqnarray}

The most important point to emphasize here is that {\itshape the introduction of the imaginary unit `$i$' into each universe is being accomplished splendidly}.
$\Vec{\mathcal{F_R}}{}^A$ and $\Vec{\mathcal{F_I}}{}^A$ contain the imaginary unit `$i$' in themselves individually.
One important subject which had been argued in our previous paper~\cite{Ohwashi:2001mz} was resolved by this division.

\section{A concrete example}

In this section, we consider matrix models based on the {\itshape compact} $E_6$ from the viewpoint of a field-decomposition $( \Vec{\mathfrak{J}_H} \oplus i \Vec{H^3} ) \times \Vec{\mathcal{G}}$\, introduced in the previous section.
For simplicity, at the outset, a toy model based on the $E_6$ is considered.
However, even if other more complicated actions based on the {\itshape compact} $E_6$ are considered, the following argument is basically the same.

\subsection{A simple case}
We consider the following most simple actions using the $E_6$ invariant:
\begin{eqnarray}
S_{simple} &=& \Bigl( \  \mathcal{M}^{A} \  , \  \mathcal{M}^B \  , \  \mathcal{M}^{C} \  \Bigr) \  \  Tr_{N} \bigl( \mathbf{T}_A \mathbf{T}_B \mathbf{T}_C \bigr) \  \  \  , \label{eq:SE_6}
\end{eqnarray}
where the $\mathcal{M}^A$ are elements of the complex exceptional Jordan algebra $\Vec{\mathfrak{J}^c}$, \  \,$( X , Y , Z ) \  \  ( X,Y,Z \in \Vec{\mathfrak{J}^c} )$ is the {\itshape cubic form}.
Now, because we would like to take up only the portion essential to our argument as simply as possible, about the gauge symmetry the trace of three generators is adopted naively.
Although this means the cubic form multiplied by both $f_{ABC}$ and $d_{ABC}$, in substance it is being multiplied by only $d_{ABC}$ because the cubic form is symmetric.
The degrees of freedom of this model live in $\Vec{\mathfrak{J}^c} \times \Vec{\mathcal{G}}$ as usual. The specific components of $\mathcal{M}^A \in \Vec{\mathfrak{J}^c}$\, are (\ref{eq:components-1})\,.

Now let us attempt here to introduce the field-decomposition discussed in the previous section.
\begin{eqnarray}
\mathcal{M}^A
&\equiv&
\Vec{\mathcal{F_R}}{}^A(A,\Phi;\Psi) + \ i \ \Vec{\mathcal{F_I}}{}^A(a,\phi;\psi) \  . \\
& &
\quad \Vec{\mathcal{F_R}}{}^A, \ \Vec{\mathcal{F_I}}{}^A \in ( \Vec{\mathfrak{J}_H} \oplus i \Vec{H^3} ) \nonumber
\end{eqnarray}
First, we note only the cubic form:
\begin{eqnarray}
\lefteqn{ \Bigl( \  \mathcal{M}^{A} \  , \  \mathcal{M}^B \  , \  \mathcal{M}^{C} \  \Bigr) } \nonumber \hspace{1mm} \\
&=& \Bigl( \  \Vec{\mathcal{F_R}}{}^A \, , \, \Vec{\mathcal{F_R}}{}^B \, , \, \Vec{\mathcal{F_R}}{}^C \  \Bigr) - i \ \Bigl( \  \Vec{\mathcal{F_I}}{}^A \, , \, \Vec{\mathcal{F_I}}{}^B \  , \  \Vec{\mathcal{F_I}}{}^C \  \Bigr) \nonumber \\
& & {} + i \ \Bigl( \  \Vec{\mathcal{F_R}}{}^A \, , \, \Vec{\mathcal{F_R}}{}^B \, , \, \Vec{\mathcal{F_I}}{}^C \  \Bigr) + i \ \Bigl( \  \Vec{\mathcal{F_R}}{}^A \, , \, \Vec{\mathcal{F_I}}{}^B \, , \, \Vec{\mathcal{F_R}}{}^C \  \Bigr) + i \ \Bigl( \  \Vec{\mathcal{F_I}}{}^A \, , \, \Vec{\mathcal{F_R}}{}^B \, , \, \Vec{\mathcal{F_R}}{}^C \  \Bigr) \nonumber \\
& & {} - \Bigl( \  \Vec{\mathcal{F_I}}{}^A \, , \, \Vec{\mathcal{F_I}}{}^B \, , \, \Vec{\mathcal{F_R}}{}^C \  \Bigr) - \Bigl( \  \Vec{\mathcal{F_I}}{}^A \, , \, \Vec{\mathcal{F_R}}{}^B \, , \, \Vec{\mathcal{F_I}}{}^C \  \Bigr) - \Bigl( \  \Vec{\mathcal{F_R}}{}^A \, , \, \Vec{\mathcal{F_I}}{}^B \, , \, \Vec{\mathcal{F_I}}{}^C \  \Bigr) \  .
\end{eqnarray}
Here we should notice that it is decomposed into the linear combination of the cubic forms phase-shifted every $\frac{\pi}{2}[rad]$.
That is, on the basis of $(\Vec{\mathcal{R}}\Vec{\mathcal{R}}\Vec{\mathcal{R}})$, $(\Vec{\mathcal{R}}\Vec{\mathcal{R}}\Vec{\mathcal{I}})$ is $e^{i\frac{\pi}{2}}$-shifted, $(\Vec{\mathcal{I}}\Vec{\mathcal{I}}\Vec{\mathcal{R}})$ is $e^{i\pi}$-shifted and $(\Vec{\mathcal{I}}\Vec{\mathcal{I}}\Vec{\mathcal{I}})$ is $e^{i\frac{3\pi}{2}}$-shifted.
Therefore, the action (\ref{eq:SE_6}) is decomposed as follows:
\begin{eqnarray}
\lefteqn{ S_{simple} = \Bigl( \  \mathcal{M}^{A} \  , \  \mathcal{M}^B \  , \  \mathcal{M}^{C} \  \Bigr) \  Tr_{N} \bigl( \mathbf{T}_A \mathbf{T}_B \mathbf{T}_C \bigr) } \nonumber \hspace{-1mm} \\
&=& \Bigl( \  \Vec{\mathcal{F_R}}{}^A \, , \, \Vec{\mathcal{F_R}}{}^B \, , \, \Vec{\mathcal{F_R}}{}^C \  \Bigr) Tr_{N} \bigl( \mathbf{T}_A \mathbf{T}_B \mathbf{T}_C \bigr) - i \Bigl( \  \Vec{\mathcal{F_I}}{}^A \, , \, \Vec{\mathcal{F_I}}{}^B \, , \, \Vec{\mathcal{F_I}}{}^C \  \Bigr) Tr_{N} \bigl( \mathbf{T}_A \mathbf{T}_B \mathbf{T}_C \bigr) \nonumber \\
& & {} + 3i \Bigl( \  \Vec{\mathcal{F_R}}{}^A \, , \, \Vec{\mathcal{F_R}}{}^B \, , \, \Vec{\mathcal{F_I}}{}^C \  \Bigr) Tr_{N} \bigl( \mathbf{T}_A \mathbf{T}_B \mathbf{T}_C \bigr) - 3 \Bigl( \  \Vec{\mathcal{F_I}}{}^A \, , \, \Vec{\mathcal{F_I}}{}^B \, , \, \Vec{\mathcal{F_R}}{}^C \  \Bigr) Tr_{N} \bigl( \mathbf{T}_A \mathbf{T}_B \mathbf{T}_C \bigr) \\
&\equiv& {} S_{\Vec{\mathcal{R}}} - i \, S_{\Vec{\mathcal{I}}} + S_{\Vec{int.}} \  ,
\end{eqnarray}
where we put as
\begin{eqnarray}
S_{\Vec{\mathcal{R}}} &=& \Bigl( \  \Vec{\mathcal{F_R}}{}^A \, , \, \Vec{\mathcal{F_R}}{}^B \, , \, \Vec{\mathcal{F_R}}{}^C \  \Bigr) Tr_{N} \bigl( \mathbf{T}_A \mathbf{T}_B \mathbf{T}_C \bigr) \  , \\
S_{\Vec{\mathcal{I}}} &=& \Bigl( \  \Vec{\mathcal{F_I}}{}^A \, , \, \Vec{\mathcal{F_I}}{}^B \, , \, \Vec{\mathcal{F_I}}{}^C \  \Bigr) Tr_{N} \bigl( \mathbf{T}_A \mathbf{T}_B \mathbf{T}_C \bigr) \  , \\
S_{\Vec{int.}} &=& \left[ 3i \Bigl( \  \Vec{\mathcal{F_R}}{}^A \, , \, \Vec{\mathcal{F_R}}{}^B \, , \, \Vec{\mathcal{F_I}}{}^C \  \Bigr) - 3 \Bigl( \  \Vec{\mathcal{F_I}}{}^A \, , \, \Vec{\mathcal{F_I}}{}^B \, , \, \Vec{\mathcal{F_R}}{}^C \  \Bigr) \right] Tr_{N} \bigl( \mathbf{T}_A \mathbf{T}_B \mathbf{T}_C \bigr) \  .
\end{eqnarray}
Further, after decomposing the action in terms of the variables defined in (\ref{eq:f-component}), we get 
\begin{eqnarray}
S_{\Vec{\mathcal{R}}} &=& \biggl[ \ \frac{3}{2} ( A_1{}^A A_2{}^B A_3{}^C ) - \frac{3}{2} \sum_{I=1}^{3} \left\{ \overline{(\Phi_I{}^A+i\Psi_I{}^A  e_4)} A_I{}^B (\Phi_I{}^C+i\Psi_I{}^C  e_4) \right\} \nonumber \\
& & {} + 3 \Vec{Re^c} \{ (\Phi_1{}^A+i\Psi_1{}^A  e_4) (\Phi_2{}^B+i\Psi_2{}^B  e_4) (\Phi_3{}^C+i\Psi_3{}^C  e_4) \} \,\biggr] \ Tr_{N} \Bigl( \mathbf{T}_A \left\{ \mathbf{T}_B , \mathbf{T}_C \right\} \Bigr) \  , \\
S_{\Vec{\mathcal{I}}} &=& \biggl[ \ \frac{3}{2} ( a_1{}^A a_2{}^B a_3{}^C ) - \frac{3}{2} \sum_{I=1}^{3} \left\{ \overline{(\phi_I{}^A+i\psi_I{}^A  e_4)} a_I{}^B (\phi_I{}^C+i\psi_I{}^C  e_4) \right\} \nonumber \\
& & {} + 3 \Vec{Re^c} \{ (\phi_1{}^A+i\psi_1{}^A  e_4) (\phi_2{}^B+i\psi_2{}^B  e_4) (\phi_3{}^C+i\psi_3{}^C  e_4) \} \,\biggr] \ Tr_{N} \Bigl( \mathbf{T}_A \left\{ \mathbf{T}_B , \mathbf{T}_C \right\} \Bigr) \  , \\
S_{\Vec{int.}} &=& \Bigg[ \ 3i \ \sum_{I=1}^{3} \bigg\{ \frac{1}{2} ( A_I{}^A A_{I+1}{}^B a_{I+2}{}^C ) - \frac{1}{2} \left( \overline{(\Phi_I{}^A+i\Psi_I{}^A  e_4)} A_I{}^B (\phi_I{}^C+i\psi_I{}^C  e_4) \right) \nonumber \\
& & {} - \frac{1}{2} \left( \overline{ (\Phi_I{}^A+i\Psi_I{}^A  e_4)} a_I{}^B (\Phi_I{}^C+i\Psi_I{}^C  e_4) \right) - \frac{1}{2} \left( \overline{ (\phi_I{}^A+i\psi_I{}^A  e_4)} A_I{}^B (\Phi_I{}^C+i\Psi_I{}^C  e_4) \right) \nonumber \\
& & {} + \Vec{Re^c} \left( (\Phi_I{}^A+i\Psi_I{}^A  e_4) (\Phi_{I+1}{}^B+i\Psi_{I+1}{}^B  e_4) (\phi_{I+2}{}^C+i\psi_{I+2}{}^C  e_4) \right) \bigg\} \nonumber \\
& & {} \ -3 \ \sum_{I=1}^{3} \bigg\{ \ A \leftrightarrow a \ , \ \Phi \leftrightarrow \phi \ , \ \Psi \leftrightarrow \psi \ \bigg\} \ \Bigg] \ Tr_{N} \Bigl( \mathbf{T}_A \left\{ \mathbf{T}_B , \mathbf{T}_C \right\} \Bigr) \  ,
\end{eqnarray}
where the index $I$ \,is $\Vec{mod \  3}$.

Now, although the classical solution of this simple model may be considered, it is not carried out here.
The argument which we would like to have in this section is a generalization common to all models based on the {\itshape compact} $E_6$.
It is the following when the matter which the above naive example shows is summarized.
\begin{itemize}
  \item By the division into two fundamental figures $\Vec{\mathcal{F_R}}{}^A$ and $\Vec{\mathcal{F_I}}{}^A$, we can decompose the action based on the {\itshape compact} $E_6$ into three parts: $S_{\Vec{\mathcal{R}}}$, $S_{\Vec{\mathcal{I}}}$ which have the completely the same structure of the fields, and $S_{\Vec{int.}}$ which represents the interactions of two fundamental figures.
  \item Given that the expectation value of one fundamental figure is vanishing $\Vec{\mathcal{F_I}}{}^A=0$ in the classical limit, the classical theory is only the other action $S_{\Vec{\mathcal{R}}}$ with the constraints.
  \item Therefore, when considering the classical theory, the argument can be closed in only one universe, but when considering the quantum theory, we have to take the existence of the two universes into account.
\end{itemize}

Now, from an analogy with the projective geometry as discussed in the previous section ([{\itshape Consideration} $5$]), we would like to demand that $\Vec{\mathcal{F_R}}{}^A$ and $\Vec{\mathcal{F_I}}{}^A$ possess a transformation group independently.
In other words, we would like to demand that $S_{\Vec{\mathcal{R}}}$ and $S_{\Vec{\mathcal{I}}}$ themselves possess a symmetry individually.
In the following, we will see that our fundamental figures $\Vec{\mathcal{F_R}}{}^A$, $\Vec{\mathcal{F_I}}{}^A$ are filling this demand.
The resulting symmetry is the $Sp(4,\Vec{H}) / \Vec{Z}_2$.
It can be comprehended clearly by using {\itshape Yokota mapping} $\mathcal{Y}$.

\subsection{Symmetry of the pairs \\ \qquad ---\ \normalsize{The benefit of Yokota's golden mapping $\mathcal{Y}$}\ ---}
As mentioned above, $S_{\Vec{\mathcal{R}}}$ and $S_{\Vec{\mathcal{I}}}$ themselves are the $Sp(4,\Vec{H}) / \Vec{Z}_2$ matrix models.
This is ensured by the following {\itshape Yokota mapping} $\mathcal{Y}$.
In this subsection, based on the review \cite{Yokota:1990e6}, we introduce the Yokota mapping $\mathcal{Y}$ briefly (See Reference \cite{Yokota:1990e6} for details.).

We consider the $4 \times 4$ complex quaternionic Jordan type algebra $\Vec{\mathfrak{J}}(4,\Vec{H^c})$ with the Jordan multiplication, the inner product and the hermitian product. The {\itshape Yokota mapping} is defined as a mapping $\mathcal{Y} : \, \Vec{\mathfrak{J}^c} = \Vec{\mathfrak{J}_H^c} \oplus (\Vec{H^c})\Vec{{}^3} \longrightarrow \, \Vec{\mathfrak{J}}(4,\Vec{H^c})_0 = \{ x \in \Vec{\mathfrak{J}}(4,\Vec{H^c}) \, | \, \  tr(x) = 0 \}$ \,by
\begin{eqnarray}
\mathcal{Y} \left( M + \Vec{v} \right) &=& 
\left(
  \begin{array}{cc}
  \frac{1}{2} tr(M) & i \Vec{v} \\
  i \Vec{v}^{\sharp} & M - \frac{1}{2} tr(M) \Vec{I}_3 \\
  \end{array}
\right) \label{eq:yokota} \  ,
\end{eqnarray}
where \,$M \in \Vec{\mathfrak{J}_H^c}$, \,$\Vec{v} \in (\Vec{H^c})\Vec{{}^3}$ \,and \,$\Vec{I}_3$ is the $3 \times 3$ unit matrix.

If this is represented by using our notation (\ref{eq:f-component}) concretely, the result is the following:
\begin{eqnarray}
\lefteqn{
\mathcal{Y}
\left(
\left(
      \begin{array}{ccc}
      A_1{}^A+ia_1{}^A & \Phi_3{}^A+i\phi_3{}^A & \tilde{\Phi}_2{}^A+i\tilde{\phi}_2{}^A \\
      \tilde{\Phi}_3{}^A+i\tilde{\phi}_3{}^A & A_2{}^A+ia_2{}^A & \Phi_1{}^A+i\phi_1{}^A \\
      \Phi_2{}^A+i\phi_2{}^A & \tilde{\Phi}_1{}^A+i\tilde{\phi}_1{}^A & A_3{}^A+ia_3{}^A
      \end{array}
\right) 
\right. \nonumber
}\hspace{0mm} \\
& &
{} +
\left.
\left(
      \begin{array}{ccc}
      0 & (-\psi_3{}^A+i\Psi_3{}^A) e_4 & -(-\psi_2{}^A+i\Psi_2{}^A) e_4 \\
      -(-\psi_3{}^A+i\Psi_3{}^A) e_4 & 0 & (-\psi_1{}^A+i\Psi_1{}^A) e_4 \\
      (-\psi_2{}^A+i\Psi_2{}^A) e_4 & -(-\psi_1{}^A+i\Psi_1{}^A) e_4 & 0
      \end{array}
\right)
\right) \nonumber \\
&=&
\left(
      \begin{array}{cccc}
      \frac{1}{2}(A_1{}^A+A_2{}^A+A_3{}^A) & -\Psi_1{}^A & -\Psi_2{}^A & -\Psi_3{}^A \\
      -\tilde{\Psi}_1{}^A & \frac{1}{2}(A_1{}^A-A_2{}^A-A_3{}^A) & \Phi_3{}^A & \tilde{\Phi}_2{}^A \\
      -\tilde{\Psi}_2{}^A & \tilde{\Phi}_3{}^A & \frac{1}{2}(A_2{}^A-A_1{}^A-A_3{}^A) & \Phi_1{}^A \\
      -\tilde{\Psi}_3{}^A & \Phi_2{}^A & \tilde{\Phi}_1{}^A & \frac{1}{2}(A_3{}^A-A_1{}^A-A_2{}^A)
      \end{array}
\right) \nonumber \\
& &
{} + \ i
\left(
\begin{array}{cccc}
      \frac{1}{2}(a_1{}^A+a_2{}^A+a_3{}^A) & -\psi_1{}^A & -\psi_2{}^A & -\psi_3{}^A \\
      -\tilde{\psi}_1{}^A & \frac{1}{2}(a_1{}^A-a_2{}^A-a_3{}^A) & \phi_3{}^A & \tilde{\phi}_2{}^A \\
      -\tilde{\psi}_2{}^A & \tilde{\phi}_3{}^A & \frac{1}{2}(a_2{}^A-a_1{}^A-a_3{}^A) & \phi_1{}^A \\
      -\tilde{\psi}_3{}^A & \phi_2{}^A & \tilde{\phi}_1{}^A & \frac{1}{2}(a_3{}^A-a_1{}^A-a_2{}^A)
      \end{array}
\right) \  . \nonumber \\
\end{eqnarray}
Here, we should notice that the real part of this $4 \times 4$ matrix is expressed only by the fields of $\Vec{\mathcal{F_R}}{}^A$, and the imaginary part is expressed only by the fields of $\Vec{\mathcal{F_I}}{}^A$.
Therefore, the restriction of Yokota mapping $\mathcal{Y}$ to the {\itshape real figure} $\Vec{\mathcal{F_R}}{}^A(A,\Phi;\Psi)$ (or {\itshape the imaginary figure} $\Vec{\mathcal{F_I}}{}^A(a,\phi;\psi)$) induces a $R$-linear isomorphism $\mathcal{Y} : \, \Vec{\mathcal{F_R}}{}^A \ (or \ \Vec{\mathcal{F_I}}{}^A) \longmapsto \, \mathcal{Y} \left( \Vec{\mathcal{F_R}}{}^A \right) \ \left(or \ \mathcal{Y} \left( \Vec{\mathcal{F_I}}{}^A \right) \right) \, \in \, \Vec{\mathfrak{J}}(4,\Vec{H})_0 \,$.
This is important.

Now, before the significant properties about the Yokota mapping $\mathcal{Y}$ defined by (\ref{eq:yokota}) are shown, Two mappings `the $\beta$-conjugation' and `the $\gamma$-conjugation' are defined.
The $\beta$-conjugation is merely the complex conjugation with respect to `$i$' which we have denoted as $(\cdots)^\ast$ until now.
\begin{eqnarray}
\beta ( a + ib ) &\equiv& a - ib \quad\quad (\ \equiv\ ( a + ib )^\ast \ ) \quad\quad\quad a,b \in \Vec{\mathfrak{C}} \  .
\end{eqnarray}
It is more convenient to use this mapping $\beta$, without using the asterisk$(\ast)$, in the following arguments.
Another $\gamma$-conjugation is defined for the first time here, and is the following complex conjugation with respect to `$e_4$'.
\begin{eqnarray}
\gamma ( w + ze_4 ) &\equiv& w - ze_4 \quad\quad\quad\quad\quad w,z \in \Vec{H^c} \  .
\end{eqnarray}
In the Appendix (\ref{ssec_e4}), we have mentioned that $e_4$ plays a role of the imaginary unit.
Of course if $\beta$ and $\gamma$ act on a matrix, it means that they act on all elements of that matrix.
The following is an example,
\begin{eqnarray*}
\gamma
\left(
      \begin{array}{ccc}
      x_1 & \xi_3 & \bar{\xi}_2 \\
      \bar{\xi}_3 & x_2 & \xi_1 \\
      \xi_2 & \bar{\xi}_1 & x_3
      \end{array}
\right)
&=&
\left(
      \begin{array}{ccc}
      x_1 & \gamma\xi_3 & \overline{\gamma\xi}_2 \\
      \overline{\gamma\xi}_3 & x_2 & \gamma\xi_1 \\
      \gamma\xi_2 & \overline{\gamma\xi}_1 & x_3
      \end{array}
\right) \  .
\end{eqnarray*}
By using these, the important properties about the {\itshape Yokota mapping} $\mathcal{Y}$ are expressed in the following.

\begin{quotation}
[{\itshape Lemma}\,$1$]
{} \,$\mathcal{Y}$ is a $C$-linear isomorphism and satisfies:
\begin{eqnarray}
\mathcal{Y}(X) \circ \mathcal{Y}(Y) &=& \mathcal{Y} \left( \, \gamma(X \times Y) \, \right) + \frac{1}{4} \left( \, \gamma(X) \, , \, Y \, \right) \, \Vec{I} \ , \quad (Yokota \ identity) \label{eq:lemma1-1} \\
\left( \, \mathcal{Y}(X) \, , \, \mathcal{Y}(Y) \, \right) &=& \left( \, \gamma(X) \, , \, Y \, \right) \ , \qquad\qquad\qquad\qquad\qquad\qquad X,Y \in \Vec{\mathfrak{J}^c} \label{eq:lemma1-2} \ .
\end{eqnarray}

[{\itshape Lemma}\,$2$]
{} \,$\mathcal{Y}$ is an isometry with respect to the hermitian product:
\begin{eqnarray}
< \, \mathcal{Y}(X) \, , \, \mathcal{Y}(Y) \, > &=& < \, X \, , \, Y \, > \ , \qquad\qquad\qquad\qquad\qquad X,Y \in \Vec{\mathfrak{J}^c} \label{eq:lemma2} \ .
\end{eqnarray}

[{\itshape Theorem}\,]
{} \,The subgroup $(E_6)^{\beta\gamma}=\{ \alpha \in E_{6} \, | \, \beta \gamma \alpha = \alpha \beta \gamma \}$ of $E_6$ is isomorphic to \\ \quad\qquad\qquad\qquad the following group:
\begin{eqnarray}
(E_6)^{\beta\gamma} &\cong& Sp(4,\Vec{H})/\Vec{Z}_2 \ , \qquad\qquad \Vec{Z}_2 = \{\Vec{I},-\Vec{I}\} \label{eq:theorem} \ .
\end{eqnarray}
In order to obtain the {\itshape Yokota identity} (\ref{eq:lemma1-1}), it is good to calculate from the first term $\mathcal{Y} \left( \, \gamma((M_{1}+\Vec{v}_{1}) \times (M_{2}+\Vec{v}_{2})) \, \right)$ of the right-hand side.
The identity (\ref{eq:lemma1-2}) is obtained by taking the trace of both sides in (\ref{eq:lemma1-1}).
The metric formula (\ref{eq:lemma2}) is calculated by elements concretely.
The isomorphism of Lie groups (\ref{eq:theorem}) is ensured by the existence of the following surjective group homomorphic mapping $f$, which is defined by using the {\itshape Yokota mapping} $\mathcal{Y}$.
\begin{eqnarray}
\lefteqn{
f : \  Sp(4,\Vec{H}) \longrightarrow \, (E_6)^{\beta\gamma}
}\hspace{30mm} \nonumber \\
f(V) \, X &=& \mathcal{Y}^{-1} \Big( \, V \big( \mathcal{Y}(X) \big) V^{\sharp} \, \Big) \label{eq:map-f} \ , \\
& & \qquad\qquad X \in \Vec{\mathfrak{J}^c} \, , \,\ \  V \in Sp(4,\Vec{H}) \, , \,\ \  f(V) \in (E_6)^{\beta\gamma} \nonumber \ .
\end{eqnarray}
Because of $Ker\ f = \{\Vec{I},-\Vec{I}\}$,\, the isomorphism (\ref{eq:theorem}) is concluded from the group homomorphism theorem.
\end{quotation}

Now, the point to observe here is that, fortunately thanks to [{\itshape Lemma}\,$2$], the compactness condition of the $E_6$ is conserved even after the map.
Moreover, because $(\Vec{\mathfrak{J}_H} \oplus i \Vec{H^3})$ is a representation space of the $(E_6)^{\beta\gamma}$ operation (See Appendix \ref{Sp_4}), we can recognize from [{\itshape Theorem}\,] that $S_{\Vec{\mathcal{R}}}$ and $S_{\Vec{\mathcal{I}}}$ discussed in the previous subsection are forming the $Sp(4,\Vec{H})/\Vec{Z}_2$ matrix model individually:
\begin{eqnarray}
\left\{
\begin{array}{l}
\Bigl( \alpha' \Vec{\mathcal{F_R}}{}^A , \alpha' \Vec{\mathcal{F_R}}{}^B , \alpha' \Vec{\mathcal{F_R}}{}^C \Bigr) Tr_{N} \bigl( \mathbf{T}_A \mathbf{T}_B \mathbf{T}_C \bigr) \ =\  \Bigl( \Vec{\mathcal{F_R}}{}^A , \Vec{\mathcal{F_R}}{}^B , \Vec{\mathcal{F_R}}{}^C \Bigr) Tr_{N} \bigl( \mathbf{T}_A \mathbf{T}_B \mathbf{T}_C \bigr) \ =\  S_{\Vec{\mathcal{R}}} \\
\Bigl( \, \alpha' \Vec{\mathcal{F_I}}{}^A ,\, \alpha' \Vec{\mathcal{F_I}}{}^B ,\, \alpha' \Vec{\mathcal{F_I}}{}^C \, \Bigr) Tr_{N} \bigl( \mathbf{T}_A \mathbf{T}_B \mathbf{T}_C \bigr) \ =\  \Bigl( \, \Vec{\mathcal{F_I}}{}^A ,\, \Vec{\mathcal{F_I}}{}^B ,\, \Vec{\mathcal{F_I}}{}^C \, \Bigr) Tr_{N} \bigl( \mathbf{T}_A \mathbf{T}_B \mathbf{T}_C \bigr) \ =\  S_{\Vec{\mathcal{I}}}
\end{array}
\right. ,
\end{eqnarray}
where $\alpha' \in (E_6)^{\beta\gamma}$.
Therefore eventually, it follows from what has been said thus far that {\itshape we can interpret the model based on the compact $E_6$ as the interacting $Sp(4,\Vec{H})/\Vec{Z}_2$ pair models}.
This picture is applicable to all the models based on the {\itshape compact} $E_6$, which may be not only matrix models but also field theories.
In the following, this point is discussed a little further.

\subsection{The extension to other models based on the {\itshape compact} $E_6$}
It is straightforward to extend the argument concerning the simple model given above to other models based on the {\itshape compact} $E_6$.
The decomposition into the interacting $Sp(4,\Vec{H})/\Vec{Z}_2$ {\itshape pair} models is applicable to all the models based on the {\itshape compact} $E_6$, which may be not only matrix models but also field theories.
In other words, it is applicable to all the complicated actions which are invariant under the $E_6$.
The reason is that $\Vec{\mathcal{F_R}}{}^A$ and $\Vec{\mathcal{F_I}}{}^A$ belong to the representation space of $(E_6)^{\beta\gamma}$.
Let us see it in more detail as follows.

An important point to emphasize is that $(\Vec{\mathfrak{J}_H} \oplus i \Vec{H^3})$ is an eigenspace with the eigenvalue $1$ to the $R$-linear mapping $(\beta \gamma)$ on the $\Vec{\mathfrak{J}^c}$.
\begin{eqnarray}
\beta \gamma \  \Vec{\mathcal{F_R}}{}^A &=& 1 \cdot \, \Vec{\mathcal{F_R}}{}^A \qquad\qquad \Vec{\mathcal{F_R}}{}^A \in (\Vec{\mathfrak{J}_H} \oplus i \Vec{H^3}) \  .
\end{eqnarray}
Generally, for \,$\alpha \in E_6$ which is chosen freely, \,$\alpha \Vec{\mathcal{F_R}}{}^A \notin (\Vec{\mathfrak{J}_H} \oplus i \Vec{H^3})$. \\
However, for \,$\alpha' \in (E_6)^{\beta\gamma} \subset E_6$, \,$\alpha' \Vec{\mathcal{F_R}}{}^A \in (\Vec{\mathfrak{J}_H} \oplus i \Vec{H^3})$ \ because of \,$(\beta \gamma \alpha') = (\alpha' \beta \gamma)$. \\
This means the model is `{\itshape closed}'.
\begin{eqnarray}
\lefteqn{
\left\{
  \begin{array}{l}
  \beta \gamma \alpha' \  \Vec{\mathcal{F_R}}{}^A \ = \ \beta \gamma \ (\alpha' \Vec{\mathcal{F_R}}{}^A) \\
  \alpha' \beta \gamma \  \Vec{\mathcal{F_R}}{}^A = \alpha' \ (\beta \gamma \Vec{\mathcal{F_R}}{}^A) \ = \  \alpha' \Vec{\mathcal{F_R}}{}^A \\
  \end{array}
\right.
}\hspace{40mm} \nonumber \\
\Longrightarrow \qquad \beta \gamma \ (\alpha' \Vec{\mathcal{F_R}}{}^A) &=& 1 \cdot \, (\alpha' \Vec{\mathcal{F_R}}{}^A) \qquad\quad i.e. \quad \alpha' \Vec{\mathcal{F_R}}{}^A \in (\Vec{\mathfrak{J}_H} \oplus i \Vec{H^3}) \  .
\end{eqnarray}
These fact is concluded for $\Vec{\mathcal{F_I}}{}^A$ completely equally.

In this way, no matter what theory is constructed by using the action which is invariant under the {\itshape compact} $E_6$, we can have a picture of interacting $Sp(4,\Vec{H})/\Vec{Z}_2$ {\itshape pair} universe. \\
For example, $\bigl( det(\mathcal{M}^{A}) \  det(\mathcal{M}^B) \  det(\mathcal{M}^{C}) \  f_{ABC} \bigr)$ is one invariant with respect to $E_6$.
Of course we may use $d_{ABC}$ instead of $f_{ABC}$.
Furthermore, it is not necessary to be cubic.
The quantity multiplied by $f_{AB}{}^{E}f_{CDE}$ is applicable too.
Needless to say, it is a completely different problem whether such mathematical quantities have the physical meaning.
In the next section, we apply this picture to our previous model.

\section{Application to our previous model}

In this section, we apply the field-decomposition $( \Vec{\mathfrak{J}_H} \oplus i \Vec{H^3} ) \times \Vec{\mathcal{G}}$ to our previous model.
After that, the opinion on the doubling problem which has been left in the previous paper~\cite{Ohwashi:2001mz} is argued.

\subsection{Interacting bi-Chern-Simons type model based on $Sp(4,\Vec{H})/\Vec{Z}_2$}
Let us attempt to introduce the field-decomposition of $\mathcal{M}^A = \Vec{\mathcal{F_R}}{}^A + i \Vec{\mathcal{F_I}}{}^A$ to the previous model,
\begin{eqnarray}
S &=& \Bigl( \  \mathcal{P}^2 ( \mathcal{M}^{[A} ) \  , \  \mathcal{P} ( \mathcal{M}^B ) \  , \  \mathcal{M}^{C]} \  \Bigr) \  \  f_{ABC} \  \  \  , \label{eq:E_6p2}
\end{eqnarray}
where $f_{ABC} = 2 \, Tr_{N} ( \mathbf{T}_A [ \mathbf{T}_B , \mathbf{T}_C ] )$.
First, we note only the cubic form. 
It is helpful to use the relation $(\,\mathcal{P}X,\mathcal{P}Y,\mathcal{P}Z\,)=(\,X,Y,Z\,)$.
The result is the following.
\begin{eqnarray}
\lefteqn{ \Bigl( \mathcal{P}^2 ( \mathcal{M}^A ) , \mathcal{P} ( \mathcal{M}^B ) , \mathcal{M}^{C} \Bigr) } \nonumber \hspace{-2mm} \\
&=& \Bigl( \mathcal{P}^2 (\Vec{\mathcal{F_R}}{}^A) , \mathcal{P} (\Vec{\mathcal{F_R}}{}^B) , \Vec{\mathcal{F_R}}{}^C \Bigr) - i \Bigl( \mathcal{P}^2 (\Vec{\mathcal{F_I}}{}^A) , \mathcal{P} (\Vec{\mathcal{F_I}}{}^B) , \Vec{\mathcal{F_I}}{}^C \Bigr) \nonumber \\
& & {} + i \Bigl( \mathcal{P}^2 (\Vec{\mathcal{F_R}}{}^A) , \mathcal{P} (\Vec{\mathcal{F_R}}{}^B) , \Vec{\mathcal{F_I}}{}^C \Bigr) + i \Bigl( \mathcal{P}^2 (\Vec{\mathcal{F_R}}{}^B) , \mathcal{P} (\Vec{\mathcal{F_R}}{}^C) , \Vec{\mathcal{F_I}}{}^A \Bigr) + i \Bigl( \mathcal{P}^2 (\Vec{\mathcal{F_R}}{}^C) , \mathcal{P} (\Vec{\mathcal{F_R}}{}^A) , \Vec{\mathcal{F_I}}{}^B \Bigr) \nonumber \\
& & {} - \Bigl( \mathcal{P}^2 (\Vec{\mathcal{F_I}}{}^A) , \mathcal{P} (\Vec{\mathcal{F_I}}{}^B) , \Vec{\mathcal{F_R}}{}^C \Bigr) - \Bigl( \mathcal{P}^2 (\Vec{\mathcal{F_I}}{}^B) , \mathcal{P} (\Vec{\mathcal{F_I}}{}^C) , \Vec{\mathcal{F_R}}{}^A \Bigr) - \Bigl( \mathcal{P}^2 (\Vec{\mathcal{F_I}}{}^C) , \mathcal{P} (\Vec{\mathcal{F_I}}{}^A) , \Vec{\mathcal{F_R}}{}^B \Bigr)
\end{eqnarray}
Like before, this is a linear combination of the cubic forms phase-shifted every $\frac{\pi}{2}[rad]$.
We use the {\itshape weight}-1 anti-symmetrization as the operation of the anti-symmetrization with respect to the indeces $A,B,C$, because contracting indices with a totally anti-symmetric tensor $f_{ABC}$ results in the ordinary summation.
Therefore, the action (\ref{eq:E_6p2}) is decomposed as follows:
\begin{eqnarray}
\lefteqn{ S = \Bigl( \  \mathcal{P}^2 ( \mathcal{M}^{A} ) \  , \  \mathcal{P} ( \mathcal{M}^B ) \  , \  \mathcal{M}^{C} \  \Bigr) \  \  f_{ABC} } \nonumber \hspace{-1mm} \\
&=& \Bigl( \mathcal{P}^2 (\Vec{\mathcal{F_R}}{}^A) , \mathcal{P} (\Vec{\mathcal{F_R}}{}^B) , \Vec{\mathcal{F_R}}{}^C \Bigr) \  f_{ABC} - i \Bigl( \mathcal{P}^2 (\Vec{\mathcal{F_I}}{}^A) , \mathcal{P} (\Vec{\mathcal{F_I}}{}^B) , \Vec{\mathcal{F_I}}{}^C \Bigr) \  f_{ABC} \nonumber \\
& & {} + 3i \Bigl( \mathcal{P}^2 (\Vec{\mathcal{F_R}}{}^A) , \mathcal{P} (\Vec{\mathcal{F_R}}{}^B) , \Vec{\mathcal{F_I}}{}^C \Bigr) \  f_{ABC} - 3 \Bigl( \mathcal{P}^2 (\Vec{\mathcal{F_I}}{}^A) , \mathcal{P} (\Vec{\mathcal{F_I}}{}^B) , \Vec{\mathcal{F_R}}{}^C \Bigr) \  f_{ABC} \\
&\equiv& {} S_{\Vec{\mathcal{R}}} - i \, S_{\Vec{\mathcal{I}}} + S_{\Vec{int.}} \  ,
\end{eqnarray}
where we put as
\begin{eqnarray}
S_{\Vec{\mathcal{R}}} &=& \Bigl( \mathcal{P}^2 (\Vec{\mathcal{F_R}}{}^A) , \mathcal{P} (\Vec{\mathcal{F_R}}{}^B) , \Vec{\mathcal{F_R}}{}^C \Bigr) \  f_{ABC} \nonumber \\
&=& \frac{1}{4} \, f_{ABC} \, \epsilon^{IJK} \, ( A_I{}^A A_J{}^B A_K{}^C ) + \frac{3}{2} \, f_{ABC} \, \epsilon^{IJK} \, \{ \overline{(\Phi_I{}^A+i\Psi_I{}^A  e_4)} A_J{}^B (\Phi_K{}^C+i\Psi_K{}^C  e_4) \} \nonumber \\
& & {} - 3 \, f_{ABC} \, \Vec{Re^c} \{ (\Phi_3{}^A+i\Psi_3{}^A  e_4) (\Phi_2{}^B+i\Psi_2{}^B  e_4) (\Phi_1{}^C+i\Psi_1{}^C  e_4) \} \nonumber \\
& & {} + f_{ABC} \, \sum_{I=1}^{3} \left[ \Vec{Re^c} \{ (\Phi_I{}^A+i\Psi_I{}^A  e_4) (\Phi_I{}^B+i\Psi_I{}^B  e_4) (\Phi_I{}^C+i\Psi_I{}^C  e_4) \} \right] \\
S_{\Vec{\mathcal{I}}} &=& \Bigl( \mathcal{P}^2 (\Vec{\mathcal{F_I}}{}^A) , \mathcal{P} (\Vec{\mathcal{F_I}}{}^B) , \Vec{\mathcal{F_I}}{}^C \Bigr) \  f_{ABC} \nonumber \\
&=& \frac{1}{4} \, f_{ABC} \, \epsilon^{IJK} \, ( a_I{}^A a_J{}^B a_K{}^C ) + \frac{3}{2} \, f_{ABC} \, \epsilon^{IJK} \, \{ \overline{(\phi_I{}^A+i\psi_I{}^A  e_4)} a_J{}^B (\phi_K{}^C+i\psi_K{}^C  e_4) \} \nonumber \\
& & {} - 3 \, f_{ABC} \, \Vec{Re^c} \{ (\phi_3{}^A+i\psi_3{}^A  e_4) (\phi_2{}^B+i\psi_2{}^B  e_4) (\phi_1{}^C+i\psi_1{}^C  e_4) \} \nonumber \\
& & {} + f_{ABC} \, \sum_{I=1}^{3} \left[ \Vec{Re^c} \{ (\phi_I{}^A+i\psi_I{}^A  e_4) (\phi_I{}^B+i\psi_I{}^B  e_4) (\phi_I{}^C+i\psi_I{}^C  e_4) \} \right] \\
S_{\Vec{int.}} &=& \left[ 3i \Bigl( \mathcal{P}^2 (\Vec{\mathcal{F_R}}{}^A) , \mathcal{P} (\Vec{\mathcal{F_R}}{}^B) , \Vec{\mathcal{F_I}}{}^C \Bigr) - 3 \Bigl( \mathcal{P}^2 (\Vec{\mathcal{F_I}}{}^A) , \mathcal{P} (\Vec{\mathcal{F_I}}{}^B) , \Vec{\mathcal{F_R}}{}^C \Bigr) \right] \  f_{ABC} \nonumber \\
&=& 3i \, f_{ABC} \, \Big[ \  \frac{1}{4} \ \epsilon^{IJK} ( A_I{}^A A_J{}^B a_K{}^C ) \nonumber \\
& & {} \ \qquad\qquad + \frac{1}{2} \  \epsilon_{IJK} \  \{ \overline{(\Phi_I{}^A+i\Psi_I{}^A  e_4)} a_J{}^B (\Phi_K{}^C+i\Psi_K{}^C  e_4) \} \nonumber \\
& & {} \ \qquad\qquad + \frac{1}{2} \  \epsilon_{IJK} \  \{ \overline{(\Phi_I{}^A+i\Psi_I{}^A  e_4)} A_J{}^B (\phi_K{}^C+i\psi_K{}^C  e_4) \} \nonumber \\
& & {} \ \qquad\qquad + \frac{1}{2} \  \epsilon_{IJK} \  \{ \overline{(\phi_I{}^A+i\psi_I{}^A  e_4)} A_J{}^B (\Phi_K{}^C+i\Psi_K{}^C  e_4) \} \nonumber \\
& & {} \ \qquad\qquad - \Vec{Re^c} \{ (\Phi_3{}^A+i\Psi_3{}^A  e_4) (\Phi_2{}^B+i\Psi_2{}^B  e_4) (\phi_1{}^C+i\psi_1{}^C  e_4) \} \nonumber \\
& & {} \ \qquad\qquad - \Vec{Re^c} \{ (\Phi_3{}^A+i\Psi_3{}^A  e_4) (\phi_2{}^B+i\psi_2{}^B  e_4) (\Phi_1{}^C+i\Psi_1{}^C  e_4) \} \nonumber \\
& & {} \ \qquad\qquad - \Vec{Re^c} \{ (\phi_3{}^A+i\psi_3{}^A  e_4) (\Phi_2{}^B+i\Psi_2{}^B  e_4) (\Phi_1{}^C+i\Psi_1{}^C  e_4) \} \nonumber \\
& & {} \ \qquad\qquad + \sum_{I=1}^{3} \left[ \Vec{Re^c} \{ (\Phi_I{}^A+i\Psi_I{}^A  e_4) (\Phi_I{}^B+i\Psi_I{}^B  e_4) (\phi_I{}^C+i\psi_I{}^C  e_4) \} \right] \  \Big] \nonumber \\
& & {} -3 \, f_{ABC} \, \Big[ \ A \leftrightarrow a \ , \ \Phi \leftrightarrow \phi \ , \ \Psi \leftrightarrow \psi \ \Big] \  .
\end{eqnarray}

The first terms of $S_{\Vec{\mathcal{R}}}$ and $S_{\Vec{\mathcal{I}}}$ indicate the matrix Chern-Simons theory individually.
Other terms in $S_{\Vec{\mathcal{R}}}$ and $S_{\Vec{\mathcal{I}}}$ mean the interaction between Chern-Simons fields and off-diagonal fields inside each.
As a whole, $S_{\Vec{\mathcal{R}}}$ and $S_{\Vec{\mathcal{I}}}$ are being based on $(E_6)^{\beta\gamma}$ independently.
Furthermore, $S_{\Vec{int.}}$ signifies the interaction between $\Vec{\mathcal{F_R}}{}^A$ and $\Vec{\mathcal{F_I}}{}^A$.
Therefore, as a consequence of our division, we result in the picture of interacting bi-Chern-Simons type model based on $Sp(4,\Vec{H})/\Vec{Z}_2$.
This result is a little interesting.
The resulting picture resembles the bi-Chern-Simons gravity (See~\cite{Borowiec:2003rd} as an example.).
However, we would like to emphasize that in our case Chern-Simons fields are being coupled with off-diagonal fields spontaneously.
Furthermore, our model is different from the usual bi-Chern-Simons gravity in that it has two Chern-Simons fields which are $\frac{\pi}{2}[rad]$-phase-shifted.
This fact is caused by the $\frac{\pi}{2}[rad]$-phase-shift of pair internal structure of space-time (i.e. fundamental figures $\Vec{\mathcal{F_R}}{}^A$, $i \Vec{\mathcal{F_I}}{}^A$).

\subsection{Author's view on the doubling problem}

So far, we have argued about the interacting $Sp(4,\Vec{H})/\Vec{Z}_2$ {\itshape pair} matrix models.
However, when this is seen from our real physical situation, we must belong to one of the matrix models.
When this is seen from the standpoint of our analysis, another universe needs to be disappearing in our ground state.
On the logic of the mathematics which describes physics, another universe must be treated as the virtual universe.
Of course this is not to say that another universe does not exist.
Although it is completely a fancy talk, if $\frac{\pi}{2}[rad]$-phase-conversion of the macroscopic objects can be performed by some methods, we will see the existence of another universe certainly.
Usually, that existence will contribute only in the quantum effects.
Therefore, we assume that the fields of $\Vec{\mathcal{F_I}}{}^A$ are vanishing in any classical solution.

Under the assumption of $\Vec{\mathcal{F_I}}{}^A=0$ in the ground state, the equations of motion of $S$ are summarized to the following five relations.
\begin{eqnarray}
\left\{
\begin{array}{r}
\epsilon^{IJK} \  \Bigl( \  \frac{1}{2} [ \mathbf{A}_J , \mathbf{A}_K ] + \sum_{\hat{i}=0}^{3} [ \mathbf{\Phi}_{\hat{i}J} , \mathbf{\Phi}_{\hat{i}K} ] - \sum_{\hat{i}=0}^{3} [ \mathbf{\Psi}_{\hat{i}J} , \mathbf{\Psi}_{\hat{i}K} ] \  \Bigr) = 0 \  , \\
\epsilon^{IJK} \  \Bigl( \  2 [ \mathbf{A}_J , \mathbf{\Phi}_{0K} ] + [ \mathbf{\Phi}_{0J} , \mathbf{\Phi}_{0K} ] - \sum_{i=1}^{3} [ \mathbf{\Phi}_{iJ} , \mathbf{\Phi}_{iK} ] + \sum_{\hat{i}=0}^{3} [ \mathbf{\Psi}_{\hat{i}J} , \mathbf{\Psi}_{\hat{i}K} ] \  \Bigr) = 0 \  , \\
\epsilon^{IJK} \ \Bigl( [ \mathbf{A}_J , \mathbf{\Psi}_{0K} ] - [ \mathbf{\Phi}_{0J} , \mathbf{\Psi}_{0K} ] \Bigr) + [ \mathbf{\Psi}_{i(I+1)} , \mathbf{\Phi}_{i(I+2)} ] - [ \mathbf{\Phi}_{i(I+1)} , \mathbf{\Psi}_{i(I+2)} ] + 2 [ \mathbf{\Phi}_{iI} , \mathbf{\Psi}_{iI} ] = 0 \  , \\
\epsilon^{IJK} \Bigl( [ \mathbf{A}_J , \mathbf{\Phi}_{iK} ] - [ \mathbf{\Phi}_{0J} , \mathbf{\Phi}_{iK} ] \Bigr) + [ \mathbf{\Psi}_{i(I+1)} , \mathbf{\Psi}_{0(I+2)} ] - [ \mathbf{\Psi}_{0(I+1)} , \mathbf{\Psi}_{i(I+2)} ] + 2 [ \mathbf{\Psi}_{0I} , \mathbf{\Psi}_{iI} ] \qquad {} \\
{} + \epsilon_{ijk} \Bigl( [ \mathbf{\Phi}_{j(I+1)} , \mathbf{\Phi}_{k(I+2)} ] + [ \mathbf{\Psi}_{j(I+1)} , \mathbf{\Psi}_{k(I+2)} ] - [ \mathbf{\Phi}_{jI} , \mathbf{\Phi}_{kI} ] - [ \mathbf{\Psi}_{jI} , \mathbf{\Psi}_{kI} ] \Bigr) = 0 \  , \\
\epsilon^{IJK} \Bigl( [ \mathbf{A}_J , \mathbf{\Psi}_{iK} ] - [ \mathbf{\Phi}_{0J} , \mathbf{\Psi}_{iK} ] \Bigr) + [ \mathbf{\Phi}_{i(I+1)} , \mathbf{\Psi}_{0(I+2)} ] - [ \mathbf{\Psi}_{0(I+1)} , \mathbf{\Phi}_{i(I+2)} ] +2 [ \mathbf{\Psi}_{0I} , \mathbf{\Phi}_{iI} ] \qquad {} \\
{} - \epsilon_{ijk} \Bigl( [ \mathbf{\Phi}_{j(I+1)} , \mathbf{\Psi}_{k(I+2)}] + [ \mathbf{\Psi}_{j(I+1)} , \mathbf{\Phi}_{k(I+2)} ] - [ \mathbf{\Phi}_{jI} , \mathbf{\Psi}_{kI} ] - [ \mathbf{\Psi}_{jI} , \mathbf{\Phi}_{kI} ] \Bigr) = 0 \  ,
\end{array}
\right.
\end{eqnarray}
where $I,J,K=1,2,3$, \,$i,j,k=1,2,3$. We can regard these relations as the constraints on $\Vec{\mathcal{F_R}}{}^A$ or the equations of motion of $S_{\Vec{\mathcal{R}}}$.
It is often helpful to use the following expressions in the calculation:
\begin{eqnarray}
\left\{
\begin{array}{l}
for \  \  i' = 1,2,3 \  , \  ( j',k'=1,2,3,4,5,6,7 ) \  ; \qquad i \equiv i' \  , \  ( i,j,k=1,2,3 ) \  ; \\
{} \ \quad \sigma_{i'j'k'} \, \varphi_{j'}{}^A \varphi_{k'}{}^B \,=\  \epsilon_{ijk} \left( \, \Phi_{j}{}^A \Phi_{k}{}^B + \Psi_{j}{}^A \Psi_{k}{}^B \, \right) + \Psi_{i}{}^A \Psi_{0}{}^B - \Psi_{0}{}^A \Psi_{i}{}^B \  , \\
for \  \  i' = 4 \  , \  ( j',k'=1,2,3,4,5,6,7 ) \  ; \\
{} \ \quad \sigma_{i'j'k'} \, \varphi_{j'}{}^A \varphi_{k'}{}^B \,=\  \sum_{j=1}^{3} \left( \ i \, \Psi_{j}{}^A \Phi_{j}{}^B - i \, \Phi_{j}{}^A \Psi_{j}{}^B \ \right) \  , \\
for \  \  i' = 5,7 \  , \  ( j',k'=1,2,3,4,5,6,7 ) \  ; \qquad i \equiv i'-4 \  , \  ( i,j,k=1,2,3 ) \  ; \\
{} \ \quad \sigma_{i'j'k'} \, \varphi_{j'}{}^A \varphi_{k'}{}^B \,=\  \epsilon_{ijk} \left( \, -i \, \Phi_{j}{}^A \Psi_{k}{}^B -i \, \Psi_{j}{}^A \Phi_{k}{}^B \, \right) +i \, \Phi_{i}{}^A \Psi_{0}{}^B -i \, \Psi_{0}{}^A \Phi_{i}{}^B \  , \\
for \  \  i' = 6 \  , \  ( j',k'=1,2,3,4,5,6,7 ) \  ; \qquad i \equiv i'-4 \  , \  ( i,j,k=1,2,3 ) \  ; \\
{} \ \quad \sigma_{i'j'k'} \, \varphi_{j'}{}^A \varphi_{k'}{}^B \,=\  \epsilon_{ijk} \left( \, i \, \Phi_{j}{}^A \Psi_{k}{}^B +i \, \Psi_{j}{}^A \Phi_{k}{}^B \, \right) -i \, \Phi_{i}{}^A \Psi_{0}{}^B +i \, \Psi_{0}{}^A \Phi_{i}{}^B \  ,
\end{array}
\right.
\end{eqnarray}
where we are assuming $\phi_{i}{}^A$ and $\psi_{i}{}^A$ are vanishing.

Because the relations of fields were derived, as one toy classical solution, let us perform the same discussion which argued in the previous paper.
Since the procedure is completely the same, it is not argued deeply.
It is the following when only the main point is shown. \\
To begin with, we represent the matrix elements $(\Vec{A}_I)^P{}_Q$, where $P$ stands for the index of `row' and $Q$ stands for the index of `column', of $N \times N$ square matrices $\Vec{A}_I$ as $A_I{}^P_Q$. Then, let us view $\Vec{\mathcal{G}}$ as a product space which is made up of four parts. Accordingly, we can give the one-to-one correspondence between $P,Q$ and $(p_1 p_2 p_3 \tilde{P}),(q_1 q_2 q_3 \tilde{Q})$,
\begin{eqnarray}
A_I{}^P_Q &\equiv& A_I{}^{p_1}_{q_1}{}^{p_2}_{q_2}{}^{p_3}_{q_3}{}^{\tilde{P}}_{\tilde{Q}} \  ,
\end{eqnarray}
where $(p_I,q_I = -L_I,\cdots,0,\cdots,L_I) \, {}_{(I=1,2,3)} \, , \, (\tilde{P},\tilde{Q} = 1,\cdots,M)$, \,so that 
\begin{eqnarray}
N = \bigl( \prod_{I=1}^{3} (2 L_I + 1) \bigr) \  M \  .
\end{eqnarray}
Next, let us focus on one of the solutions of $S_{\Vec{\mathcal{R}}}$, given by 
\begin{eqnarray}
\left\{
    \begin{array}{c}
    A_I{}^{p_1}_{q_1}{}^{p_2}_{q_2}{}^{p_3}_{q_3}{}^{\tilde{P}}_{\tilde{Q}} = P_I{}^{p_1}_{q_1}{}^{p_2}_{q_2}{}^{p_3}_{q_3} \  \delta^{\tilde{P}}_{\tilde{Q}} \\
    P_I{}^{p_1}_{q_1}{}^{p_2}_{q_2}{}^{p_3}_{q_3} = p_I \delta^{p_1}_{q_1} \delta^{p_2}_{q_2} \delta^{p_3}_{q_3}
    \end{array}
\right.
\end{eqnarray}
with other fields vanishing. Now, we expand the theory around this solution,
\begin{eqnarray}
A_I{}^{p_1}_{q_1}{}^{p_2}_{q_2}{}^{p_3}_{q_3}{}^{\tilde{P}}_{\tilde{Q}} &=& P_I{}^{p_1}_{q_1}{}^{p_2}_{q_2}{}^{p_3}_{q_3} \  \delta^{\tilde{P}}_{\tilde{Q}} \  + \  \tilde{A}_I{}^{p_1}_{q_1}{}^{p_2}_{q_2}{}^{p_3}_{q_3}{}^{\tilde{P}}_{\tilde{Q}} \  ,
\end{eqnarray}
and then consider the mapping into the space of functional by using the usual matrix compactification procedure based on the complex Fourier series expansion, 
\begin{eqnarray}
tr_{N \times N} \Bigl( F [ P , G ] \Bigr) &=& \frac{1}{T} \oint \! dt \  tr_{M \times M} \Bigl( F(t) ( -i \frac{\partial G(t)}{\partial t} ) \Bigr) \label{eq:T^1} \  .
\end{eqnarray}
Under these conditions, the action of the theory becomes 
\begin{eqnarray}
S &=& S_{\Vec{\mathcal{R}}} -i \tilde{S}_{\Vec{\mathcal{I}}} + \tilde{S}_{\Vec{int.}} \\
&=& - \frac{3}{2 \Lambda} \oint_{T^3} \! d^3x tr_{M \times M} \Bigl( \epsilon^{IJK} ( \Vec{\tilde{A}}_I \partial_J \Vec{\tilde{A}}_K + \frac{2i}{3} \Vec{\tilde{A}}_I \Vec{\tilde{A}}_J \Vec{\tilde{A}}_K ) + (Couplings\ with\ \Vec{\tilde{\Phi}},\Vec{\tilde{\Psi}}) \Bigr) \nonumber \\
& & {} -i \tilde{S}_{\Vec{\mathcal{I}}} + \tilde{S}_{\Vec{int.}}
\end{eqnarray}
in the $L_I \to \infty$ limits, where $\Lambda=(T_1 T_2 T_3)$, $T_I = l^{(I)} (2L_I+1)$, with $T_I$ held fixed and $l^{(I)} \to 0$. The dimensional scales $l^{(I)}$ are introduced in order to adjust the physical dimensions. \\
Here, the most noteworthy point is that, unlike the previous case, we have succeeded in reducing degrees of freedom by half near the classical limit.

Eventually, the author's opinion about the doubling problem is the following.
\begin{itemize}
  \item The doubling problem of the degrees of freedom is avoidable on the classical level by hypothesizing that we belong to one side of two structures of the universe.
  \item At the classical level, the structure of our universe is completely closed in the theory based on $Sp(4,\Vec{H})/\Vec{Z}_2$.
  \item When we consider the quantum theory, another structure of the universe must also be taken into account and the theory based on the {\itshape compact} $E_6$ must be used.
\end{itemize}

\section{Towards the unknown geometry}\label{geometry}

As mentioned in the previous paper~\cite{Ohwashi:2001mz}, one of the purposes of our study is to pursue the geometry of the universe.
It is known from of old that Jordan type algebras $\Vec{\mathfrak{J}}(n,\Vec{K})$ ($\Vec{K}=\Vec{R}, \Vec{C}, \Vec{H}, \mathfrak{C}$\,), which are $\Vec{K}$-Hermite matrices, can define {\itshape pure} projective spaces (See Appendix~\ref{KP_m}).
Also, it is known from of old that one can proceed the same argument as the projective geometry by using the `{\itshape split-octonions}'.
A good review has been discussed by Catto~\cite{Catto:2003hb}.
In the case of the {\itshape compact} $E_6$, however, the structure is a bit different.
The {\itshape compact} $E_6$ does not possess the structure of usual projective geometry.
However, we might be able to construct a deformed geometry. 
For example, that might be an unknown non-associative geometry deriving from the non-associativity of the octonion.
At least, it is quite likely that it is a kind of non-Desarguesian geometry.
So, we would like to see about the relevance between our model and the geometry a little here from the viewpoint of the `{\itshape matrix model}'.
A more detailed argument will be discussed in elsewhere.

\subsection{Naive consideration}
The idea about which we have argued in this paper was what assumes two fundamental figures $\Vec{\mathcal{F_R}}{}^A$ and $\Vec{\mathcal{F_I}}{}^A$, as internal structures of space-time from `analogy' with the projective geometry.
At first, because the matrix models about these fundamental figures have $Sp(4,\Vec{H})/\Vec{Z}_2$ symmetry, we have an impulse to embed the projective space $\Vec{H}P_3$ into $\Vec{\mathcal{F_R}}{}^A$ (or $\Vec{\mathcal{F_I}}{}^A$) immediately.
However, this is unrealizable with the difference of the trace unfortunately.
\begin{eqnarray}
\mathcal{Y} \left( \Vec{\mathcal{F_R}}{}^A \right) &\in& \Vec{\mathfrak{J}}(4,\Vec{H})_0 = \{ x \in \Vec{\mathfrak{J}}(4,\Vec{H}) \, | \, \  tr(x) = 0 \} \\
\Vec{H}P_3 &=& \{ x \in \Vec{\mathfrak{J}}(4,\Vec{H}) \, | \, \  x^2 = x , \ tr(x) = 1 \}
\end{eqnarray}
Therefore, we have to grope for other possibilities.
First of all, it is more desirable to exist as the geometry of the whole theory which $\mathcal{M}^A \in \Vec{\mathfrak{J}^c}$ including both satisfies, rather than the geometry which $\Vec{\mathcal{F_R}}{}^A$ and $\Vec{\mathcal{F_I}}{}^A$ satisfy separately.
So, it seems better to proceed with our considering along the line of the `analogy' with the projective geometry as before.
Because the projective geometry is powerful, one assumption that `the geometry behind physics is similar to the projective geometry, and $\Vec{\mathcal{F_R}}{}^A$ and $\Vec{\mathcal{F_I}}{}^A$ correspond to {\itshape fundamental figures} in the projective geometry' itself provide us a vague image of the universe.

Now, because $\Vec{\mathcal{F_R}}{}^A$ differs $\frac{\pi}{2}[rad]$ from $\Vec{\mathcal{F_I}}{}^A$ in phase, we can regard them as intersecting orthogonally.
Therefore, these $\Vec{\mathcal{F_R}}{}^A$ and $\Vec{\mathcal{F_I}}{}^A$ do not have a common set originally.
We should not overlook the fact that $\Vec{\mathcal{F_R}}{}^A$ and $\Vec{\mathcal{F_I}}{}^A$ will not be associated without coupling with the Gauge symmetry ($f_{ABC}$ etc.).
In other words, they are related each other by constructing the matrix model using a product space $\mathcal{M}^A \times \Vec{\mathcal{G}}$ for the first time.
This means that the existence of each point itself is the cause of the connection between $\Vec{\mathcal{F_R}}{}^A$ and $\Vec{\mathcal{F_I}}{}^A$, because, in many matrix models, the diagonal parts of $N \times N$ matrices are regarded as space-time points usually.
If it is seen from the viewpoint of the projective geometry, this suggests that each point of the space-time is playing a role of `{\itshape cut}' or `{\itshape center of projection}.'
For example, if we regard each point as a {\itshape cut}, the diagonal of the matrix will correspond to `{\itshape point range}.'
In this case, an image as shown in (Figure~\ref{fig_range.eps}) is obtained.
This is an image which {\itshape global} $E_6$ probably pictures.
Also, if we regard each point as {\itshape center of projection}, we can say that $\Vec{\mathcal{F_R}}{}^A$ and $\Vec{\mathcal{F_I}}{}^A$ are in a kind of `{\itshape perspective relation}' (Figure~\ref{fig_center.eps}).
This is an image which {\itshape local} $E_6$ probably pictures.
%
%
The matrix model which we have considered until now is {\itshape global} $E_6$ matrix models.
However, if possible, the author thinks that {\itshape local} $E_6$ matrix models are more promising.
It is a future subject whether the model of {\itshape local} type can be made.
We will now see the idea of `analogy' with projective geometry a little further in the following subsection.

\subsection{A grope of unknown geometry}
One of the reasons why we believe that the cubic form $( X , Y , Z )$\, is more essential than the trilinear form \,$tr ( X , Y , Z )$\, is because it is a kind of determinant.
Now, as an easiest example, let us naively regard $\mathcal{M}^{A}$, $\mathcal{M}^{B}$ and $\mathcal{M}^{C}$ as the `points' in a certain unknown space, and advance our arguments on the `analogy' with the projective geometry.
Since one line $\ell$ can be defined by deciding two points in the projective geometry, we will assume naively the Freudenthal multiplication defined as a operation between two points to be that line $\ell$:
\begin{eqnarray}
\ell^{BC} &=& \mathcal{M}^{B} \times \mathcal{M}^{C} \  .
\end{eqnarray}
Usually, the `{\itshape incidence}' $(\!(P,\ell)\!)$ of the projective geometry is a certain relation (operation) defined between a point and a line.
If $(\!(P,\ell)\!) \neq 0$, it says that the point $P$ is {\itshape not} on the line $\ell$, and if $(\!(P,\ell)\!) = 0$, it says that the point $P$ is on the line $\ell$.
So, if we put $(\!(P,\ell)\!) \equiv \frac{1}{2} tr(P \ell)+\frac{1}{2} tr(\ell P)$, the {\itshape incidence} $(\!(\mathcal{M}^{A},\ell^{BC})\!)$ is as follows:
\begin{eqnarray}
(\!(\mathcal{M}^{A},\ell^{BC})\!) \  = \  \frac{1}{2} tr(\mathcal{M}^{A} \ell^{BC})+\frac{1}{2} tr(\ell^{BC} \mathcal{M}^{A}) \  = \  tr(\mathcal{M}^{A} \circ \ell^{BC}) \  = \  (\mathcal{M}^{A},\,\mathcal{M}^{B},\,\mathcal{M}^{C}) \  .
\end{eqnarray}
Namely, the {\itshape cubic form} corresponds to the {\itshape incidence}.

\begin{figure}[h]                                                         %
\begin{center}                                                            %
\begin{minipage}{.28\linewidth}                                           %
\includegraphics[width=\linewidth]{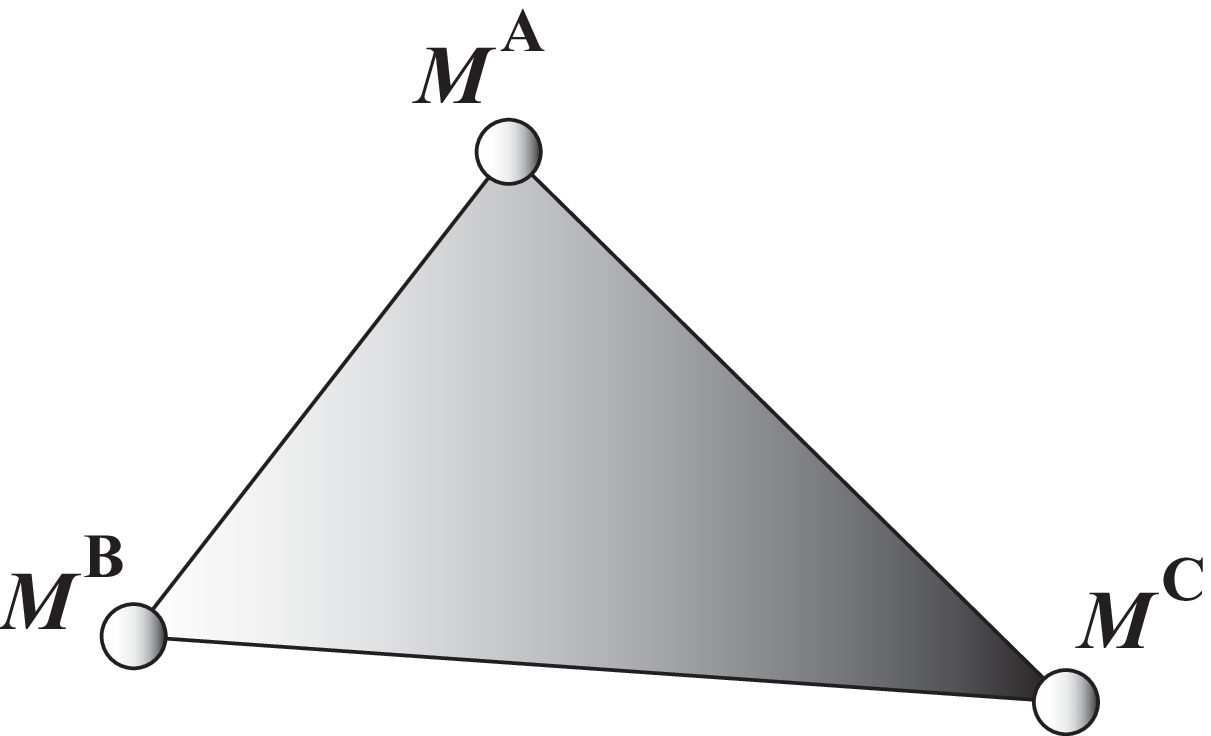}                       %
\caption{Volume}                                                          %
\label{fig_volume1.eps}                                                   %
\end{minipage}                                                            %
\hspace{20mm}                                                             %
\begin{minipage}{.40\linewidth}                                           %
\includegraphics[width=\linewidth]{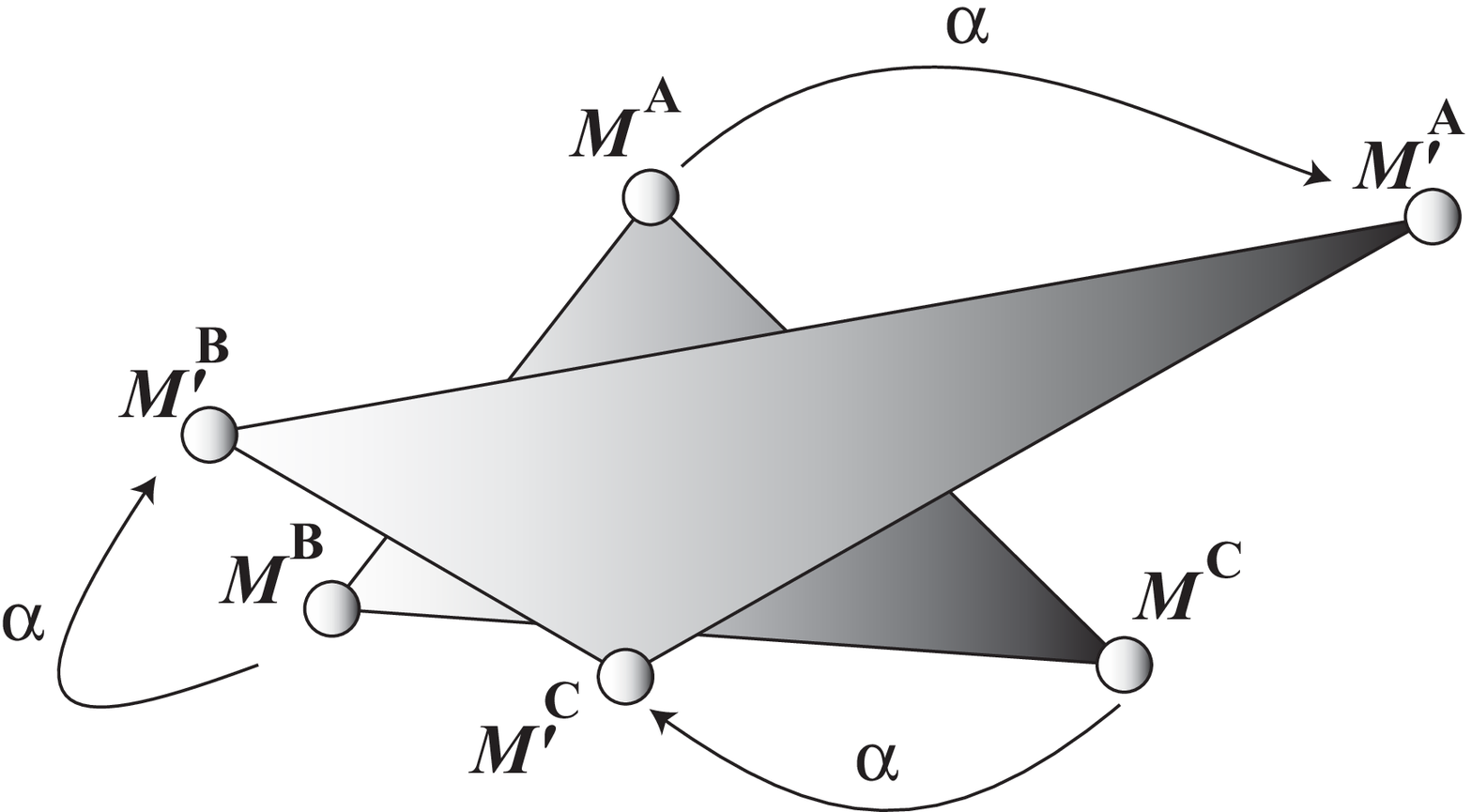}                       %
\caption{Volume-preserving}                                               %
\label{fig_volume2.eps}                                                   %
\end{minipage}                                                            %
\end{center}                                                              %
\end{figure}

Now, how is the situation of $(\!(P,\ell)\!) \neq 0$ and $(\!(P,\ell)\!) = 0$ understood in this case?
Usually, the determinant of the complex exceptional Jordan algebra is defined as follows using the cubic form,
\begin{eqnarray}
det ( X ) &\equiv& \frac{1}{3} ( X , X , X ) \  .
\end{eqnarray}
By the way, usually, the term `determinant' will immediately remind us the scalar density as the invariant volume element of the space.
The universe we consider here is based on $E_6$, and we are assuming that our model is connected with a certain unknown geometry.
So, let us assume that the volume density in that geometry is given by the determinant as usual.
When it does so, if the model is invariant under the $E_6$ mapping, naturally it is desired that the invariant volume element of corresponding geometry to be also invariant under the $E_6$ mapping.
This means that the following relation is concluded with respect to the determinant:
\begin{eqnarray}
det ( \alpha X ) &=& det ( X ) \label{det-1} \  .
\end{eqnarray}
Now, let us represent $X$ as $X = a\mathcal{M}^{A} + b\mathcal{M}^{B} +c\mathcal{M}^{C}$ using arbitrary $a$, $b$ and $c$, and substitute this into (\ref{det-1}) here.
If the coefficient of $abc$ is compared, we result in that above expression (\ref{det-1}) is equivalent to the following expression (\ref{det-2})\footnote{Hence, we can define the {\itshape compact} $E_6$ by $E_6 = \{ \alpha \in Iso_C ( \Vec{\mathfrak{J}^c} ) \, | \, det ( \alpha X ) = det ( X ) , \langle \alpha X , \alpha Y \rangle = \langle X , Y \rangle \}$, instead of (\ref{E6_def1}).}.
\begin{eqnarray}
(\alpha \mathcal{M}^{A}, \alpha \mathcal{M}^{B}, \alpha \mathcal{M}^{C}) = (\mathcal{M}^{A},\mathcal{M}^{B},\mathcal{M}^{C}) \label{det-2}
\end{eqnarray}
Therefore, because (\ref{det-1}) and (\ref{det-2}) include the equivalent information in themselves, if $det(X)$ has the meaning of the volume density in the unknown geometry, it is natural to consider that the cubic form $(\mathcal{M}^{A},\mathcal{M}^{B},\mathcal{M}^{C})$ is also expressing a certain `volume.'
Thus, let us assume naively that the cubic form is representing the area surrounded by the three points.
In this case, (\ref{det-2}) can be interpreted as the expression representing the {\itshape volume-preserving} (Figure~\ref{fig_volume1.eps},\,Figure~\ref{fig_volume2.eps}).
That is, although it moves from each point to another point by $E_6$ mapping, it means that the area surrounded by three points does not change.
%
%
Actually, these are being multiplied by the gauge invariants ($f_{ABC}$ etc.).
That is, $\mathcal{M}^{A} \  (A=1,2,\cdots)$ exist infinitely.
Therefore, we can imagine the single huge network floating in a internal space, as a picture of the universe being made by the {\itshape compact} $E_6$.
And if the idea of decomposing $\mathcal{M}^{A}$ into fundamental figures $\Vec{\mathcal{F_R}}{}^A$, $\Vec{\mathcal{F_I}}{}^A$ which intersect orthogonally is right, this huge polyhedron will possess double coating structure (Figure~\ref{fig_network.eps}).
%
%
Since this is a network of fundamental figures, it is a network of internal structures different from the space-time points.
Furthermore, this network is not a solid.
It is the `living' huge network which can be moved by $E_6$ mapping as shown in (Figure~\ref{fig_volume2.eps}).
Also, this reminds us of a kind of the diffeomorphism.

\begin{figure}[h]                                                         %
\begin{center}                                                            %
\begin{minipage}{.25\linewidth}                                           %
\includegraphics[width=\linewidth]{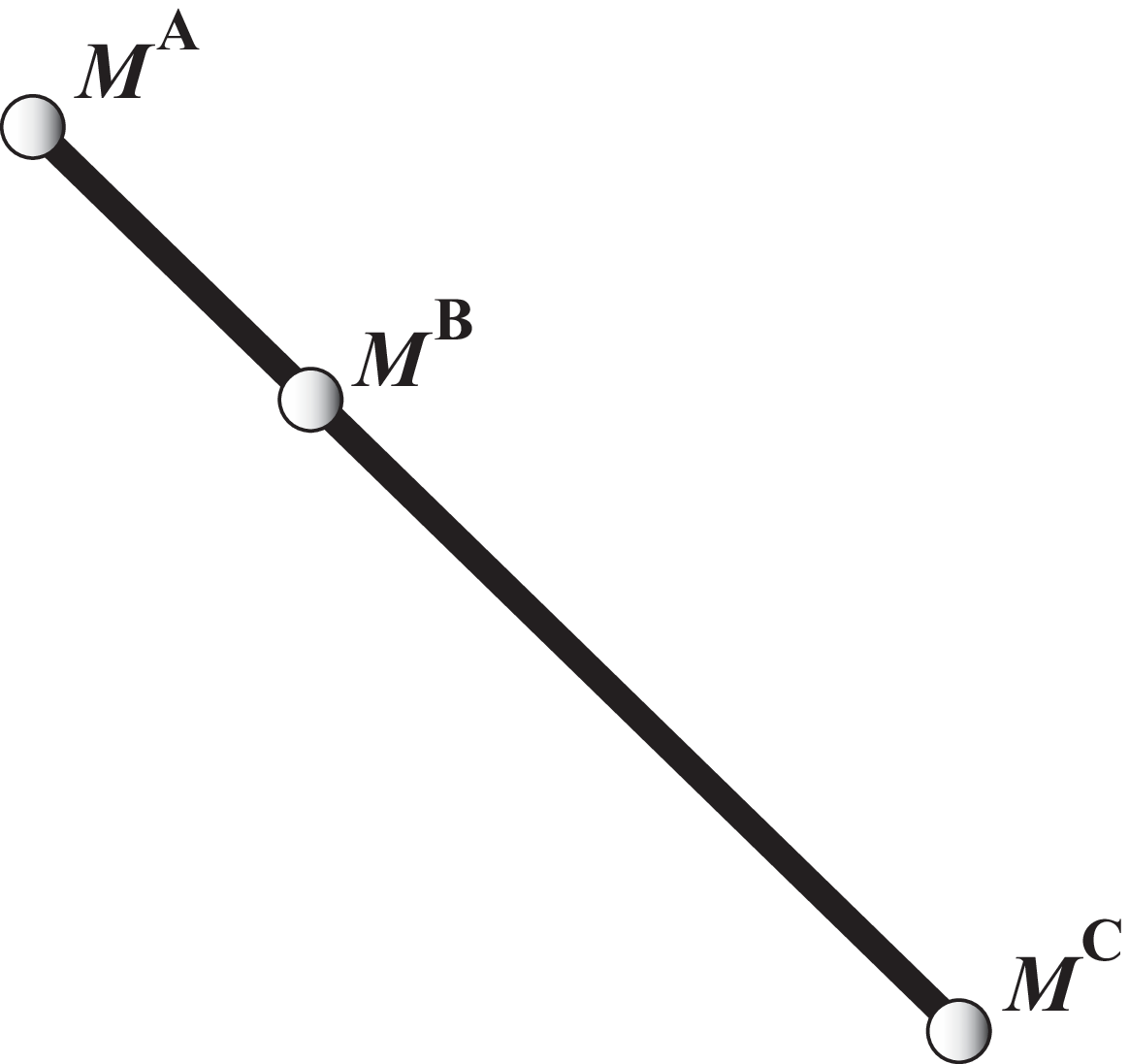}                         %
\caption{Line}                                                            %
\label{fig_line1.eps}                                                     %
\end{minipage}                                                            %
\hspace{25mm}                                                             %
\begin{minipage}{.35\linewidth}                                           %
\includegraphics[width=\linewidth]{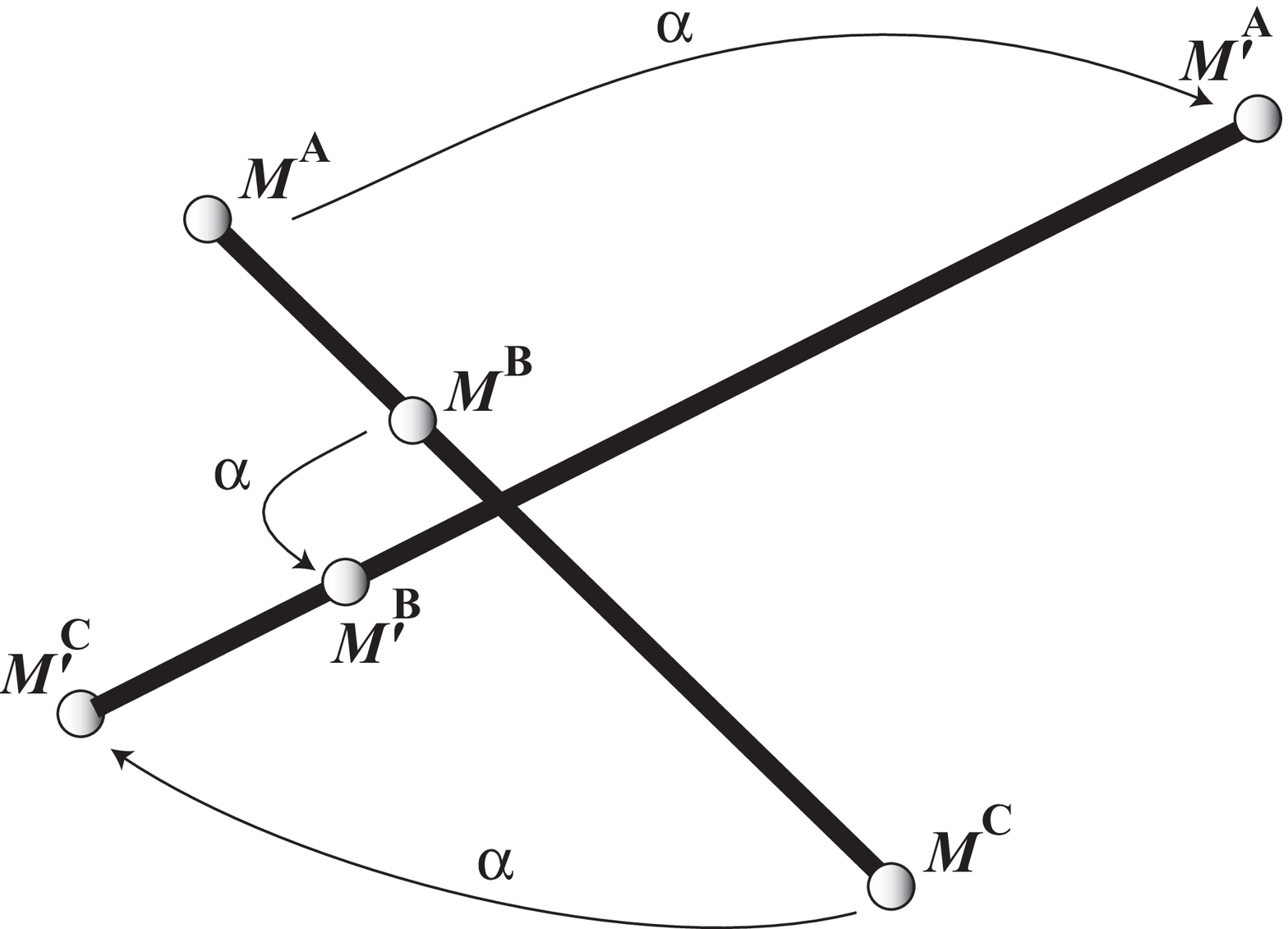}                         %
\caption{Projection of the line}                                          %
\label{fig_line2.eps}                                                     %
\end{minipage}                                                            %
\end{center}                                                              %
\end{figure}

Furthermore, the special interesting case is $(\alpha \mathcal{M}^{A}, \alpha \mathcal{M}^{B}, \alpha \mathcal{M}^{C}) = (\mathcal{M}^{A},\mathcal{M}^{B},\mathcal{M}^{C}) = 0$.
We might be able to call this case as the {\itshape topological} case because of $S=0$.
In this case, a line is mapped into a line because the volume is vanishing (Figure~\ref{fig_line1.eps},\,Figure~\ref{fig_line2.eps}).
%
%
This case corresponds to $(\!(P,\ell)\!) = 0$.
This fact suggests strongly that an unknown geometry which is quite similar to the projective geometry exists behind the cubic form because the geometry which investigates invariant property under the projective transformation is the projective geometry.
Also, this reminds us of the transition of the state vector (i.e. {\itshape ray}) in the quantum mechanics.
Of course for arbitrary $\! A,B,C$,\, if $(\mathcal{M}^{A},\mathcal{M}^{B},\mathcal{M}^{C}) = 0$, it means that all $\mathcal{M}^{A}$ are on the same line such as `beads' (Figure~\ref{fig_beads.eps}).
%
%

The above is one of the reasons to come to believe that a certain projective geometrical correspondence exists behind the model.
The image of the matrix models based on the {\itshape compact} $E_6$ is probably the set of `beads' floating in a internal space, because the Lie algebra $\mathfrak{e}_6$ of $E_6$ can be defined as follows,
\begin{eqnarray}
\mathfrak{e}_6 = \{ \delta \in Hom_C ( \Vec{\mathfrak{J}^c} , \Vec{\mathfrak{J}^c} ) \  | \  \  ( \delta X , X , X ) = 0 \  , \  \langle \delta X , Y \rangle + \langle X , \delta Y \rangle = 0 \} \  .
\end{eqnarray}
There is a possibility that the {\itshape topological} case have a close relation to the Lie algebra $\mathfrak{e}_6$.
The author cannot say having clarified for the moment, although groped about unknown geometry for a long time.
Detailed research is a future subject.

\section{Conclusion and Discussion}

In this paper, we have proposed a solution of the doubling problem which has been left in the previous paper~\cite{Ohwashi:2001mz}.
As a solution, we have constructed interacting $Sp(4,\Vec{H})/\Vec{Z}_2$ {\itshape pair} matrix models inside the {\itshape compact} $E_6$ matrix models.

Under the `analogy' with the projective geometry and {\itshape Klein's Erlangen Program}, we decomposed $\mathcal{M}^A$ into the two fundamental figures $\Vec{\mathcal{F_R}}{}^A$ and $\Vec{\mathcal{F_I}}{}^A$.
By this division, we can decompose the action based on the {\itshape compact} $E_6$ into three parts: $S_{\Vec{\mathcal{R}}}$, $S_{\Vec{\mathcal{I}}}$ which have the completely the same structure of the fields, and $S_{\Vec{int.}}$ which represents the interactions of two fundamental figures.
As a consequence, we result in the picture of interacting {\itshape pair} universe which is being $\frac{\pi}{2}[rad]$-phase-shifted each other.
This picture is applicable to all the models based on the {\itshape compact} $E_6$, which may be not only matrix models but also field theories.
An interacting bi-Chern-Simons model is provided when this result is applied to our previous matrix model~\cite{Ohwashi:2001mz}.
This result is interesting because it resembles the bi-Chern-Simons gravity (See~\cite{Borowiec:2003rd} as an example.).
Finally, the doubling problem of the degrees of freedom is avoidable on the classical level by hypothesizing that we belong to one side of two structures of the universe.
From our viewpoint, the structure of our universe is completely closed in the theory based on $Sp(4,\Vec{H})/\Vec{Z}_2$ at the classical level.
When we consider the quantum theory, another structure of the universe must also be taken into account and the theory based on the {\itshape compact} $E_6$ must be used.
In a sense, the idea itself of dividing into two universes argued in this paper is not a new view.
The view of the parallel-brane world or the multi-brane world discussed in the phenomenology in recent years is a concept just like having argued here.
However, in our case, the arbitrariness of the model is a few in the sense that it is a necessary result of deriving from the group and the structure of its representation space.

Now, the reader might think that: `if so, should not we construct one model based on $Sp(4,\Vec{H})/\Vec{Z}_2$ without constructing the model based on {\itshape compact} $E_6$ from the beginning?'
However, the author thinks that both should be taken into consideration.
The reason is the existence of the {\itshape compactness condition} of $E_6$.
As already argued, the compactness condition of $E_6$ has very good congeniality to {\itshape the postulate of positive definite metric} of the quantum theory.
Especially, It is very precious when we deal with the matrix model.
The author cannot abandon this beautiful property.
Since there is property of the isometry in {\itshape Yokota mapping} $\mathcal{Y}$, to the combination of $\Vec{\mathcal{F_R}}{}^A+i\Vec{\mathcal{F_I}}{}^A$, the compactness condition of $E_6$ is held as it is.
\begin{eqnarray}
invariant_{E_6}&=& < \, \Vec{\mathcal{F_R}}{}^A+i\Vec{\mathcal{F_I}}{}^A \, , \, \Vec{\mathcal{F_R}}{}^B+i\Vec{\mathcal{F_I}}{}^B \, > \\
&=& < \, \mathcal{Y}(\Vec{\mathcal{F_R}}{}^A+i\Vec{\mathcal{F_I}}{}^A) \, , \, \mathcal{Y}(\Vec{\mathcal{F_R}}{}^B+i\Vec{\mathcal{F_I}}{}^B) \, >
\end{eqnarray}
Therefore, the author are thinking seriously that: `although the argument might be able to be closed in only one universe when we consider the classical theory, but we have to take the existence of the two universes into account when we consider the quantum theory.'
However, when we think in this way, it turns out that another possibility also exists.
Let us discuss about this point below.

Since we are considering the model based on the {\itshape compact} $E_6$, of course, the universe as the whole is {\itshape compact}.
However, if that is held, it will be thought that not each universe needs to be compact.
Namely, the following condition should just be maintained at worst.
\begin{itemize}
  \item The probability is conserved inside the whole which united two universes.
\end{itemize}
That is, we think that the probability does not need to be conserved inside one universe itself.
Since $Sp(4,\Vec{H})$ is a compact group, it means that the above condition says that $Sp(4,\Vec{H})$ can be abandoned.
In order to solve the doubling problem, we used the {\itshape Klein's Erlangen Program} as our policy and finally reached the picture of the interacting pair universe.
However, once we reached this picture, we had inconsistency of necessarily not needing {\itshape Klein's Erlangen Program}.
That is, if they constitute the compact $E_6$ as a whole, an interpretation that the two universes do not need possessing the independent transformation group respectively exists.
Let us consider one concrete example.
Although we made a complex-octonion result in four quaternions $\Vec{H}$, the method of dividing this into four split-quaternions $\Vec{H'}$ also exists.
The formal procedure is the same as that of what argued in this paper.
The definition of our split-quaternions $\Vec{H'}$ is the following:
\begin{eqnarray}
\Vec{H'} &=& \Vec{R} \oplus \Vec{R} e'_1 \oplus \Vec{R} e'_2 \oplus \Vec{R} e'_3 \\
&=& \{ r_0 + r_1 e'_1 + r_2 e'_2 + r_3 e'_3 \  | \  r_0,r_1,r_2,r_3 \in \Vec{R} \} \\
&\equiv& \{ r_0 + r_1 (ie_1) + r_2 e_2 + r_3 (ie_3) \  | \  r_0,r_1,r_2,r_3 \in \Vec{R} \} \\
& & \left\{
    \begin{array}{l}
    e'_1 \equiv ie_1   \\
    e'_2 \equiv e_2   \\
    e'_3 \equiv ie_3   \\
    \end{array}
    \right. \\
& & (e'_1)^2=(e'_3)^2=1 \ , \quad (e'_2)^2=-1 \\
& & e'_1e'_2=-e'_2e'_1=e'_3,\ \ e'_2e'_3=-e'_3e'_2=e'_1,\ \ e'_3e'_1=-e'_1e'_3=-e_2 \  .
\end{eqnarray}
We must note that for $x,y \in \Vec{H'}$, even if $x \neq 0, \,y \neq 0$, there is a possibility that $xy=0$.
Also, we must note that $x\tilde{x} \ge 0$ may not be concluded.
When we use $\Vec{H'}$, it is more natural to divide each universe into two portion further.
Therefore, we must deal with four portions of the universe simultaneously.
Although such each universe is quite similar to the AdS-space $O(2,3) \cong Sp(2,\Vec{H'})/\Vec{Z}_2$,\, it is distinct from the AdS-space.
It is because {\itshape compact} $E_6$ should not contain $Sp(4,\Vec{H'})$.
The picture of the universe in this case is apparently strange because although each universe has a structure which is very near the {\itshape negative curvature}, they are compact as a whole due to the compactness condition of $E_6$.
From our viewpoint in this paper, in other words, we can say that: `when considering the classical theory, the argument can be closed inside the open universe, but when considering the quantum theory, we have to use the whole universe which is compact.'
Physically, such picture is extremely interesting.
The further research is one of the most exciting subjects.

As said at the beginning, Einstein gravity is a low-energy effective theory, because it is not perturbatively renormalizable.
If the energy scale is raised, many irrelevant operators ought to contribute.
Therefore, there is no definite promise that the unified theory at the fundamental scale is surely described with the Riemannian geometry itself.
In the author's viewpoint, the mathematics which the universe should possess in the fundamental scale is the geometry similar to the projective geometry.
About this point, a few were discussed in the previous section.
Although the {\itshape pure} projective geometries have been classified mathematically, it is quite likely that we can construct a deformed geometry such as the non-Desarguesian geometry or the non-associative geometry.
From the viewpoint of the projective geometry, the Riemannian geometry should be understood inside the space where the transformation group acts.
Therefore, in the matrix model, a possibility of realizing the diffeomorphism inside the space of matrix-eigenvalues (i.e. the information space) is high.
Because the symmetry which the action of the matrix model has can be seen corresponding to a kind of transformation group, the action itself may be unrelated to diffeomorphism.
Fortunately, since the treatment of the {\itshape information geometry} is possible to the symplectic groups, there is a possibility that we can discuss the metric geometry in the space of matrix-eigenvalues certainly.
Also, it may be interesting to examine the interacting bi-Chern-Simons model.
The grope of the geometry is an important future subject.

Now, what must not be forgotten is that another problem exists.
As has been pointed out, there is a problem how to deal with the fermion in the model using Jordan algebras.
The naive idea which introduces anticommuting c-numbers into the theory is that we use the complex exceptional Jordan {\itshape super}-algebra instead of the complex exceptional Jordan algebra.
However, the Jordan {\itshape super}-algebra is extremely artificial algebra.
The author is still hesitant about taking the plunge into that direction.
The author cannot abandon a concern that our definition of the fermion itself is not probably perfect.
At least, it is quite likely that `the prescription of the formal functional integral to the fermion of integrating with the anticommuting c-numbers is provisional handling and finally the right way of handling exists independently'.
This point will be argued someday.

The consideration which we have so far performed is mainly about $\mathcal{M}^A$.
The technical discussions using the gauge symmetry about the $N \times N$ matrix have not been performed.
Therefore, we can perform various analyses further.
They are future subjects.
We are groping for the new directivity of physics.
Of course it may be unrealistic to expect some big result in such a trial immediately.
However, compared with our previous paper~\cite{Ohwashi:2001mz}, we think that we are moving forward in contents a little.
The author thinks that the essential problem is that we do not have the mathematics which describe our universe properly, rather than the problem of physics itself.
For example, there is a problem of whether the model based on {\itshape compact} $E_6$ includes the information about a certain fundamental object.
As discussed in the previous paper, the model based on $E_6$ can have an effective theory which is similar to the string theory.
However, the author thinks that the fundamental object might not be the string, because the open universe can have a fundamental object which is different from the string.
This point will be discussed on another occasion.

%
%
%
\subsection*{Acknowledgments}

I would like to thank Y.H.~Ohtsuki and I.~Yokota for encouragements. I am also grateful to Y.~Machida for inviting me to The 12th and The 13th Numazu Meeting (March 9-11, 2004 and March, 8-10 2005, Shizuoka, Japan). The existence of that meeting had supported that I continue this research.

\section*{Appendices}
\appendix
In this paper, repeated indices are generally summed, unless otherwise indicated. 
We use the notation `$tr$' for the trace of the Jordan type algebra and `$Tr_{N}$' for the $N \times N$ matrix.
The definitions of $f_{ABC}$ and $d_{ABC}$ are the following,
\begin{eqnarray}
Tr_{N} ( \mathbf{T}_A \mathbf{T}_B ) \, = \, \frac{1}{2} \delta_{AB} \ ,\quad f_{ABC} \, = \, 2 \  Tr_{N} ( \mathbf{T}_A [ \mathbf{T}_B , \mathbf{T}_C ] ) \ ,\quad d_{ABC} \, = \, 2 \  Tr_{N} ( \mathbf{T}_A \{ \mathbf{T}_B , \mathbf{T}_C \} ) \  .
\end{eqnarray}
The conventions are basically the same as our previous paper~\cite{Ohwashi:2001mz}.
\section{Complex Graves-Cayley Algebra $\Vec{\mathfrak{C}^c}$}\label{sec_B}

\subsection{Graves-Cayley algebra $\Vec{\mathfrak{C}}$}
\begin{figure}[htbp]                                                      %
\begin{center}                                                            %
\includegraphics[width=.5\linewidth]{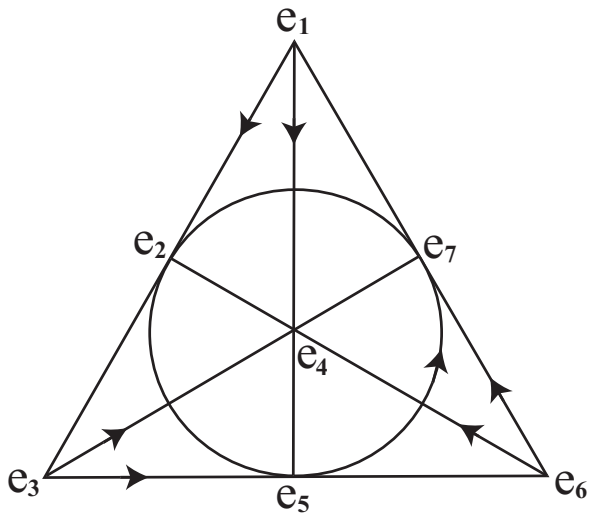}                    %
\end{center}                                                              %
\caption{Multiplication diagram for the octonion.}                        %
\label{fig_octonion.eps}                                                  %
\end{figure}

Let $\Vec{\mathfrak{C}} = \sum_{\tilde{i}=0}^{7} \, \Vec{R} \, e_{\tilde{i}}$ be the Graves-Cayley algebra: \ $\Vec{\mathfrak{C}}$ is an 8-dimensional $\Vec{R}$-vector space with multiplication defined such that $e_0 = 1$ is the unit, \  $e_i{}^2 = -1 \  ( i = 1,\cdots,7 )$, \ $e_i e_j = - e_j e_i \  ( 1 \le i \neq j \le 7 )$ \ and $e_1 e_2 = e_3$, $e_1 e_4 = e_5$, $e_2 e_5 = e_7$, {\itshape etc}. The element `$a$' of $\Vec{\mathfrak{C}}$ is called the {\itshape octonion} (or {\itshape Graves-Cayley number}). The multiplication rule among the bases of $\Vec{\mathfrak{C}}$ can be represented in a diagram as in Fig\,\ref{fig_octonion.eps}.
Thus, if we take $e_1$, $e_2$ and $e_3$, for example, we have 
\begin{equation}
    \begin{array}{c}
    e_1e_2=e_3 \  , \  e_2e_3=e_1 \  , \  e_3e_1=e_2 \  , \\
    e_3e_2=-e_1 \  , \  e_2e_1=-e_3 \  , \  e_1e_3=-e_2 \  .
    \end{array}
\end{equation}
The same is true for the other six lines. What should be noted here is that this algebra is {\itshape non-associative} as well as non-commutative. It is often very useful to introduce the notation 
\begin{eqnarray}
e_ie_j &=& - \delta_{ij} + \sum_{k=1}^{7} \sigma_{ijk} e_k \  , \quad ( i,j,k = 1,\cdots,7 )
\end{eqnarray}
where the $\sigma_{ijk}$ are totally anti-symmetric with respect to their indices, with values $1$, $0$, $-1$\,. \ For instance, \,$\sigma_{ijk} = +1$ \,for \,$ijk = 123, \  356, \  671, \  145, \  347, \  642, \  257$\,.

In $\Vec{\mathfrak{C}}$, \,the conjugate $\bar{a}$ \,and the real part $\Vec{Re} ( a )$ \,are defined respectively as follows:
\begin{eqnarray}
a &\equiv& a_0 + \sum_{i=1}^{7} a_i e_i \  , \\
\vspace{1cm} \nonumber \\
\bar{a} &=& \overline{a_0 + \sum_{i=1}^{7} a_i e_i} \\
&\equiv& a_0 - \sum_{i=1}^{7} a_i e_i \  , \\
\vspace{1cm} \nonumber \\
\Vec{Re} ( a ) &\equiv& \frac{1}{2} ( a + \bar{a} ) \quad\quad \in \  \Vec{R} \\
&=& a_0 \\
&=& \Vec{Re} ( \bar{a} ) \  .
\end{eqnarray}
Moreover, \,the inner product $( a , b ) \  \  ( a,b \in \Vec{\mathfrak{C}} )$ \,is defined by 
\begin{eqnarray}
( a , b ) &\equiv& a_0 b_0 + \sum_{i=1}^{7} a_i b_i \quad \in \  \Vec{R} \\
&=& ( b , a ) \  .
\end{eqnarray}
Therefore, we have 
\begin{eqnarray}
( a , a ) &=& (a_0)^2 + \sum_{i=1}^{7} (a_i)^2 \quad \in \  \Vec{R} \\
&\ge& 0 \  .
\end{eqnarray}
$| a | = \sqrt{( a , a )}$ \,is called the {\itshape length}\,(or {\itshape norm}) of \,$a$.

\subsection{$\Vec{C}$ and $\Vec{H}$ \  in $\Vec{\mathfrak{C}}$}\label{ssec_e4}
\subsubsection{Complex number field in $\Vec{\mathfrak{C}}$}\label{sssec_e4}
The Graves-Cayley algebra $\Vec{\mathfrak{C}}$ contains the field of complex numbers $\Vec{C}$: 
\begin{eqnarray}
\Vec{C} &=& \{ r_0 + r_4 e_4 \  | \  r_0,r_4 \in \Vec{R} \} \  , \\
a &=& a_0 + \sum_{i=1}^{7} a_i e_i \\
&=& (a_0 + a_4 e_4) + (a_1 - a_5 e_4) e_1 + (a_2 + a_6 e_4) e_2 + (a_3 - a_7 e_4) e_3 \\
&=& c_0 + c_1 e_1 + c_2 e_2 + c_3 e_3 \\
& &
\quad\quad\quad\quad\quad c_k \in \Vec{C} \  . \quad (k = 0,1,2,3) \nonumber
\end{eqnarray}
It must be noted that these complex numbers, which have the imaginary unit `$e_4$', are independent of those introduced in the following subsection, whose imaginary unit is \,`$i$'\,.

\subsubsection{Quaternion field in $\Vec{\mathfrak{C}}$}\label{sssec_quaternion}
The Graves-Cayley algebra $\Vec{\mathfrak{C}}$ contains the field of quaternions $\Vec{H}$ as well: 
\begin{eqnarray}
\Vec{H} &=& \{ r_0 + r_1 e_1 + r_2 e_2 + r_3 e_3 \  | \  r_0,r_1,r_2,r_3 \in \Vec{R} \} \  , \\
a &=& a_0 + \sum_{i=1}^{7} a_i e_i \\
&=& (a_0 + a_1 e_1 + a_2 e_2 + a_3 e_3) + (a_4 + a_5 e_1 - a_6 e_2 + a_7 e_3) e_4 \\
&=& q_0 + q_4 e_4 \\
& &
\quad\quad\quad\quad q_k \in \Vec{H} \  . \quad (k = 0,4) \nonumber
\end{eqnarray}

\subsection{Complex Graves-Cayley algebra $\Vec{\mathfrak{C}^c}$}
Let $\Vec{\mathfrak{C}^c}$, called the complex Graves-Cayley algebra, be the complexification of $\Vec{\mathfrak{C}}$: 
\begin{eqnarray}
\Vec{\mathfrak{C}^c} = \{ a + ib \  | \  a,b \in \Vec{\mathfrak{C}} \ , \  i^2=-1 \} \  .
\end{eqnarray}
Here, we should note that \,`$i$' \,is introduced as an imaginary unit distinct from `$e_4$', which is that of the complex number field embedded in $\Vec{\mathfrak{C}}$, as mentioned in the previous subsection. This `$i$' commutes with all the $e_{\tilde{i}} \  ( \tilde{i} = 0,\cdots,7 )$\,.

In $\Vec{\mathfrak{C}^c}$, \,the conjugate $\bar{x}$ \,and the real part $\Vec{Re^c} ( x )$ \,are defined respectively as follows: 
\begin{eqnarray}
x &=& \Bigl( a_0 + \sum_{i=1}^{7} a_i e_i \Bigr) + i \Bigl( b_0 + \sum_{i=1}^{7} b_i e_i \Bigr) \nonumber \\
&=& (a_0 + ib_0) + \sum_{i=1}^{7} (a_i + ib_i) e_i \nonumber \\
&\equiv& x_0 + \sum_{i=1}^{7} x_i e_i \  . \\
\vspace{1cm} \nonumber \\
\bar{x} &\equiv& \bar{a} + i\bar{b} \\
&=& \Bigl( a_0 - \sum_{i=1}^{7} a_i e_i \Bigr) + i \Bigl( b_0 - \sum_{i=1}^{7} b_i e_i \Bigr) \nonumber \\
&=& (a_0 + ib_0) - \sum_{i=1}^{7} (a_i + ib_i) e_i \nonumber \\
&\equiv& x_0 - \sum_{i=1}^{7} x_i e_i \  . \\
\vspace{1cm} \nonumber \\
\Vec{Re^c} ( x ) &\equiv& \frac{1}{2} ( x + \bar{x} ) \quad\quad \in \  \Vec{C} \\
&=& x_0 \\
&=& a_0 + ib_0 \\
&=& \Vec{Re^c} ( \bar{x} ) \  .
\end{eqnarray}
Moreover, \,for any two elements $x = a + ib$ and $y = c + id$ \,of $\Vec{\mathfrak{C}^c}$, \,the inner product $( x , y )$ \,is defined by 
\begin{eqnarray}
( x , y ) &\equiv& x_0 y_0 + \sum_{i=1}^{7} x_i y_i \quad \in \  \Vec{C} \\
&=& (a_0 + ib_0) (c_0 + id_0) + \sum_{i=1}^{7} (a_i + ib_i) (c_i + id_i) \\
&=& ( y , x ) \  .
\end{eqnarray}
Therefore, we have 
\begin{eqnarray}
( x , x ) &=& (x_0)^2 + \sum_{i=1}^{7} (x_i)^2 \quad \in \  \Vec{C} \\
&=& \bar{x} x \  = \  x \bar{x} \  .
\end{eqnarray}
Furthermore, \,in $\Vec{\mathfrak{C}^c}$, \,the hermitian product $\langle x , y \rangle \  \  ( x,y \in \Vec{\mathfrak{C}^c} )$ \,is defined by 
\begin{eqnarray}
\langle x , y \rangle &\equiv& ( x^\ast , y ) \quad \in \  \Vec{C} \\
&=& (a_0 - ib_0) (c_0 + id_0) + \sum_{i=1}^{7} (a_i - ib_i) (c_i + id_i) \  ,
\end{eqnarray}
where $(\cdots)^\ast$, \,called the complex conjugation with respect to `$i$', \,is defined by the mapping
\begin{eqnarray}
( a + ib )^\ast &\equiv& a - ib \  , \\
& &
\quad\quad\quad\quad a,b \in \Vec{\mathfrak{C}} \  . \nonumber
\end{eqnarray}
Therefore we have 
\begin{eqnarray}
x^\ast &=& x_0^\ast + \sum_{i=1}^{7} x_i^\ast e_i \  .
\end{eqnarray}
Naturally, \,we must not confuse this complex conjugation $(\cdots)^\ast$ \,with the octonionic conjugation $\overline{(\cdots)}$\,. \,An example is 
\begin{eqnarray}
\Vec{Re^c} ( x ) = \Vec{Re^c} ( \bar{x} ) \neq \Vec{Re^c} ( x^\ast ) \  .
\end{eqnarray}
Consequently, \,for any element $x = a + ib$ \,of $\Vec{\mathfrak{C}^c}$, \,we have 
\begin{eqnarray}
\langle x , x \rangle &=& \Bigl( (a_0)^2 + \sum_{i=1}^{7} (a_i)^2 \Bigr) + \Bigl( (b_0)^2 + \sum_{i=1}^{7} (b_i)^2 \Bigr) \\
\vspace{1cm} \nonumber \\
&=& ( a , a ) + ( b , b ) \quad \in \  \Vec{R} \\
&\ge& 0 \  .
\end{eqnarray}

\subsection{Some helpful formulas involving elements of $\Vec{\mathfrak{C}^c}$}
We can use the following formulas for any $w,x,y,z \in \Vec{\mathfrak{C}^c}$: 
\begin{eqnarray}
(x^\ast)^\ast &=& x \  , \\
( x + y )^\ast &=& x^\ast + y^\ast \  , \\
(xy)^\ast &=& x^\ast y^\ast \  , \\
\overline{(\bar{x})} &=& x \  , \\
\overline{( x + y )} &=& \bar{x} + \bar{y} \  , \\
\overline{(xy)} &=& \bar{y} \bar{x} \  , \\
( x , y ) &=& \frac{1}{2} ( \bar{x} y + \bar{y} x ) \  = \  \frac{1}{2} ( x \bar{y} + y \bar{x} ) \\
&=& \Vec{Re^c} ( \bar{x} y ) \  = \  \Vec{Re^c} ( x \bar{y} ) \  , \\
( x , y z ) &=& ( y , x \bar{z} ) \  = \  ( z , \bar{y} x ) \  , \\
( x , y ) z &=& \frac{1}{2} \{ \bar{x} ( y z ) + \bar{y} ( x z ) \} \  = \  \frac{1}{2} \{ ( z y ) \bar{x} + ( z x ) \bar{y} \} \  , \\
( w , x )( y , z ) &=& \frac{1}{2} \{ ( w y , x z ) + ( x y , w z ) \} \  = \  \frac{1}{2} \{ ( y w , z x ) + ( y x , z w ) \} \  , \\
\Vec{Re^c} ( x y ) &=& x_0 y_0 - x_i y_i \\
&=& \frac{1}{2} ( xy + \bar{y}\bar{x} ) \  = \  \frac{1}{2} ( \bar{x}\bar{y} + yx ) \\
&=& \Vec{Re^c} ( y x ) \  , \\
\Vec{Re^c} ( x y z ) &\equiv& \Vec{Re^c} ( x (y z) ) \  = \  \Vec{Re^c} ( (x y) z ) \\
&=& x_0 y_0 z_0 - x_0 y_i z_i - x_i y_0 z_i - x_i y_i z_0 - x_i y_j z_k \sigma_{ijk} \\
&=& \frac{1}{2} ( x(yz) + (\bar{z}\bar{y})\bar{x} ) \  = \  \frac{1}{2} ( (xy)z + \bar{z}(\bar{y}\bar{x}) ) \\
&=& \Vec{Re^c} ( y z x ) \  = \  \Vec{Re^c} ( z x y ) \\
&=& \Vec{Re^c} ( z y x ) - 2 x_i y_j z_k \sigma_{ijk} \  , \quad (i,j,k = 1,\cdots,7) \\
( x y ) z &=& \Vec{Re^c} ( x y z ) \nonumber \\
& & {} + \Bigl( \  x_0 y_0 z_l + x_0 y_l z_0 + x_l y_0 z_0 - x_i y_i z_l \nonumber \\
& & {} + x_0 y_i z_j \sigma_{ijl} + x_i y_0 z_j \sigma_{ijl} + x_i y_j z_0 \sigma_{ijl} \nonumber \\
& & {} + x_i y_j z_k \sigma_{ijm} \sigma_{klm} \  \Bigr) e_l \  , \\
x ( y z ) &=& \Vec{Re^c} ( x y z ) \nonumber \\
& & {} + \Bigl( \  x_0 y_0 z_l + x_0 y_l z_0 + x_l y_0 z_0 - x_l y_i z_i \nonumber \\
& & {} + x_0 y_i z_j \sigma_{ijl} + x_i y_0 z_j \sigma_{ijl} + x_i y_j z_0 \sigma_{ijl} \nonumber \\
& & {} - x_i y_j z_k \sigma_{jkm} \sigma_{ilm} \  \Bigr) e_l \  , \\
( x y ) \bar{y} &=& x ( y \bar{y} ) \  = \  ( y \bar{y} ) x \  = \  y ( \bar{y} x ) \  , \\
( x y ) \bar{x} &=& x ( y \bar{x} ) \  , \quad ( x y ) x \  = \  x ( y x ) \  , \\
( x x ) y &=& x ( x y ) \  , \quad x ( y y ) \  = \  ( x y ) y \  , \\
{ [x,y,z] } &\equiv& ( x y ) z - x ( y z ) \\
&=& x_i y_j z_k \  ( \sigma_{ijm} \sigma_{klm} + \sigma_{jkm} \sigma_{ilm} + \delta_{kj} \delta_{il} - \delta_{kl} \delta_{ij} ) \  e_l \\
&=& x_i y_j z_k \  ( \sigma_{ijm} \sigma_{klm} + \sigma_{jkm} \sigma_{ilm} + \sigma_{kim} \sigma_{jlm} ) \  e_l \\
&\equiv& x_i y_j z_k \  ( \rho_{ijkl} ) \  e_l \  , \\
& &
\quad\quad\quad\quad (\mbox{with} \  \rho_{ijkl} \  \mbox{completely anti-symmetric}) \nonumber \\
{ [x,y,z] } &=& { [y,z,x] } \  = \  { [z,x,y] } \\
&=& { -[z,y,x] } \  = \  { -[y,x,z] } \  = \  { -[x,z,y] } \\
&=& { -[x,y,\bar{z}] } \  = \  { [\bar{z},\bar{y},\bar{x}] } \  = \  - \overline{ [x,y,z] } \  , \\
\Vec{Re^c} ( \  { [x,y,z] } \  ) &=& 0 \  , \\
( x y ) z + \overline{x ( y z )} &=& 2 \Vec{Re^c} ( x y z ) + { [x,y,z] } \  , \\
\overline{( x y ) z} + x ( y z ) &=& 2 \Vec{Re^c} ( x y z ) - { [x,y,z] } \  , \\
x ( y z ) + ( y z ) x &=& ( x y ) z + y ( z x ) \  , \\
x ( y z ) + x ( z y ) &=& ( x y ) z + ( x z ) y \  , \\
( x y ) z + ( y x ) z &=& x ( y z ) + y ( x z ) \  .
\end{eqnarray}

\section{Complex Exceptional Jordan Algebra $\Vec{\mathfrak{J}^c}$}\label{sec_C}

\subsection{Exceptional Jordan algebra $\Vec{\mathfrak{J}}$}
We define $\Vec{\mathfrak{J}}$ \,as the exceptional Jordan algebra consisting of all $3 \times 3$ hermitian matrices $A$ \,with entries in the Graves-Cayley algebra $\Vec{\mathfrak{C}}$: 
\begin{eqnarray}
\Vec{\mathfrak{J}} = \{ A \in M(3,\Vec{\mathfrak{C}}) \  | \  A^\ddagger = A \} \  . \quad \bigl( A^\ddagger \equiv (\bar{A})^T \bigr)
\end{eqnarray}
The specific components of $A$ \,can be written as follows: 
\begin{eqnarray}
A
&=&
\left(
      \begin{array}{ccc}
      Q_1 & \phi_3 & \bar{\phi}_2 \\
      \bar{\phi}_3 & Q_2 & \phi_1 \\
      \phi_2 & \bar{\phi}_1 & Q_3
      \end{array}
\right) \  , \\
& &
\quad\quad\quad Q_I \in \Vec{R} \  , \  \  \phi_I \in \Vec{\mathfrak{C}} \  . \quad (I=1,2,3) \nonumber
\end{eqnarray}
Therefore, \,$\Vec{\mathfrak{J}}$ \,is a 27-dimensional $\Vec{R}$-vector space.

\subsection{Complex exceptional Jordan algebra $\Vec{\mathfrak{J}^c}$}
Let $\Vec{\mathfrak{J}^c}$, called the complex exceptional Jordan algebra, be the complexification of $\Vec{\mathfrak{J}}$: 
\begin{eqnarray}
\Vec{\mathfrak{J}^c} = \{ A + iB \  | \  A,B \in \Vec{\mathfrak{J}} \ , \  i^2=-1 \} \  .
\end{eqnarray}
Therefore, the specific components of $X \in \Vec{\mathfrak{J}^c}$ \,can be written as follows: 
\begin{eqnarray}
X
&=&
\left(
      \begin{array}{ccc}
      Q_1 & \phi_3 & \bar{\phi}_2 \\
      \bar{\phi}_3 & Q_2 & \phi_1 \\
      \phi_2 & \bar{\phi}_1 & Q_3
      \end{array}
\right)
+
i
\left(
      \begin{array}{ccc}
      P_1 & \pi_3 & \bar{\pi}_2 \\
      \bar{\pi}_3 & P_2 & \pi_1 \\
      \pi_2 & \bar{\pi}_1 & P_3
      \end{array}
\right) \\
& &
\quad\quad\quad Q_I, P_I \in \Vec{R} \  , \  \  \phi_I, \pi_I \in \Vec{\mathfrak{C}} \  \quad (I=1,2,3) \nonumber \\
&=&
\left(
      \begin{array}{ccc}
      x_1 & \xi_3 & \bar{\xi}_2 \\
      \bar{\xi}_3 & x_2 & \xi_1 \\
      \xi_2 & \bar{\xi}_1 & x_3
      \end{array}
\right) \\
& &
\quad\quad\quad x_I \in \Vec{C} \  , \  \  \xi_I \in \Vec{\mathfrak{C}^c} \  \quad (I=1,2,3) \nonumber \\
&\equiv& X(x,\xi) \\
& &
\qquad\qquad\qquad
\left\{
    \begin{array}{l}
    x_I = Q_I + i P_I \\
    \xi_I = \phi_I + i \pi_I \\
    \bar{\xi}_I = \bar{\phi}_I + i \bar{\pi}_I \  .
    \end{array}
\right. 
\nonumber
\end{eqnarray}
Accordingly, we can also define this $\Vec{\mathfrak{J}^c}$ as 
\begin{eqnarray}
\Vec{\mathfrak{J}^c} = \{ X \in M(3,\Vec{\mathfrak{C}^c}) \  | \  X^\ddagger = X \} \  . \quad \bigl( X^\ddagger \equiv (\bar{X})^T \bigr)
\end{eqnarray}

\subsection{Two kinds of hermitian adjoints}
In $\Vec{\mathfrak{C}^c}$, \,there exist two kinds of conjugation, \ complex conjugation $(\cdots)^\ast$, \,and octonionic conjugation $\overline{(\cdots)}$\,. \,As a result, in $\Vec{\mathfrak{J}^c}$ \,there exist two kinds of hermitian adjoints: 
\begin{eqnarray}
X^\dagger &\equiv& (X^\ast)^T \  , \\
X^\ddagger &\equiv& (\bar{X})^T \  .
\end{eqnarray}

\subsection{Operations}
For any $X,Y,Z \in \Vec{\mathfrak{J}^c}$, \,given by 
\begin{eqnarray}
X
=
\left(
      \begin{array}{ccc}
      x_1 & \xi_3 & \bar{\xi}_2 \\
      \bar{\xi}_3 & x_2 & \xi_1 \\
      \xi_2 & \bar{\xi}_1 & x_3
      \end{array}
\right) \  , \quad
Y
=
\left(
      \begin{array}{ccc}
      y_1 & \eta_3 & \bar{\eta}_2 \\
      \bar{\eta}_3 & y_2 & \eta_1 \\
      \eta_2 & \bar{\eta}_1 & y_3
      \end{array}
\right) \  , \quad
Z
=
\left(
      \begin{array}{ccc}
      z_1 & \zeta_3 & \bar{\zeta}_2 \\
      \bar{\zeta}_3 & z_2 & \zeta_1 \\
      \zeta_2 & \bar{\zeta}_1 & z_3
      \end{array}
\right) \  , \nonumber \\
\end{eqnarray}
various operations are defined as follows.

\subsubsection{Trace \  $tr ( X )$ :}
\begin{eqnarray}
tr ( X ) &\equiv& x_1 + x_2 + x_3 \  .
\end{eqnarray}

\subsubsection{Jordan multiplication \  $X \circ Y$ :}
\begin{eqnarray}
X \circ Y &\equiv& \frac{1}{2} ( XY + YX ) \\
             &=& \frac{1}{2} \{ X , Y \} \\
             &=& Y \circ X \  .
\end{eqnarray}

\subsubsection{Inner product \  $( X , Y ) \  \in \Vec{C}$ :}
\begin{eqnarray}
( X , Y ) &\equiv& tr ( X \circ Y ) \\
          &=& \frac{1}{2} tr ( XY ) + \frac{1}{2} tr ( YX ) \\
          &=& ( Y , X ) \  .
\end{eqnarray}

\subsubsection{Hermitian product \  $\langle X , Y \rangle \  \in \Vec{C}$ :}
\begin{eqnarray}
\langle X , Y \rangle &\equiv& ( X^\ast , Y ) \  . \\
\vspace{1cm} \nonumber \\
\Bigl( \  0 &\le& \langle X , X \rangle \ \  \in \ \Vec{R} \  \Bigr)
\end{eqnarray}

\subsubsection{Freudenthal multiplication \  $X \times Y$ :}
\begin{eqnarray}
X \times Y &\equiv& X \circ Y - \frac{1}{2} tr ( X ) Y - \frac{1}{2} tr ( Y ) X + \frac{1}{2} tr ( X ) tr ( Y ) I - \frac{1}{2} ( X , Y ) I \\
&=& Y \times X \  . \\
& &
\qquad\qquad\qquad\qquad\qquad (\mbox{where} \  I \  \mbox{is the unit matrix}) \nonumber
\end{eqnarray}

\subsubsection{Trilinear form \  $tr ( X , Y , Z ) \  \in \Vec{C}$ :}
\begin{eqnarray}
tr ( X , Y , Z ) &\equiv& ( X , Y \circ Z ) \\
                            &=& tr ( X \circ ( Y \circ Z ) ) \\
&=& \frac{1}{4} tr ( X(YZ) ) + \frac{1}{4} tr ( X(ZY) ) \nonumber \\
& & {} + \frac{1}{4} tr ( (YZ)X ) + \frac{1}{4} tr ( (ZY)X ) \\
&=& \frac{1}{2} ( X,YZ ) + \frac{1}{2} ( X,ZY ) \\
&=& tr ( Y , Z , X ) \  = \  tr ( Z , X , Y ) \\
&=& tr ( Z , Y , X ) \  = \  tr ( Y , X , Z ) \  = \  tr ( X , Z , Y ) \\
&=& ( X \circ Y , Z ) \  .
\end{eqnarray}

\subsubsection{Cubic form \  $( X , Y , Z ) \  \in \Vec{C}$ :}
\begin{eqnarray}
( X , Y , Z ) &\equiv& ( X , Y \times Z ) \\
                            &=& tr ( X \circ ( Y \times Z ) ) \\
&=& tr ( X , Y , Z ) \nonumber \\
& & {} - \frac{1}{2} tr ( X ) \  ( Y , Z ) - \frac{1}{2} tr ( Y ) \  ( Z , X ) - \frac{1}{2} tr ( Z ) \  ( X , Y ) \nonumber \\
& & {} + \frac{1}{2} tr ( X ) \  tr ( Y ) \  tr ( Z ) \\
&=& ( Y , Z , X ) \  = \  ( Z , X , Y ) \\
&=& ( Z , Y , X ) \  = \  ( Y , X , Z ) \  = \  ( X , Z , Y ) \\
&=& ( X \times Y , Z ) \  .
\end{eqnarray}

\subsubsection{Determinant \  $det ( X ) \  \in \Vec{C}$ :}
\begin{eqnarray}
det ( X ) &\equiv& \frac{1}{3} ( X , X , X ) \\
&=& \frac{1}{6} tr ( X(XX) ) + \frac{1}{6} tr ( (XX)X ) - \frac{1}{2} tr ( X^2 ) tr ( X ) + \frac{1}{6} {tr ( X )}^3 \  .
\end{eqnarray}

\subsubsection{Cycle mapping \  $\mathcal{P} ( X )$ :}
\bigskip
For any $X \in \Vec{\mathfrak{J}^c}$, \,given by 
\begin{eqnarray}
X &=&
\left(
      \begin{array}{ccc}
      x_1 & \xi_3 & \bar{\xi}_2 \\
      \bar{\xi}_3 & x_2 & \xi_1 \\
      \xi_2 & \bar{\xi}_1 & x_3
      \end{array}
\right) \  ,
\end{eqnarray}
the cycle mapping $\mathcal{P} ( X )$ \,is defined by 
\begin{eqnarray}
\mathcal{P} ( X )
&\equiv&
\left(
      \begin{array}{ccc}
      x_2 & \xi_1 & \bar{\xi}_3 \\
      \bar{\xi}_1 & x_3 & \xi_2 \\
      \xi_3 & \bar{\xi}_2 & x_1
      \end{array}
\right) \  .
\end{eqnarray}
In other words, the cycle mapping consists of cyclic permutation with respect to the indices $I = 1, 2, 3$\,. \,Therefore, we have 
\begin{eqnarray}
\mathcal{P}^2 ( X )
&=&
\left(
      \begin{array}{ccc}
      x_3 & \xi_2 & \bar{\xi}_1 \\
      \bar{\xi}_2 & x_1 & \xi_3 \\
      \xi_1 & \bar{\xi}_3 & x_2
      \end{array}
\right) \  , \\
& & {} \nonumber \\
\mathcal{P}^3 ( X ) &=& 1 \cdot X \  = \  X \  .
\end{eqnarray}

\subsection{Some helpful formulas involving elements of $\Vec{\mathfrak{J}^c}$}
We can use the following formulas for any $X,Y,Z \in \Vec{\mathfrak{J}^c}$: 
\begin{eqnarray}
I \circ I &=& I \  , \\
I \circ X &=& X \  , \\
I \times I &=& I \  , \\
I \times X &=& \frac{1}{2} ( tr(X) I - X ) \  , \\
(I,I) &=& 3 \  , \\
(X,I) &=& tr(X,I,I) \  = \  (X,I,I) \  = \  tr(X) \  , \\
(X,Y) &=& tr(X,Y,I) \  , \\
(X,YZ) &=& (Y,ZX) \  = \  (Z,XY) \  , \\
tr(X \times Y) &=& \frac{1}{2} tr(X) tr(Y) - \frac{1}{2} (X,Y) \  , \\
X \circ (X \times X) &=& det(X) \  I \  . \\
& &
\qquad\qquad (\mbox{where} \  I \  \mbox{is the unit matrix}) \nonumber
\end{eqnarray}

\subsection{Explicit expressions in terms of components}
We have the following: 
\begin{eqnarray}
X \circ X &=& XX \\
&=&
\left(
      \begin{array}{ccc}
      (x_1)^2+\xi_2\bar{\xi}_2+\xi_3\bar{\xi}_3 & \overline{\xi_1\xi_2}+(x_1+x_2)\xi_3 & \xi_3\xi_1+(x_3+x_1)\bar{\xi}_2 \\
      \xi_1\xi_2+(x_1+x_2)\bar{\xi}_3 & (x_2)^2+\xi_3\bar{\xi}_3+\xi_1\bar{\xi}_1 & \overline{\xi_2\xi_3}+(x_2+x_3)\xi_1 \\
      \overline{\xi_3\xi_1}+(x_3+x_1)\xi_2 & \xi_2\xi_3+(x_2+x_3)\bar{\xi}_1 & (x_3)^2+\xi_1\bar{\xi}_1+\xi_2\bar{\xi}_2
      \end{array}
\right) \  , \nonumber \\
\\
X \times X &=& 
\left(
      \begin{array}{ccc}
      x_2x_3-\xi_1\bar{\xi}_1 & \overline{\xi_1\xi_2}-x_3\xi_3 & \xi_3\xi_1-x_2\bar{\xi}_2 \\
      \xi_1\xi_2-x_3\bar{\xi}_3 & x_3x_1-\xi_2\bar{\xi}_2 & \overline{\xi_2\xi_3}-x_1\xi_1 \\
      \overline{\xi_3\xi_1}-x_2\xi_2 & \xi_2\xi_3-x_1\bar{\xi}_1 & x_1x_2-\xi_3\bar{\xi}_3
      \end{array}
\right) \  , \\
\vspace{1cm} \nonumber \\
tr ( X Y ) &=& \sum_{I=1}^{3} \Bigl( x_I y_I + ( \bar{\xi}_I \eta_I + \xi_I \bar{\eta}_I ) \Bigr) \  , \\
\vspace{1cm} \nonumber \\
tr ( X ( Y Z ) ) &=& \sum_{I=1}^{3} \Bigl( \  x_I y_I z_I + x_I ( (\bar{\eta}_{I+1} \zeta_{I+1}) + (\eta_{I+2} \bar{\zeta}_{I+2}) ) \nonumber \\
& & {} + y_I ( (\xi_{I+1} \bar{\zeta}_{I+1}) + (\bar{\xi}_{I+2} \zeta_{I+2}) ) + z_I ( (\bar{\xi}_{I+1} \eta_{I+1}) + (\xi_{I+2} \bar{\eta}_{I+2}) ) \nonumber \\
& & {} + ( \xi_I (\eta_{I+1} \zeta_{I+2}) + \overline{(\zeta_{I+1} \eta_{I+2}) \xi_I} ) \  \Bigr) \  , \\
\vspace{1cm} \nonumber \\
tr ( ( X Y ) Z ) &=& \sum_{I=1}^{3} \Bigl( \  x_I y_I z_I + x_I ( (\bar{\eta}_{I+1} \zeta_{I+1}) + (\eta_{I+2} \bar{\zeta}_{I+2}) ) \nonumber \\
& & {} + y_I ( (\xi_{I+1} \bar{\zeta}_{I+1}) + (\bar{\xi}_{I+2} \zeta_{I+2}) ) + z_I ( (\bar{\xi}_{I+1} \eta_{I+1}) + (\xi_{I+2} \bar{\eta}_{I+2}) ) \nonumber \\
& & {} + ( (\xi_I \eta_{I+1}) \zeta_{I+2} + \overline{\zeta_{I+1} (\eta_{I+2} \xi_I)} ) \  \Bigr) \  , \\
\vspace{1cm} \nonumber \\
( X , Y ) &=& \sum_{I=1}^{3} \Bigl( x_I y_I + 2( \xi_I , \eta_I ) \Bigr) \  , \\
\vspace{1cm} \nonumber \\
\langle X , Y \rangle &=& \sum_{I=1}^{3} \Bigl( x_I^\ast y_I + 2\langle \xi_I , \eta_I \rangle \Bigr) \  , \\
\vspace{1cm} \nonumber \\
tr ( X , Y , Z ) &=& \sum_{I=1}^{3} \Bigl( \  x_I y_I z_I + x_I ( (\eta_{I+1},\zeta_{I+1}) + (\eta_{I+2},\zeta_{I+2}) ) \nonumber \\
& & {} + y_I ( (\zeta_{I+1},\xi_{I+1}) + (\zeta_{I+2},\xi_{I+2}) ) + z_I ( (\xi_{I+1},\eta_{I+1}) + (\xi_{I+2},\eta_{I+2}) ) \nonumber \\
& & {} + \Vec{Re^c} ( \xi_I \eta_{I+1} \zeta_{I+2} + \xi_I \zeta_{I+1} \eta_{I+2} ) \  \Bigr) \  , \\
\vspace{1cm} \nonumber \\
( X , Y , Z ) &=& \sum_{I=1}^{3} \Bigl( \  \frac{1}{2} ( x_I y_{I+1} z_{I+2} + x_I y_{I+2} z_{I+1} ) - ( x_I (\eta_I,\zeta_I) + y_I (\zeta_I,\xi_I) + z_I (\xi_I,\eta_I) ) \nonumber \\
& & {} + \Vec{Re^c} ( \xi_I \eta_{I+1} \zeta_{I+2} + \xi_I \zeta_{I+1} \eta_{I+2} ) \  \Bigr) \  .
\end{eqnarray}
Here, \,the index $I$ \,is defined $\Vec{mod \  3}$. \ Therefore, we have 
\begin{eqnarray}
det ( X ) &=& x_1 x_2 x_3 - x_1 \xi_1 \bar{\xi}_1 - x_2 \xi_2 \bar{\xi}_2 - x_3 \xi_3 \bar{\xi}_3 + 2 \Vec{Re^c} ( \xi_1 \xi_2 \xi_3 ) \  .
\end{eqnarray}

\section{Complex Quaternion Field $\Vec{H^c}$}

We summarize the complex quaternion field $\Vec{H^c}$ in parallel to the complex Graves-Cayley algebra $\Vec{\mathfrak{C}^c}$ in the Appendix~\ref{sec_B}.

\subsection{Quaternion field $\Vec{H}$}
As mentioned in Appendix~\ref{sssec_quaternion}, the quaternion field $\Vec{H}$ is defined as follows:
\begin{eqnarray}
\Vec{H} &=& \Vec{R} \oplus \Vec{R} e_1 \oplus \Vec{R} e_2 \oplus \Vec{R} e_3 \\
&=& \{ r_0 + r_1 e_1 + r_2 e_2 + r_3 e_3 \  | \  r_0,r_1,r_2,r_3 \in \Vec{R} \} \\
& & (e_1)^2=(e_2)^2=(e_3)^2=-1 \\
& & e_1e_2=-e_2e_1=e_3,\ \ e_2e_3=-e_3e_2=e_1,\ \ e_3e_1=-e_1e_3=e_2 \  .
\end{eqnarray}
This $\Vec{H}$ is a non-commutative field. It is often very useful to introduce the notation 
\begin{eqnarray}
e_ie_j &=& - \delta_{ij} + \sum_{k=1}^{3} \epsilon_{ijk} e_k \  , \quad ( i,j,k = 1,\cdots,3 )
\end{eqnarray}
where the $\epsilon_{ijk} \,(\epsilon_{123}=+1)$ are totally anti-symmetric with respect to their indices, with values $1$, $0$, $-1$\,.

In $\Vec{H}$, \,the conjugate $\tilde{q}$ \,and the length $| q |$ \,are defined respectively as follows:
\begin{eqnarray}
q &=& q_0 + q_1 e_1 + q_2 e_2 + q_3 e_3 \\
\tilde{q} &\equiv& q_0 - q_1 e_1 - q_2 e_2 - q_3 e_3 \  , \\
\vspace{0.3cm} \nonumber \\
|q| &\equiv& \sqrt{(q_0)^2+(q_1)^2+(q_2)^2+(q_3)^2} \\
&=& \sqrt{q\tilde{q}} \ \ = \ \sqrt{\tilde{q}q} \  .
\end{eqnarray}

\subsection{Complex number field in $\Vec{H}$}\label{ssec_e1}
The quaternion field $\Vec{H}$ contains the field of complex numbers $\Vec{C}$: 
\begin{eqnarray}
\Vec{C} &=& \{ r_0 + r_1 e_1 \  | \  r_0,r_1 \in \Vec{R} \} \  , \\
q &=& q_0 + q_1 e_1 + q_2 e_2 + q_3 e_3 \\
&=& (q_0 + q_1 e_1) + (q_2 + q_3 e_1) e_2 \\
&=& c_0 + c_2 e_2 \\
& &
\quad\quad\quad\quad c_k \in \Vec{C} \  . \quad (k = 0,2) \nonumber
\end{eqnarray}
We must not confuse \,`$e_1$' \,with `$i$' and `$e_4$' which were introduced as imaginary units in the Appendix~\ref{sec_B}. 

\subsection{Complex quaternion field $\Vec{H^c}$}
Let $\Vec{H^c}$, called the complex quaternion field, be the complexification of $\Vec{H}$:
\begin{eqnarray}
\Vec{H^c} = \{ q + ip \  | \  q,p \in \Vec{H} \ , \  i^2=-1 \} \  .
\end{eqnarray}
Here, we should note again that \,`$i$' \,is introduced as an imaginary unit distinct from `$e_4$' and `$e_1$', which are those of the complex number fields embedded in $\Vec{\mathfrak{C}}$~(\ref{sssec_e4}) and $\Vec{H}$~(\ref{ssec_e1}). This `$i$' commutes with all the $e_{\tilde{i}} \  ( \tilde{i} = 0,\cdots,7 )$\,.

In $\Vec{H^c}$, \,the conjugate $\tilde{z}$ \,and the length $| z |$ \,are defined respectively as follows: 
\begin{eqnarray}
z &=& \Bigl( q_0 + \sum_{i=1}^{3} q_i e_i \Bigr) + i \Bigl( p_0 + \sum_{i=1}^{3} p_i e_i \Bigr) \nonumber \\
&=& (q_0 + ip_0) + \sum_{i=1}^{3} (q_i + ip_i) e_i \nonumber \\
&\equiv& z_0 + \sum_{i=1}^{3} z_i e_i \  . \\
\vspace{1cm} \nonumber \\
\tilde{z} &\equiv& \tilde{q} + i\tilde{p} \\
&=& \Bigl( q_0 - \sum_{i=1}^{3} q_i e_i \Bigr) + i \Bigl( p_0 - \sum_{i=1}^{3} p_i e_i \Bigr) \nonumber \\
&=& (q_0 + ip_0) - \sum_{i=1}^{3} (q_i + ip_i) e_i \nonumber \\
&\equiv& z_0 - \sum_{i=1}^{3} z_i e_i \  . \\
\vspace{0.3cm} \nonumber \\
|z| &\equiv& \sqrt{|q|^2+|p|^2} \\
&=& \sqrt{(q_0)^2+(q_1)^2+(q_2)^2+(q_3)^2+(p_0)^2+(p_1)^2+(p_2)^2+(p_3)^2} \\
&=& \sqrt{q\tilde{q}+p\tilde{p}} \ \ = \ \sqrt{\tilde{q}q+\tilde{p}p} \\
&=& \sqrt{\frac{1}{2} \Bigl( z^\ast \tilde{z} + z \tilde{z}^\ast \Bigr)} \ \ = \ \sqrt{\frac{1}{2} \Bigl( \tilde{z} z^\ast + \tilde{z}^\ast z \Bigr)} \  ,
\end{eqnarray}
where $(\cdots)^\ast$, \,called the complex conjugation with respect to `$i$', \,is defined by the mapping
\begin{eqnarray}
( q + ip )^\ast &\equiv& q - ip \  , \\
& &
\quad\quad\quad\quad q,p \in \Vec{H} \,(\Vec{\mathfrak{C}}) \  . \nonumber
\end{eqnarray}
Therefore we have 
\begin{eqnarray}
z^\ast &=& z_0^\ast + \sum_{i=1}^{3} z_i^\ast e_i \  .
\end{eqnarray}
We must not confuse this complex conjugation $(\cdots)^\ast$ \,with the quaternionic conjugation $\widetilde{(\cdots)}$ and the octonionic conjugation $\overline{(\cdots)}$\,.

\subsection{Some helpful formulas involving elements of $\Vec{H^c}$}
We can use the following formulas for any $w,z \in \Vec{H^c}$: 
\begin{eqnarray}
(z^\ast)^\ast &=& z \  , \\
( w + z )^\ast &=& w^\ast + z^\ast \  , \\
(wz)^\ast &=& w^\ast z^\ast \  , \\
\widetilde{(\tilde{z})} &=& z \  , \\
\widetilde{( w + z )} &=& \tilde{w} + \tilde{z} \  , \\
\widetilde{(wz)} &=& \tilde{z} \tilde{w} \  , \\
| w + z | &\le& |w| + |z| \  , \\
| wz | &\le& 2 |w| |z| \  , \\
z \, e_4 &=& e_4 \, \tilde{z} \  , \\
\vspace{0.1cm} \nonumber \\
x &=& x_0 + \sum_{i=1}^{7} x_i e_i \ = \ w + z \, e_4 \quad \in \Vec{\mathfrak{C}^c} \\
\bar{x} &=& x_0 - \sum_{i=1}^{7} x_i e_i \ = \ \tilde{w} - z \, e_4 \quad \in \Vec{\mathfrak{C}^c} \\
\vspace{0.1cm} \nonumber \\
{} & &
\left\{
\begin{array}{llll}
w_0=x_0 & w_1=x_1 & w_2=x_2 & w_3=x_3 \\
z_0=x_4 & z_1=x_5 & z_2=-x_6 & z_3=x_7 \ \ . \\
\end{array}
\right. \nonumber
\end{eqnarray}

\section{$n \times n$ Complex Quaternionic Jordan Type Algebra $\Vec{\mathfrak{J}}(n,\Vec{H})\Vec{^c}$}

We summarize the complex quaternionic Jordan type algebra $\Vec{\mathfrak{J}}(n,\Vec{H})\Vec{^c}$ in parallel to the complex exceptional Jordan algebra $\Vec{\mathfrak{J}^c}$ in the Appendix~\ref{sec_C}.

\subsection{Quaternionic Jordan algebra $\Vec{\mathfrak{J}_H}$}
We define $\Vec{\mathfrak{J}_H}$ \,as the quaternionic Jordan algebra consisting of all $3 \times 3$ hermitian matrices $a$ \,with entries in the quaternion field $\Vec{H}$: 
\begin{eqnarray}
\Vec{\mathfrak{J}_H} = \{ a \in M(3,\Vec{H}) \  | \  a^\sharp = a \} \  . \quad \bigl( a^\sharp \equiv (\tilde{a})^T \bigr)
\end{eqnarray}
The specific components of $a$ \,can be written as follows: 
\begin{eqnarray}
a
&=&
\left(
      \begin{array}{ccc}
      Q_1 & \phi_3 & \tilde{\phi}_2 \\
      \tilde{\phi}_3 & Q_2 & \phi_1 \\
      \phi_2 & \tilde{\phi}_1 & Q_3
      \end{array}
\right) \  , \\
& &
\quad\quad\quad Q_I \in \Vec{R} \  , \  \  \phi_I \in \Vec{H} \  . \quad (I=1,2,3) \nonumber
\end{eqnarray}
Therefore, \,$\Vec{\mathfrak{J}_H}$ \,is a 15-dimensional $\Vec{R}$-vector space.

\subsection{Complex quaternionic Jordan algebra $\Vec{\mathfrak{J}_H^c}$}
Let $\Vec{\mathfrak{J}_H^c}$, called the complex quaternionic Jordan algebra, be the complexification of $\Vec{\mathfrak{J}_H}$: 
\begin{eqnarray}
\Vec{\mathfrak{J}_H^c} = \{ a + ib \  | \  a,b \in \Vec{\mathfrak{J}_H} \ , \  i^2=-1 \} \  .
\end{eqnarray}
Therefore, the specific components of $x \in \Vec{\mathfrak{J}_H^c}$ \,can be written as follows: 
\begin{eqnarray}
x
&=&
\left(
      \begin{array}{ccc}
      Q_1 & \phi_3 & \tilde{\phi}_2 \\
      \tilde{\phi}_3 & Q_2 & \phi_1 \\
      \phi_2 & \tilde{\phi}_1 & Q_3
      \end{array}
\right)
+
i
\left(
      \begin{array}{ccc}
      P_1 & \pi_3 & \tilde{\pi}_2 \\
      \tilde{\pi}_3 & P_2 & \pi_1 \\
      \pi_2 & \tilde{\pi}_1 & P_3
      \end{array}
\right) \\
& &
\quad\quad\quad Q_I, P_I \in \Vec{R} \  , \  \  \phi_I, \pi_I \in \Vec{H} \  \quad (I=1,2,3) \nonumber \\
&=&
\left(
      \begin{array}{ccc}
      x_1 & \xi_3 & \tilde{\xi}_2 \\
      \tilde{\xi}_3 & x_2 & \xi_1 \\
      \xi_2 & \tilde{\xi}_1 & x_3
      \end{array}
\right) \\
& &
\quad\quad\quad x_I \in \Vec{C} \  , \  \  \xi_I \in \Vec{H^c} \  \quad (I=1,2,3) \\
& &
\qquad\qquad\qquad
\left\{
    \begin{array}{l}
    x_I = Q_I + i P_I \\
    \xi_I = \phi_I + i \pi_I \\
    \tilde{\xi}_I = \tilde{\phi}_I + i \tilde{\pi}_I \  .
    \end{array}
\right. 
\nonumber
\end{eqnarray}
Accordingly, we can also define this $\Vec{\mathfrak{J}_H^c}$ as 
\begin{eqnarray}
\Vec{\mathfrak{J}_H^c} = \{ x \in M(3,\Vec{H^c}) \  | \  x^\sharp = x \} \  . \quad \bigl( x^\sharp \equiv (\tilde{x})^T \bigr)
\end{eqnarray}

\subsection{$n \times n$ quaternionic Jordan type algebra $\Vec{\mathfrak{J}}(n,\Vec{H})$}
Moreover, we define $\Vec{\mathfrak{J}}(n,\Vec{H})$ \,as the quaternionic Jordan type algebra consisting of all $n \times n$ hermitian matrices $a$ \,with entries in the quaternion field $\Vec{H}$: 
\begin{eqnarray}
\Vec{\mathfrak{J}}(n,\Vec{H}) = \{ a \in M(n,\Vec{H}) \  | \  a^\sharp = a \} \  . \quad \bigl( a^\sharp \equiv (\tilde{a})^T \bigr)
\end{eqnarray}
Therefore, \,$\Vec{\mathfrak{J}}(n,\Vec{H})$ \,is a $(2n^2-n)$-dimensional $\Vec{R}$-vector space. \\
Of course,
\begin{eqnarray}
\Vec{\mathfrak{J}_H} &=& \Vec{\mathfrak{J}}(3,\Vec{H}) \  .
\end{eqnarray}

\subsection{$n \times n$ complex quaternionic Jordan type algebra $\Vec{\mathfrak{J}}(n,\Vec{H})\Vec{^c}$}
Let $\Vec{\mathfrak{J}}(n,\Vec{H})\Vec{^c}$, called the complex quaternionic Jordan type algebra, be the complexification of $\Vec{\mathfrak{J}}(n,\Vec{H})$: 
\begin{eqnarray}
\Vec{\mathfrak{J}}(n,\Vec{H})\Vec{^c} = \{ a + ib \  | \  a,b \in \Vec{\mathfrak{J}}(n,\Vec{H}) \ , \  i^2=-1 \} \  .
\end{eqnarray}
Accordingly, we can also define this $\Vec{\mathfrak{J}}(n,\Vec{H})\Vec{^c}$ as 
\begin{eqnarray}
\Vec{\mathfrak{J}}(n,\Vec{H})\Vec{^c} &=& \{ x \in M(n,\Vec{H^c}) \  | \  x^\sharp = x \} \  \quad \bigl( x^\sharp \equiv (\tilde{x})^T \bigr) \\
&\stackrel{\mathrm{def}}{\equiv}& \Vec{\mathfrak{J}}(n,\Vec{H^c}) \  .
\end{eqnarray}
Therefore,
\begin{eqnarray}
\Vec{\mathfrak{J}_H^c} &=& \Vec{\mathfrak{J}}(3,\Vec{H})\Vec{^c} \ \  = \ \  \Vec{\mathfrak{J}}(3,\Vec{H^c}) \  .
\end{eqnarray}

\subsection{$\Vec{\mathfrak{J}}(n,\Vec{H})_0$ \, and \, $\Vec{\mathfrak{J}}(n,\Vec{H})_0\Vec{^c}$ \quad --- Traceless subspace --- }
Let $\Vec{\mathfrak{J}}(n,\Vec{H})_0$ be the vector space of all $ a \in \Vec{\mathfrak{J}}(n,\Vec{H}) $ such that $tr(a) = 0$~: 
\begin{eqnarray}
\Vec{\mathfrak{J}}(n,\Vec{H})_0 &=& \{ a \in \Vec{\mathfrak{J}}(n,\Vec{H}) \, | \, \  tr(a) = 0 \} \  ,
\end{eqnarray}
and let $\Vec{\mathfrak{J}}(n,\Vec{H})_0\Vec{^c}$ be the complexification of $\Vec{\mathfrak{J}}(n,\Vec{H})_0$ as usual: 
\begin{eqnarray}
\Vec{\mathfrak{J}}(n,\Vec{H})_0\Vec{^c} &=& \{ a + ib \  | \  a,b \in \Vec{\mathfrak{J}}(n,\Vec{H})_0 \ , \  i^2=-1 \} \  .
\end{eqnarray}
Accordingly, we can also define this $\Vec{\mathfrak{J}}(n,\Vec{H})_0\Vec{^c}$ as 
\begin{eqnarray}
\Vec{\mathfrak{J}}(n,\Vec{H})_0\Vec{^c} &=& \{ x \in \Vec{\mathfrak{J}}(n,\Vec{H})\Vec{^c} \, | \, \  tr(x) = 0 \} \\
&=& \{ x \in \Vec{\mathfrak{J}}(n,\Vec{H^c}) \, | \, \  tr(x) = 0 \} \\
&\stackrel{\mathrm{def}}{\equiv}& \Vec{\mathfrak{J}}(n,\Vec{H^c})_0 \  .
\end{eqnarray}

\subsection{Two kinds of hermitian adjoints}
In $\Vec{H^c}$, \,there exist two kinds of conjugation, \ complex conjugation $(\cdots)^\ast$, \,and quaternionic conjugation $\widetilde{(\cdots)}$\,. \,As a result, in $\Vec{\mathfrak{J}}(n,\Vec{H^c})$ \,there exist two kinds of hermitian adjoints: 
\begin{eqnarray}
x^\dagger &\equiv& (x^\ast)^T \  , \\
x^\sharp &\equiv& (\tilde{x})^T \  .
\end{eqnarray}

\subsection{Operations}
For any $x,y \in \Vec{\mathfrak{J}}(n,\Vec{H^c})$, \,operations are defined as usual.

\subsubsection{Jordan multiplication \  $x \circ y$ :}
\begin{eqnarray}
x \circ y &\equiv& \frac{1}{2} ( xy + yx ) \\
             &=& \frac{1}{2} \{ x , y \} \\
             &=& y \circ x \  .
\end{eqnarray}

\subsubsection{Inner product \  $( x , y ) \  \in \Vec{C}$ :}
\begin{eqnarray}
( x , y ) &\equiv& tr ( x \circ y ) \\
          &=& \frac{1}{2} tr ( xy ) + \frac{1}{2} tr ( yx ) \\
          &=& ( y , x ) \  .
\end{eqnarray}

\subsubsection{Hermitian product \  $\langle x , y \rangle \  \in \Vec{C}$ :}
\begin{eqnarray}
\langle x , y \rangle &\equiv& ( x^\ast , y ) \  . \\
\vspace{1cm} \nonumber \\
\Bigl( \  0 &\le& \langle x , x \rangle \ \  \in \ \Vec{R} \  \Bigr)
\end{eqnarray}

\section{Projective Spaces $\Vec{K}P_m$}\label{KP_m}

It is known from of old that Jordan type algebras $\Vec{\mathfrak{J}}(n,\Vec{K})$ ($\Vec{K}=\Vec{R}, \Vec{C}, \Vec{H}, \mathfrak{C}$\,), which are $\Vec{K}$-Hermite matrices, can define ({\itshape pure}-)projective spaces. Here, we summarize the definition of each projective space together.

\subsection{Projective lines $\Vec{K}P_1$}
The projective line $\Vec{K}P_1$ over $\Vec{K}=\Vec{R}, \Vec{C}, \Vec{H}, \mathfrak{C}$ \,is a set defined as follows:
\begin{eqnarray}
\Vec{K}P_1 &=& \{ A \in \Vec{\mathfrak{J}}(2,\Vec{K}) \, | \, \  A^2 = A , \ tr(A) = 1 \} \  .
\end{eqnarray}

\subsection{Projective planes $\Vec{K}P_2$}
The projective plane $\Vec{K}P_2$ over $\Vec{K}=\Vec{R}, \Vec{C}, \Vec{H}, \mathfrak{C}$ \,is a set defined as follows:
\begin{eqnarray}
\Vec{K}P_2 &=& \{ A \in \Vec{\mathfrak{J}}(3,\Vec{K}) \, | \, \  A^2 = A , \ tr(A) = 1 \} \  .
\end{eqnarray}

\subsection{Projective spaces $\Vec{K}P_m$ \ ($m \ge 3$)}
The projective space $\Vec{K}P_m$ ($m \ge 3$) over $\Vec{K}=\Vec{R}, \Vec{C}, \Vec{H}$ \,is a set defined as follows:
\begin{eqnarray}
\Vec{K}P_m &=& \{ A \in \Vec{\mathfrak{J}}(m+1,\Vec{K}) \, | \, \  A^2 = A , \ tr(A) = 1 \} \  .
\end{eqnarray}
Here, the reason we do not include $\Vec{K}=\mathfrak{C}$ \,is that the Desargues' Theorem is not met as known well, because $\mathfrak{C}$ does not satisfy the associative law. Therefore, there is no projective space which contains $\mathfrak{C}P_2$ as far as we obey the viewpoint of normal {\itshape pure} projective geometry.

\subsection{Comparison with other definition}
Usually, a different definition about the projective spaces over $\Vec{K}=\Vec{R}, \Vec{C}, \Vec{H}$ \,is used. In general, we define `the identification space $(\Vec{K}^{m+1}-\{0\})/\sim$ \,by the equivalence relation $\Vec{a} \sim \Vec{a} \lambda \ (\lambda \in \Vec{K})$ \,in $\Vec{K}^{m+1} - \{0\} \ (m \ge 0)$\,' \ as \,`$\Vec{K}P^m$.' \ The relation between these two different definitions can be understood by defining the following map $f:\,\Vec{K}^{m+1} - \{0\} \,\to\, \Vec{K}P_m$\ ,
\begin{eqnarray}
f(\Vec{a})=f
\left(
\begin{array}{c}
a_1 \\
\vdots \\
a_{m+1}
\end{array}
\right)
 \ = \ 
\frac{1}{\sum_{k=1}^{m+1} |a_k|^2}
\left(
\begin{array}{cccc}
a_1 \tilde{a}_1 & a_1 \tilde{a}_2 & \cdots & a_1 \tilde{a}_{m+1} \\
a_2 \tilde{a}_1 & a_2 \tilde{a}_2 & \cdots & a_2 \tilde{a}_{m+1} \\
\vdots & \vdots & \ddots & \vdots \\
a_{m+1} \tilde{a}_1 & a_{m+1} \tilde{a}_2 & \cdots & a_{m+1} \tilde{a}_{m+1} \\
\end{array}
\right) \  ,
\end{eqnarray}
where $\tilde{a}_i$ denotes the conjugate of $a_i \in \Vec{K}$. \,This map $f$ induces a bijection $\bar{f}:\ \Vec{K}P^m \,\to\, \Vec{K}P_m$ . Therefore, \,$\Vec{K}P^m$ is isomorphic to $\Vec{K}P_m$\,.

\section{Symplectic Group $Sp(4,\Vec{H})$}\label{Sp_4}

The symplectic group $Sp(4,\Vec{H})$ is defined as follows:
\begin{eqnarray}
Sp(4,\Vec{H}) &=& \{ V \in M(4,\Vec{H}) \, | \, \  VV^{\sharp} = \Vec{I} \} \  .
\end{eqnarray}
This group $Sp(4,\Vec{H})$ acts on $\Vec{\mathfrak{J}}(4,\Vec{H})$ by the mapping $f:\,Sp(4,\Vec{H}) \times \Vec{\mathfrak{J}}(4,\Vec{H}) \,\to\, \Vec{\mathfrak{J}}(4,\Vec{H})$, 
\begin{eqnarray}
f(V,A) &=& VAV^{\sharp} \  ,
\end{eqnarray}
where $A \in \Vec{\mathfrak{J}}(4,\Vec{H})$.
Therefore, this action induces an automorphism of $\Vec{\mathfrak{J}}(4,\Vec{H})$ and an isometry of $\Vec{\mathfrak{J}}(4,\Vec{H})$ (and $\Vec{\mathfrak{J}}(4,\Vec{H})_0$):
\begin{eqnarray}
V(A_1 \circ A_2)V^{\sharp} &=& V A_1 V^{\sharp} \circ V A_2 V^{\sharp} \  , \\
( V A_1 V^{\sharp} , V A_2 V^{\sharp} ) &=& (A_1 , A_2) \  .
\end{eqnarray}

These relations can be naturally extended to \Vec{\mathfrak{J}}(4,\Vec{H})\Vec{^c}. The group $Sp(4,\Vec{H})$ acts on $\Vec{\mathfrak{J}}(4,\Vec{H^c})$ by the mapping $g:\,Sp(4,\Vec{H}) \times \Vec{\mathfrak{J}}(4,\Vec{H^c}) \,\to\, \Vec{\mathfrak{J}}(4,\Vec{H^c})$, 
\begin{eqnarray}
g(V,B) &=& VBV^{\sharp} \  ,
\end{eqnarray}
where $B \in \Vec{\mathfrak{J}}(4,\Vec{H^c})$.
Therefore,
\begin{eqnarray}
V(B_1 \circ B_2)V^{\sharp} &=& V B_1 V^{\sharp} \circ V B_2 V^{\sharp} \  , \\
( V B_1 V^{\sharp} , V B_2 V^{\sharp} ) &=& (B_1 , B_2) \  , \\
< V B_1 V^{\sharp} , V B_2 V^{\sharp} > &=& <B_1 , B_2> \  .
\end{eqnarray}

Furthermore, we can define $\Vec{H}P_3$\, using this $Sp(4,\Vec{H})$:
\begin{eqnarray}
\Vec{H}P_3 &=& \{ x \in \Vec{\mathfrak{J}}(4,\Vec{H}) \, | \, \  x^2 = x , \ tr(x) = 1 \} \\
&=& \{ \, VI_{1}V^{\sharp} \  | \, \  V \in Sp(4,\Vec{H}) \} \  ,
\end{eqnarray}
where $I_{1}=\left(
  \begin{array}{cccc}
  1  &  0  &  0  &  0  \\
  0  &  0  &  0  &  0  \\
  0  &  0  &  0  &  0  \\
  0  &  0  &  0  &  0  \\
  \end{array}
\right)
$.

%
%
%

%
%
%
\end{document}